\documentclass[aps,pre,superscriptaddress,citeautoscript, amsmath,amssymb,twocolumn]{revtex4-1}
\usepackage{graphicx}
\usepackage{amsmath}
\usepackage{hyperref}
\usepackage{color}
\usepackage{bigints}
\usepackage{mathrsfs}
\usepackage{algorithm}
\usepackage{algpseudocode}
\definecolor{darkblue}{RGB}{8,81,156}
\hypersetup{colorlinks,breaklinks,
            linkcolor=darkblue,urlcolor=darkblue,
           anchorcolor=darkblue,citecolor=darkblue}

\date{\today}

\allowdisplaybreaks

    \definecolor{dark-purple}{RGB}{118,42,131}
    \definecolor{dark-green}{RGB}{27,120,55}
    \definecolor{light-purple}{RGB}{231,212,232}
    \definecolor{LIGHT-PURPLE}{RGB}{194,165,207}
    \definecolor{light-green}{RGB}{168,216,183}
    \definecolor{gray}{RGB}{186,186,186}
    \definecolor{super-dark-green}{RGB}{0,69,41}
    \definecolor{super-dark-purple}{RGB}{63,0,125}
    \definecolor{super-dark-blue}{RGB}{8,48,107}
    \definecolor{super-dark-red}{RGB}{165,0,38}
    
    \definecolor{super-dark-purple}{RGB}{64,0,75}
    \definecolor{super-dark-green}{RGB}{0,68,27}

\usepackage{tabularx,booktabs} 

\usepackage{array}
\usepackage{longtable}
\newcolumntype{L}[1]{>{\raggedright\let\newline\\\arraybackslash\hspace{0pt}}p{#1}}
\newcolumntype{C}[1]{>{\centering\let\newline\\\arraybackslash\hspace{0pt}}m{#1}}
\newcolumntype{R}[1]{>{\raggedleft\let\newline\\\arraybackslash\hspace{0pt}}m{#1}}

\setcitestyle{super}

\begin{document}

\title{Studying Rare Events using Forward-Flux Sampling: Recent Breakthroughs and Future Outlook}

\author{Sarwar Hussain, Amir Haji-Akbari}
\email{amir.hajiakbaribalou@yale.edu}
\affiliation{Department of Chemical and Environmental Engineering, Yale University, New Haven, CT  06520}

\date{\today}

\begin{abstract}
\noindent
Rare events are processes that occur upon the emergence of unlikely fluctuations. Unlike what their name suggests, rare events are  fairly ubiquitous in nature, as the occurrence of many structural transformations in biology and material sciences is predicated upon crossing large free energy barriers. Probing the kinetics and uncovering the molecular mechanisms of possible barrier crossings in a system is critical to predicting and controlling its structural and functional properties. Due to their activated nature, however, rare events are exceptionally difficult to study using conventional experimental and computational techniques. In recent decades, a wide variety of specialized computational techniques-- known as advanced sampling techniques-- have been developed to systematically capture improbable fluctuations relevant to rare events.  In this perspective, we focus on a technique called forward flux sampling (Allen~\emph{et al.,}~\emph{J. Chem. Phys.}, {\bf 124}: 024102, 2006), and overview its recent methodological variants and extensions. We also provide a detailed overview of its application to study a wide variety of rare events, and map out potential avenues for further explorations.  
\end{abstract}
\maketitle


\section{Introduction}
\label{section:intro}

\noindent
Rare events are an important-- and ubiquitous-- class of phenomena that occur in both macroscopic and molecular systems, ranging from extreme weather events~\cite{FreiJClimate2001}, earthquakes~\cite{GabrielovGeophysJInt2000}, social unrest~\cite{ComptonSecurInformatics2014}, stock market crashes~\cite{GoodwinTechnolForecastSoc2010}, and electric grid failures~\cite{ChassinPhysicaA2005} in the macroscopic realm to crystal nucleation~\cite{OxtobyJPhysCondenMatter1992}, protein folding~\cite{OnuchicAnnRevPhysChem1997, OnuchicCurrOpinStrucBiol2004} and aggregation~\cite{RossNatMed2004}, and DNA hybridization~\cite{PorschkeJMolBiol1971} in the realm of atoms and molecules. What unifies all these disparate processes is that their occurrence involves infrequent-- but swift-- changes that are caused by intrinsic and infrequent fluctuations in the corresponding system.  Due to the rarity  of such fluctuations, a wide separation of timescales emerges between the time needed for the completion of the actual event and the wait time between consecutive events. This, in turn, makes the task of efficiently probing the kinetics of rare events  extremely challenging both experimentally and computationally. On one hand, most  experimental techniques lack the spatiotemporal resolution needed for capturing the highly localized and swift fluctuations relevant to rare events, and even though efforts are underway to develop high-resolution ultrafast imaging and microscopy techniques~\cite{GaffneyScience2007, PlemmonsChemMater2015, AdhikariACSApplMaterInter2017}, experiments are still not fully capable of  realtime monitoring of many microscopic rare events. Conventional computational techniques  such as molecular dynamics (MD) \cite{AlderMDJCP1959} or Monte Carlo (MC) \cite{Metropolis1953}, on the other hand, are well-suited for monitoring a rare event during its occurrence, but can be too inefficient to capture the waiting time that elapses prior to its occurrence.  As a result, specialized advanced sampling techniques have been developed over the years to conduct a targeted sampling of the infrequent fluctuations that are relevant for the occurrence and completion of rare events.

In atomic and molecular systems, a rare event typically involves a transition between  two minima in the free energy landscape that are separated by a free energy barrier. The regions of the configuration space that are within the basins of attraction of these two minima will be denoted by  $A$ and $B$ in the remainder of this paper. When it comes to a transition from $A$ to $B$, three pieces of information are usually of interest:  (i) the transition rate, $k_{AB}$, or the average number of transitions	 per unit time, (ii) the free energy landscape, and more specifically the free energy barrier that separate $A$ and $B$, and (iii) the transition mechanism. In general, advanced sampling techniques can provide some or all of these information, and are broadly classified  into two categories based on their ability to directly estimate $k_{AB}$. Some advanced sampling techniques, such as umbrella sampling~\cite{TorrieJCompPhys1977}, the Wang-Landau method~\cite{Wang_Letter_2001}, adaptive biasing force~\cite{DarveJCP2001},  metadynamics~\cite{LaioPNAS2002} and the string method~\cite{MaraglianoJChemPhys2006} are based on applying a biasing potential along a pre-specified set of collective variables in order to access unlikely regions of the configuration space. The free energy landscape within the collective variable space is then mapped out through proper reweighing approaches, such as the weighted histogram analysis method (WHAM)~\cite{KumarWHAM1992}. Bias-based techniques alter the intrinsic dynamics of the system, and can therefore provide indirect estimates of $k_{AB}$ at best, e.g.,~by assuming the validity of the transition state theory~\cite{TruhlarJPhysChem1996}.

\begin{figure}
	\centering
	\includegraphics[width=.45\textwidth]{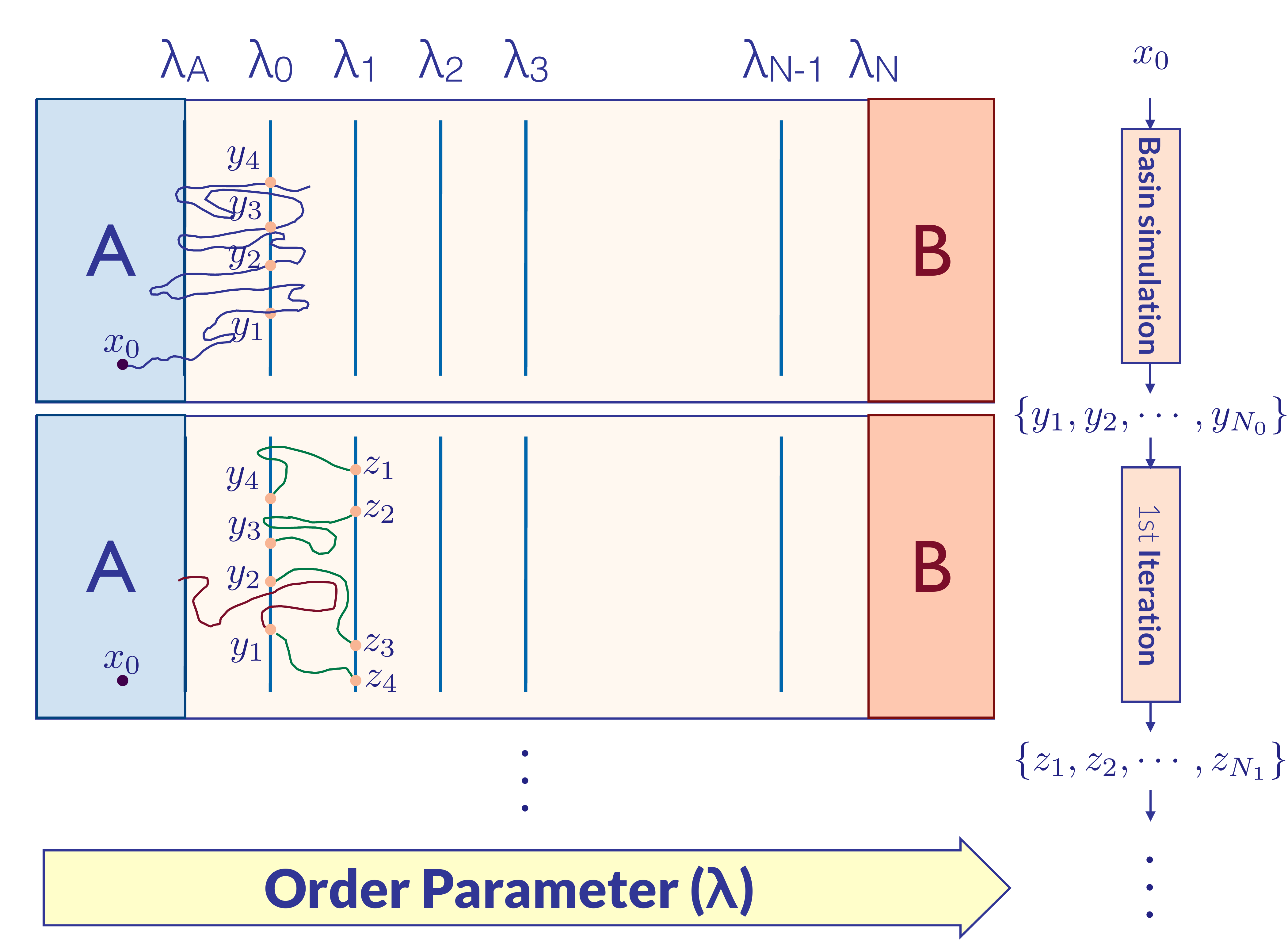}
	\caption{\label{fig:FFS-schematic}Schematic representation of the direct FFS algorithm of Ref.~\citenum{FrenkelFFS_JCP2006}. A dynamic trajectory is launched from $x_0\in\mathcal{Q}$, and $y_0,y_1,\cdots,y_{N_0}$, its first crossings of $\lambda_0$ after leaving $A$ are stored. These configurations are then passed onto the first FFS iterator. The process is continued until $\lambda_B$ is reached.  }
\end{figure}

 In the second class of methods, which are typically known as \emph{path sampling techniques}, no biasing potential is applied and instead reactive trajectories are chosen in accordance to their weight in the transition state ensemble (TSE).  Since the intrinsic dynamics of the system remains intact in path sampling methods, they can usually provide accurate information about the kinetics and microscopic mechanism of the underlying transition. Several methods belong to this category, including transition path sampling (TPS)\cite{DellagoJCP1998}, transition interface sampling (TIS)\cite{VanErpJCompPhys2005}, milestoning~\cite{FaradjianJCP2004} and forward flux sampling (FFS)~\cite{FrenkelFFS_JCP2006, AllenFrenkel2006}.  Further information about path sampling techniques can be found in several excellent reviews that have been written on this topic~\cite{BolhiusAnnRevPhysChem2002, DellagoAdvChemPhys2002, DellagoTPSReview2009}. The focus of this perspective is on the forward flux sampling algorithm, which was originally developed by Allen \emph{et al.}~\cite{FrenkelFFS_JCP2006, AllenFrenkel2006} for probing the kinetics of rare events in both equilibrium and non-equilibrium systems. In recent years, FFS has gained increased popularity  due to its simplicity and ease of implementation, as well as its applicability to systems that are out-of-equilibrium, whereas several other path sampling techniques such as TPS and TIS can only be applied to equilibrium systems with microscopically reversible dynamics~\cite{DellagoJCP1999, DellagoJCP1998}. Despite its versatility, however, the accuracy and efficiency of FFS can depend on the particulars of its implementation. Therefore, developing more robust and efficient variants and implementations of FFS has  become an intense focus of research, with earlier methodological efforts covered in several comprehensive reviews published shortly after its development~\cite{AllenRJJPCM2009, EscobedoJPCM2009}. 
In recent years, FFS has been extensively utilized for studying a wide variety of rare events, such as phase transitions in the Ising~\cite{SearJPhysChemB2006, PagePRL2006, SearEPL2008, AllenJChemPhys2008, AllenProgTheorPhys2008,  ChenPhysRevE2011, ChenPhysRevE2011p,  SearJPhysCondensMatter2012, HedgesSoftMatter2012, HedgesSoftMatter2013, ShenChaos2013, ShenEPL2015} and Potts~\cite{SandersPhysRevB2007, SearJPhysCondensMatter2007, SearJChemPhys2009, ChenPhysicaA2015} models, crystal nucleation~\cite{ValerianiJCP2005, SanzPRL2007, SanzJPhysCOndensMatter2008, vanMeelJChemPhys2008,  LiNatMater2009,  LiJChemPhys2009,  FillionJCP2010, FilionJCP2011, GalliPCCP2011, GalliNatComm2013,  RoehmHighPerfComput2013, HajiAkbariFilmMolinero2014, MithenJChemPhys2014, ThaparPRL2014, LiHydrateJPCB2014, KratzerSoftMatter2015, CabrioluPRE2015, HajiAkbariPNAS2015, GianettiPCCP2016, BiJPhysChemC2016, SossoJPhysChemLett2016, BiJChemPhys2016, DittmarJChemPhys2016,  HajiAkbariPNAS2017, DeFeverJChemPhys2017, SossoChemSci2018, AmirHajiAkbariJCP2018, JiangJChemPhys2018, JiangJChemPhys2018p, RichardJCP2018, RichardIIJCP2018}, evaporation~\cite{WangJCPB2009, MeadleyJCP2012, SumitPNAS2012, SharmaJPhysChemB2012,  AltabetPNAS2017, AltabetJChemPhys2017}, phase separation~\cite{RichardSoftMatter2016, QinActaMaterialia2018}, coalescence~\cite{RekvigJCP2007}, wetting~\cite{SavoyLangmuir2012, ShahrazLangmuir2014}, protein folding, rupture and aggregation~\cite{BorreroJCP2006, JuraszekBiophysicalJ2008, VelezJCP2009, BorreroBioPhy2010, JiangPCCP2014, LuikenJPhysChemB2015, SmitJPhysChemB2017}, DNA hybridization~\cite{OuldridgeNucleicAcidRes2013, SrinivasNucleicAcidRes2013, HinckleyJChemPhys2013, HinckleyJCP2014, MosayebiJPhysChemB2014, SchreckNucleicAcidRes2015, SulcBiophysJ2015}, polymer relaxation and translocation~\cite{HuangJChemPhys2008, HernandezOrtizJChemPhys2009, RuzickaPCCP2012, CaoJCP2015, ZhuJCP2017, RezvantalabMacromolecules2018}, ion transport~\cite{Malmir2019Arxiv} and genetic~\cite{AllenPRL2005, MorelliBiophysJ2008, MorelliJChemPhys2008, MorelliPNAS2009, MorelliBiophysJ2011, ZhangCJChemPhys2012} and magnetic switching~\cite{VoglerPRB2013, VoglerJApplPhys2015, DesplatArxiv2019}. Also, several new variants of FFS have been developed to expand its applicability~\cite{AdamsJCP2010, BeckerJChemPhys2012, HajiAkbariPNAS2015, AmirHajiAkbariJCP2018, VijaykumarJCP2018, DeFeverJChemPhys2019}, optimize reaction coordinates\cite{BorreroJCP2007}, extract free energy profiles~\cite{ValerianiJChemPhsy2007, BorreroJPCB2009, ThaparJCP2015, RichardJCP2018}, and improve its efficiency~\cite{AllenJCP2013, KleinBioRxiv2018}.
These recent developments call for an updated overview of FFS, both to discuss newer  variants and extensions, as well as the types of systems and rare events that have been studied using FFS-like approaches.

This paper is organized as follows. In Section~\ref{section:ffs-overview}, we provide a brief overview of the original FFS method (Section~\ref{section:ffs:original-method}) and its classical variants (Section~\ref{section:classical:variants}), its benchmarking and validation (Section~\ref{section:validation}), and efficiency and accuracy (Section~\ref{section:accuracy}). We dedicate Section~\ref{section:variants:new} to discussing newer variants and extensions of FFS developed for expanding its applicability (Section~\ref{section:variants:new:expand}), optimizing order parameters (Section~\ref{section:variants:new:op}), computing free energy profiles (Section~\ref{section:variants:new:free-energy}), and enhancing its efficiency (Section~\ref{section:variants:new:efficiency}). Section~\ref{section:implementation} is reserved for a discussion of software packages developed for implementing FFS. In Section~\ref{Impact of FFS}, we discuss numerous applications of FFS to study rare events, such as nucleation (Section~\ref{section:nucleation}), conformational rearrangements in biomolecules (Section~\ref{section:conformation-bio}), structural relaxation in polymers (Section~\ref{section:polymer-relaxation}), solute and ion transport through membranes (Section~\ref{section:filtration}) and rare switching events (Section~\ref{section:rare-switching}). Finally, Section~\ref{section:conclusion} is reserved for conclusions and future outlook.

\section{Overview of Forward-Flux Sampling} \label{section:ffs-overview}

\subsection{The Original Algorithm} \label{section:ffs:original-method}
\noindent
Before discussing recent methodological developments and applications, we first need to provide a brief overview of the original FFS algorithm. 
Like many other path sampling techniques, FFS~\cite{FrenkelFFS_JCP2006} is based on dividing the configuration space into non-overlapping regions separated by level sets of an \emph{order parameter (OP)}, a mathematical function, $\lambda(\cdot):\mathcal{Q}\rightarrow\mathbb{R}$  that quantifies the progress of the corresponding rare event. Here, $\mathcal{Q}$ is the configuration space. For any configuration $\textbf{x}\in\mathcal{Q}$, $\lambda(\textbf{x})$ is a measure of its proximity to $B$, with the initial and target basins $A$ and $B$ given by $A:=\{\textbf{x}:\lambda(\textbf{x})<\lambda_A\}$ and $B:=\{\textbf{x}:\lambda(\textbf{x})\ge\lambda_B\}$, respectively. In other words, $\lambda(\textbf{x})=\lambda_A$ and $\lambda(\textbf{x})=\lambda_B$ are two milestones that demarcate the boundaries of $A$ and $B$. Typically, $\lambda_A$ and $\lambda_B$ are chosen so that they are thermally accessible to the configurations within that respective basin. For instance, $\lambda_A$ can be chosen within the range $[\mu,\mu+\sigma]$ where $\mu$ and $\sigma$ are the mean and the standard deviation of the order parameter distribution in $A$.
The parts of $\mathcal{Q}$ that neither belong to $A$ nor to $B$ are further divided into $N$ non-overlapping regions separated by $N$ milestones $\lambda_A\le\lambda_0<\lambda_1<\cdots<\lambda_N=\lambda_B$ (Fig.~\ref{fig:FFS-schematic}).  It is necessary to emphasize that even though it is technically allowed for $\lambda_0$ and $\lambda_A$ to coincide, it is usually more prudent to place $\lambda_0$ away from $\lambda_A$ in order to avoid correlations and increase efficiency. For instance, $\lambda_0$ can be chosen within the 1-0.1\% tail of the order parameter distribution. We will discuss the issue of milestone placement in further detail in Section~\ref{section:variants:new:efficiency}.

After dividing the configuration space into non-overlapping regions, the rate of transition from $A$ to $B$ is computed in an iterative two-step process. First, $A$  is sampled using the unbiased intrinsic dynamics of the underlying system, e.g.,~via methods such as MD or MC, and $N_0$, the number of times that a trajectory crosses $\lambda_0$ after leaving $A$ is enumerated. At each such crossing, the corresponding configuration is stored for use in upcoming iterations ($\{y_1,\cdots,y_{N_0}\}$ in Fig.~\ref{fig:FFS-schematic}). The flux of trajectories crossing $\lambda_0$ after leaving $A$ is then computed as:
\begin{eqnarray}
\Phi_{A,0} &=& \frac{N_0}{T}
\label{eq:A-flux}
\end{eqnarray} 
with $T$ the total duration of trajectories conducted in $A$. In many applications, $\Phi_{A,0}$ is normalized using a proper measure of system size, such as volume or surface area. The next step is to compute $P(B|\lambda_0)$, the probability that a trajectory that has crossed $\lambda_0$ will cross into $B$ before returning to $A$. $P(B|\lambda_0)$ is computed from $N$ computationally tractable iterations with the $k$th iteration aimed at computing  $P(\lambda_k|\lambda_{k-1})$, the probability that a trajectory initiated from an FFS configuration stored at $\lambda_{k-1}$ reaches $\lambda_k$ before returning to $A$. $P(\lambda_k|\lambda_{k-1})$'s are typically referred to as \emph{transition probabilities}. The first transition probability, $P(\lambda_1|\lambda_0)$ is calculated by launching $M_0$ trial trajectories from the $N_0$ configurations stored at $\lambda_0$, with each trial trajectory terminated either when it reaches $\lambda_1$ or returns to $A$. Whenever a trajectory crosses $\lambda_1$, the corresponding configuration is stored for the second iteration ($\{z_1,\cdots,z_{N_1}\}$ in Fig.~\ref{fig:FFS-schematic}), and the transition probability is computed as $P(\lambda_1|\lambda_0) = {N_1}/{M_0}$ with $N_1$ the total number of successful crossings of $\lambda_1$. This procedure is repeated until all transition probabilities are computed.  $P(B|\lambda_0)$ is then estimated as:
\begin{eqnarray}
P(\lambda_B|\lambda_0) &=& \prod_{i=0}^{N-1} P(\lambda_{i+1}|\lambda_i)\label{eq:cumm-trans-prob}
\end{eqnarray}
which can then be used for properly reweighing the flux of trajectories leaving $A$ and estimating $k_{AB}$:
\begin{eqnarray}
k_{AB} &=& \Phi_{A,0} \prod_{i=0}^{N-1} P(\lambda_{i+1}|\lambda_i)
\label{eq:FFS}
\end{eqnarray}
The power of FFS arises from the fact that an arbitrarily small $P(B|\lambda_0)$ can be accurately and efficiently estimated by breaking it into larger-- and more computationally tractable-- transition probabilities. 

It is necessary to emphasize that in practice, an FFS calculation is  terminated when the transition probabilities approach unity and $P(\lambda_k|\lambda_0)=\prod_{i=0}^{k-1}P(\lambda_{i+1}|\lambda_i)$ gets saturated. This occurs when the system overcomes the free energy barrier and moves downhill along $\lambda$. It is therefore not necessary to \emph{a priori} determine $\lambda_i$'s and $\lambda_B$ prior to using FFS, as each $\lambda_k$ can be determined after finishing the iteration aimed at crossing $\lambda_{k-1}$, and the calculation is terminated when the transition probability is unity.

One of the main advantages of FFS is that it is not very sensitive to choosing an imperfect OP as long as the utilized OP is sufficiently close to the true reaction coordinate of the corresponding transition. It is therefore possible to construct several equally valid OPs for studying the same rare event within the same system. The particular mathematical form of an OP is system- and transition-dependent and can be as simple as the number of water molecules between two hydrophobic plates in the case of hydrophobic evaporation~\cite{SumitPNAS2012} and as complex as a linear combination of the distance between DNA strands and the number of in-register base pairs in the case of DNA hybridization~\cite{HinckleyJCP2014}. In our overview of FFS applications in Section~\ref{Impact of FFS}, we discuss different types of OPs that can be utilized for studying different transitions. We also discuss more rigorous approaches for constructing optimal order parameters.


\subsection{Classical Variants of FFS}\label{section:classical:variants}
\noindent The scheme described above is generally referred to as \emph{direct FFS} in which transition pathways are generated in a piecewise manner by concatenating successful trial trajectories connecting successive milestones. Allen~\emph{et al.} developed~\cite{FrenkelFFS_JCP2006} two other FFS variants in which transition paths are generated one at a time instead. The first variant is called \emph{branched growth FFS (BG-FFS)} and  is comprised of the following steps: 
\begin{enumerate}
	\item [(i)] Similar to direct FFS, the $A$ basin is exhaustively sampled and first crossings of $\lambda_0$ are recorded and enumerated. This results in $N_0$ configurations at $\lambda_0$.  The initial flux $\Phi_{A,0}$ is then  computed from Eq.~(\ref{eq:A-flux}). 
	
	\item [(ii)] From each configuration stored at $\lambda_0$, $k_0$ trajectories are initiated, which are then terminated after crossing $\lambda_1$ or returning to the basin. This process results in $N_s^{(1)}$ configurations at $\lambda_1$. If $N_s^{(1)}>0$, $k_1$ trajectories are initiated from each $N_s^{(1)}$ configuration at $\lambda_1$, which are terminated upon crossing $\lambda_2$ or returning to $A$. This process is repeated until $N_s^{(N)}$ configurations are generated at $\lambda_N$ or  no success is observed for an intermediate milestone. The rate is then computed as:
	\begin{eqnarray}
	k_{AB}^{\text{BG}} &=& \Phi_{A,0}\left\langle \frac{N_s^{(N)}}{\prod_{j=0}^{N-1}k_j}\right\rangle_{\lambda_0}
	\end{eqnarray}
	Similarly, the partial cumulative transition probability will be given by:
	\begin{eqnarray}
	P(\lambda_k|\lambda_0) &=& \left\langle \frac{N_s^{(k)}}{\prod_{j=0}^{k-1}k_j}\right\rangle_{\lambda_0}
	\end{eqnarray}
	
\end{enumerate}
The second variant is called \emph{Rosenbluth FFS (RB-FFS)} due to its conceptual similarity to the Rosenbluth algorithm used for growing polymer chains~\cite{JCPRosenbluth1955}. RB-FFS is very similar to BG-FFS in the sense that both tend to generate full transition pathways one at a time. The paths generated by RB-FFS, however, are not branched since at every milestone, $\lambda_j$, only one of the $N_s^{(j)}$ possible configurations are chosen for further growth. The generated path is then accepted or rejected in accordance with a Metropolis scheme, with a weight given by $W=\prod_{j=1}^NN_s^{(j)}$. The total and partial cumulative probabilities are then updated accordingly. 

The other classical variant of FFS introduced in Ref.~\citenum{FrenkelFFS_JCP2006} is \emph{pruning} which is also motivated by the pruning method in simulating polymers~\cite{GrassbergerPhysRevE1997}, and is aimed at avoiding full integration of trial trajectories that are destined to fail. More precisely, the trajectories that start at $\lambda_k$ are terminated with a fixed probability when they revert to $\lambda_{k-1}$, and the surviving trials are properly reweighed to account for such immature terminations. Later benchmarking, however, revealed little increase in efficiency upon using pruning. A similar variant of pruning-- called \emph{constrained branched growth FFS}-- was later introduced by Velez-Vega~\emph{et al.}~\cite{VelezVJCP2010}.

In order for these classical variants to accurately estimate $k_{AB}$, it is necessary that FFS milestones are crossed sequentially, i.e.,~that a reactive trajectory never skips milestones while crawling towards $B$.  As will be discussed in Section~\ref{section:FFS:jFFS}, this imposes a stringent smoothness condition on the utilized OP, a criterion that is  difficult to satisfy with most commonly used OPs. 
 
\subsection{Validation and Benchmarking}\label{section:validation}

\begin{figure}
\centering
\includegraphics[width=.5\textwidth]{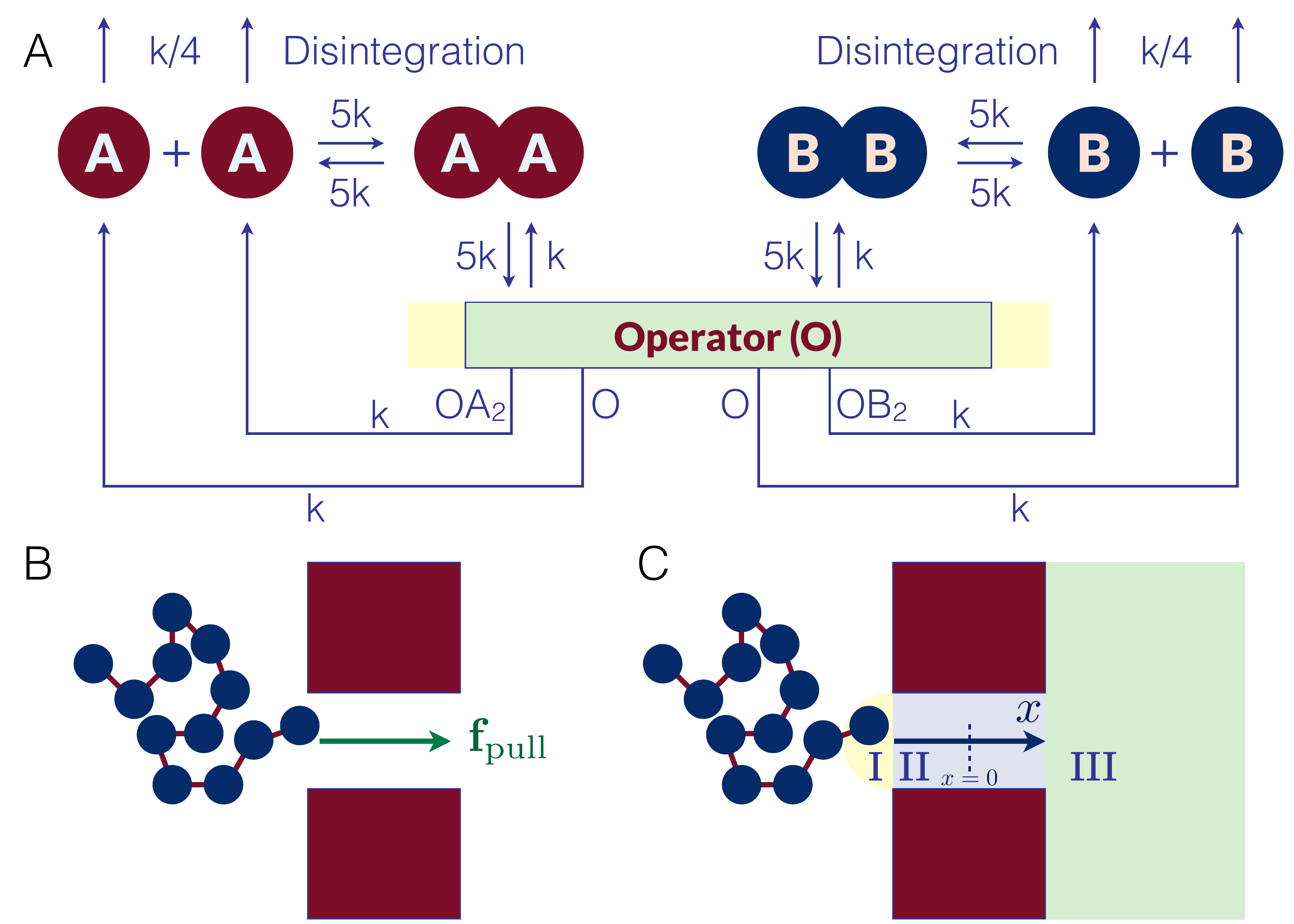}
\caption{\label{fig:validation-systems} The two systems widely studied for validating FFS variants: (A) The bistable genetic switch network of Ref.~\citenum{FrenkelFFS_JCP2006}. (B-C) The simple polymer model utilized for studying polymer translocation with (C) the three regions used for defining the order parameter.}
\end{figure}

\begin{figure}
\centering
\includegraphics[scale=0.4]{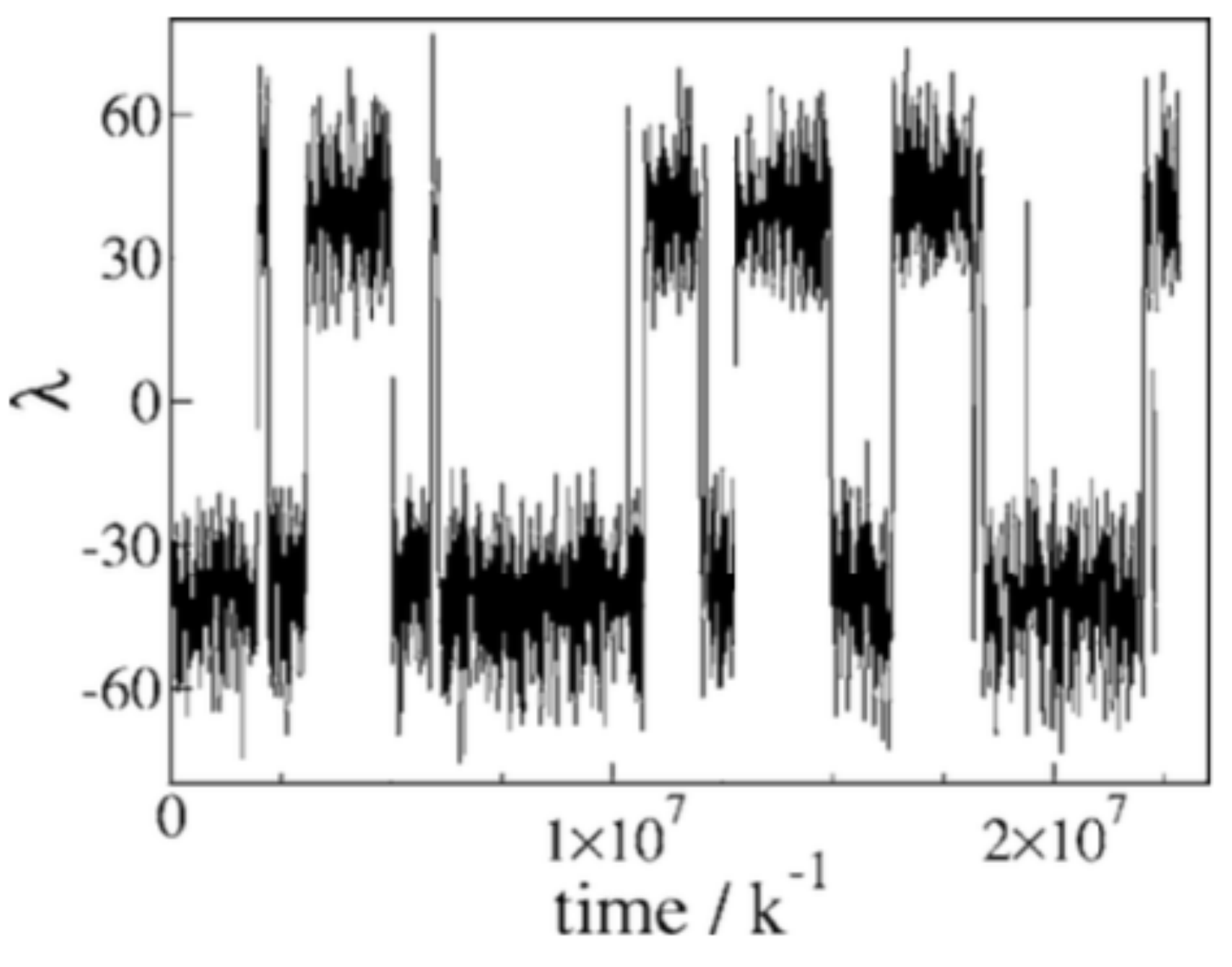} 
\caption{(Reproduced from Ref.~\citenum{FrenkelFFS_JCP2006}) Temporal evolution of the bistable genetic switch order parameter. }
\label{fig:FFS-genetic}
\end{figure}

\noindent
An important aspect of developing any new method is to benchmark and validate it using standard established computational methods. As for FFS, the validation involves comparing the rate estimate from FFS with those obtained from conventional  sampling techniques such as MC, MD and Brownian dynamics (BD). Historically, the classical FFS variants discussed above were  tested and validated based on their ability to accurately predict the switching kinetics of the bistable genetic switch system (Fig.~\ref{fig:validation-systems}A), which consists  of two proteins $\mathscr{A}$ and $\mathscr{B}$ that are encoded by two genes on a segment of DNA with a controlling sequence $\mathscr{O}$. Both $\mathscr{A}$ and $\mathscr{B}$ can dimerize to form $\mathscr{A}_2$ and $\mathscr{B}_2$, which can, in turn, compete to bind to $\mathscr{O}$. When $\mathscr{A}_2$ binds to $\mathscr{O}$ and forms $\mathscr{OA}_2$, it blocks the formation of $\mathscr{B}$. The same is true for $\mathscr{B}$ and $\mathscr{OB}_2$, which will impede the formation of $\mathscr{A}$.  The rate at which   $\mathscr{A}$ (or $\mathscr{B}$) is formed is $k$, irrespective of whether $\mathscr{O}$ is free or bound to $\mathscr{A}_2$ (or $\mathscr{B}_2$). Both $\mathscr{A}$ and $\mathscr{B}$ perish at a constant rate, $k/4$.  (See Ref.~\citenum{FrenkelFFS_JCP2006} for a complete list of reactions.)  This leads to a system with two stable states enriched in $\mathscr{A}$ or $\mathscr{B}$, and a transient state where $\mathscr{A}$ and $\mathscr{B}$ are present in equal quantities. These states can be distinguished using an order parameter given by:
$$\lambda = N_{\mathscr{B}} - N_{\mathscr{A}} = n_{\mathscr{B}} + 2n_{\mathscr{B}_2} + 2n_{\mathscr{OB}_2} - (n_{\mathscr{A}} + 2n_{\mathscr{A}_2} + 2n_{\mathscr{OA}_2})$$
which is the difference between the  total number of $\mathscr{B}$ and $\mathscr{A}$ monomers present in the system. It can be shown using kinetic Monte Carlo (KMC)~\cite{VoterKMC2007} simulations that  the actual switching events are rapid, and occur at timescales considerably shorter than the relatively long wait times between consecutive switching events (Fig.~\ref{fig:FFS-genetic}). This makes switching a quintessential rare event, and hence ideal for assessing the performance of a path sampling technique such as FFS. For all FFS variants discussed above, excellent agreement was observed between the rates estimated from FFS and KMC, with using FFS resulting in a  considerable reduction in computational cost.

Another model rare event utilized for validating FFS  is polymer  translocation through a cylindrical pore (Fig.~\ref{fig:validation-systems}B-C), which, unlike the genetic switch system, does not have two equally likely basins, but involves a unidirectional driven transition between a metastable basin and a stable basin.  The polymer chain is represented using the standard bead-spring model~\cite{BirdJNonNewtonFluid1980} with individual beads interacting via the Lennard-Jones (LJ)~\cite{LJProcRSoc1924} potential, while consecutive beads are connected via harmonic springs. The beads interact with the pore wall via the repulsive part of the LJ potential, and the first bead within the pore is always constrained to regions I, II and III of Fig.~\ref{fig:validation-systems}C. The pore switches between ON and OFF states with rate constants $k_1$ (OFF$\rightarrow$ON) and $k_{-1}$ (ON$\rightarrow$OFF). While the pore is ON, the polymer beads within its interior experience a positive pulling force $\textbf{f}_{\text{pull}}$ (Fig.~\ref{fig:validation-systems}B). The system is temporally evolved using Langevin dynamics. This simple model is primarily used for assessing the accuracy of FFS-like schemes, and is not intended for systematic investigation of any real translocation process.  The utilized order parameter is given by:
$$\lambda = \frac{n_{I} + 2n_{II} + 4n_{III}}{4N} $$
where, $n_{I}$, $n_{II}$ and $n_{III}$ are the number of monomers lying in the regions I, II and III of shown in Fig.~\ref{fig:validation-systems}C, respectively.  The form of $\lambda$ was chosen out of convenience and does not reflect the true reaction coordinate. Similar to the genetic switch system, the rates computed from different FFS variants  agree with the translocation times estimated from conventional Langevin dynamics, but can be estimated with far fewer simulation steps.

\subsection{Numerical Accuracy and Computational Efficiency}  
\label{section:accuracy}
\noindent
The proof-of-concept calculations of Section~\ref{section:validation} demonstrate the power of FFS variants in efficiently estimating the rates of rare events at a fraction of the computational costs of conventional techniques. In order to rigorously assess their efficiency, however, it is necessary to estimate the level of statistical uncertainty in the computed rates. More precisely, the efficiency of an FFS calculation, $\mathcal{E}$, can be defined as~\cite{AllenFrenkel2006}:
\begin{eqnarray}
\mathcal{E} &=& \frac{1}{\mathcal{C}\mathcal{V} }
\label{eq:efficiency}
\end{eqnarray}
Here, $\mathcal{C}=\langle n_S\rangle_{\lambda_0}$ is the \emph{computational cost} per configuration at $\lambda_0$. For any configuration $x$ at $\lambda_0$, $n_S(x)$ is defined as the total  number of (MC or MD) steps of all trajectories initiated from $x$ as well as from all its progeny at later milestones.  $\mathcal{V}$, however, is the \emph{relative} variance in the rate constant per configuration at $\lambda_0$, and is defined as: 
\begin{eqnarray}
\mathcal{V} = \frac{N_0V[k^e_{AB}]}{\langle k^e_{AB}\rangle^2}
\end{eqnarray}
Here, $k_{AB}^e$ is the rate \emph{estimate} obtained from Eq.~(\ref{eq:FFS}). $\langle k^e_{AB}\rangle$ and $V[k^e_{AB}]$ are the mean and variance of this estimate, namely the true rate, and the error bar squared in the estimated rate, respectively.  These quantities can be estimated from FFS as follows:\\

\noindent
\textbf{Computational Cost ($\mathcal{C}$)}: In estimating $\mathcal{C}$, Allen~\emph{et al.}~\cite{AllenFrenkel2006} only enumerated the total number of simulation steps across all FFS trajectories, and did not account for any overheads associated with initiating or storing those trajectories. Within this framework, $\mathcal{C}$ can be broken into its two major contributions: (i) $\mathcal{R}$, the cost of obtaining a configuration at $\lambda_0$, or the average number of steps between successive crossings of $\lambda_0$ during the exhaustive sampling of $A$, (ii) the cost of sampling the transition region $(\lambda_A<\lambda\le\lambda_B)$, or the total length of trial trajectories between successive milestones. Assuming that the length of a trajectory starting from a configuration at $\lambda_i$ and ending at $\lambda_j$ is proportional to $|\lambda_j-\lambda_i|$, the total cost of trial trajectories initiated at $\lambda_i$ will be given by:
\begin{eqnarray}
\mathcal{C}_i&=& S\left(p_i|\lambda_{i+1} - \lambda_{i}|+q_i|\lambda_i - \lambda_A|\right)
\label{eq:cost-per-iteration}
\end{eqnarray}
Here, $p_i=P(\lambda_{i+1}|\lambda_i)$, $q_i=1-p_i$, and $S$ is a system-dependent proportionality constant. For direct FFS therefore, the overall computational cost per configuration at $\lambda_0$ will be given by:
\begin{eqnarray}
\mathcal{C} &=& \mathcal{R} + \frac{1}{N_0} \sum_{i=0}^{n-1} M_i\mathcal{C}_i
\label{eq:cost-per-conf}
\end{eqnarray}
Similar expressions can be obtained  for other FFS variants, such as the branched growth and Rosenbluth methods. It is necessary to emphasize that Eq.~(\ref{eq:cost-per-iteration}) might be violated in many systems, in which case the more generalized expressions of Section~\ref{section:FFSPilot} need to be utilized.\\

\noindent
\textbf{Variance ($\mathcal{V}$)}: The rate computed from Eq.~(\ref{eq:FFS}) can be expressed as $k_{AB}^e=\Phi^e_{A,0}P^e(\lambda_B|\lambda_0)$. Here, the superscripts $e$ refer to the fact that each quantity is the statistical estimator of an unknown parameter. In principle, the uncertainty in $k_{AB}^e$ can arise  from the uncertainty in both $\Phi^e_{A,0}$ and $P^e(\lambda_B|\lambda_0)$. More often than not, however, the former is negligible in comparison to the latter. Indeed, as long as $\lambda_0$ is not too far from $\lambda_A$, basin explorations can be made exhaustive enough to estimate $\Phi_{A,0}^e$ with a high level of accuracy, at a considerably lower computational cost than what is needed for estimating $P^e$. 
Under such circumstances, $\mathcal{V}$ will be given by:
\begin{eqnarray}
\mathcal{V} &=&  \frac{N_0V[k^e_{AB}]}{E^2[k^e_{AB}]} \approx N_0 \frac{V[P_B^e]}{P_B^2}
\end{eqnarray}
Assuming that the iterations initiated from different FFS milestones are uncorrelated, $V[P_B^e]$ can be expressed as:
\begin{eqnarray}
V[P_B^e] &\approx& P_B^2\sum_{j=0}^{N-1} \frac{V[p_j^e]}{p_j^2}
\end{eqnarray}
Here, $V[p_j^e]$'s correspond to uncertainties in individual transition probabilities. It might be argued that $N_{j+1}$ is expected to be binomially distributed, which will result in a variance given by $V[p_j^e]=M_jp_j^eq_j^e$. In reality, however, the trajectories initiated from different configurations at $\lambda_j$ will have different success probabilities and only the trajectories initiated from the same configuration will be binomially distributed. Taking this granularity into consideration yields the following estimate for variance:
\begin{eqnarray}
V[p_j^e] &=& \frac{p_jq_j}{M_j}+\frac{U_j}{N_j}\left(1-\frac1{M_j}\right)
\label{eq:variance-pj}
\end{eqnarray} 
with $U_j$ called the \emph{landscape variance} and given by:
\begin{eqnarray}
U_j &=& \frac{1}{k-1}\left[\frac{V\left[N_s^{(j)}\right]}{k}-p_jq_j\right]
\end{eqnarray}
Here $k$ is the average number of trajectories initiated from a typical configuration at $\lambda_j$, and $V\left[N_s^{(j)}\right]$ is the variance in the number of successes obtained from trajectories initiated from the $j$th configuration. Similar expressions can be obtained for other FFS variants. It must, however, be noted that the uncertainty in $p_j$ is bounded from below by $U_j/N_j$, irrespective of the total number of trial trajectories. This underscores the importance of proper sampling of the starting basin and the earlier milestones, as doing so can result in a drastic reduction in the landscape variance.


\section{Newer Variants of FFS}
\label{section:variants:new}
\noindent
In recent years, several newer FFS variants have been developed to achieve one of these four broad purposes: (i) expanding the applicability of FFS-like schemes to new systems and/or order parameters (Section~\ref{section:variants:new:expand}), (ii) constructing and optimizing order parameters (Section~\ref{section:variants:new:op}), (iii) constructing free energy profiles from FFS (Section~\ref{section:variants:new:free-energy}), and (iv) facilitating the implementation and enhancing the efficiency of FFS (Section~\ref{section:variants:new:efficiency}). In this section, we will discuss developments in each of these arenas separately.

\begin{figure}
	\includegraphics[width=.45\textwidth]{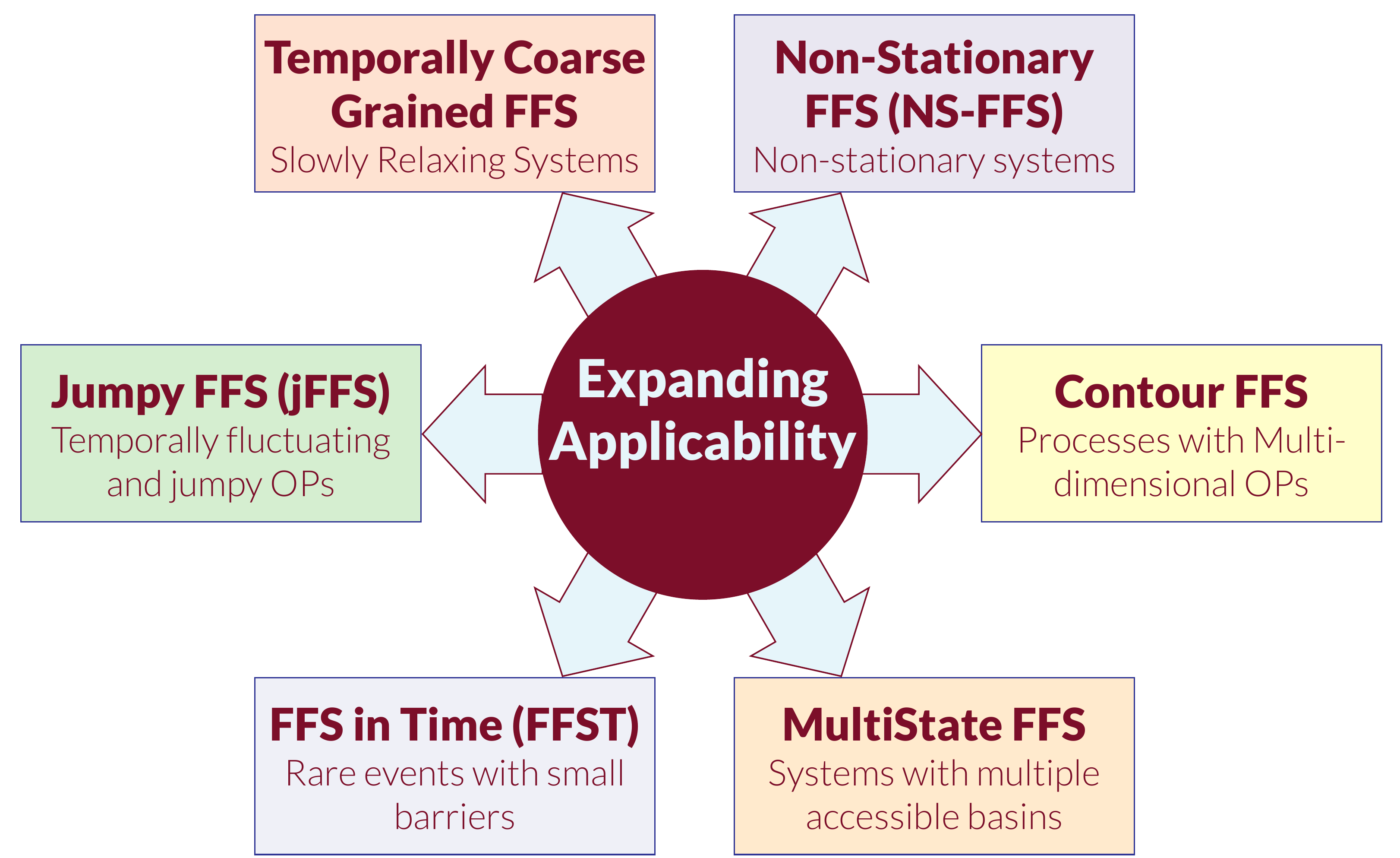}
	\caption{\label{fig:ffs-variants-applicability}New variants of FFS aimed at expanding its applicability and discussed in Section~\ref{section:variants:new}.}
\end{figure}

\subsection{Expanding Applicability}
\label{section:variants:new:expand}
\noindent
The FFS variants discussed in Section~\ref{section:classical:variants} were all developed for rare events  that occur in systems with time-invariant dynamics, small relaxation times and two metastable basins, and that can be described using smooth one-dimensional order parameters. Moreover, it is generally assumed that a clear separation of timescales exist between the wait time and the transition time. 
 The  FFS variants discussed below (Fig.~\ref{fig:ffs-variants-applicability}) are attempts at relaxing these  important constraints, and allowing for rate calculations in a broader range of systems with a wider variety of order parameters.

\subsubsection{Non-stationary FFS (NS-FFS)}
\noindent 
Non-stationary FFS was developed by Becker et al.~\cite{BeckerJChemPhys2012}  to compute  time-dependent transition rates, phase space densities, and crossing-fluxes in systems with time-dependent Hamiltonians, such as systems exposed to on-off and/or oscillating external fields.  Examples in the real life include ice nucleation during flash freezing, or transient biomolecular conformational rearrangements. As expected, such time-dependent systems will have rates that are also time-dependent. The task of defining such time-dependent rates, however, might not be easy as non-stationarity can affect the transition kinetics in nontrivial ways.  The simplest situation is when transition events  are \emph{uniformly rare} over the timescale of interest $T$. (Here, $T$ is the timescale associated with the lowest-frequency change in the Hamiltonian.) Under such circumstances, the generalized time-dependent rate,  $k_{AB}(t)$, will have the property that $k_{AB}^{-1}(t)\gg T$ for $0\le t\le T$, and $S_A(t)$, the survival probability in $A$, will always be close to unity at all times.   However, if transitions from $A$ to $B$ occur frequently enough over the time interval $(0,T)$, uniform rarity can be broken and $S_A(t)$ will no longer be close to unity. Consequently, the time-dependent rate can be defined as $k_{AB}(t) = q_{AB}(t)/S_A(t)$, with $q_{AB}(t)$  the flux of trajectories leaving $A$ and reaching $B$ at time $t$. The other possibility is for the system to exhibit macroscopic memory effects, leading to a history dependent rate constant $k_{AB}(t|t')$ defined as the transition rate at time $t$ assuming that the previous transition occurred at $t'$.

 In all these cases, capturing the time-dependent nature of the $A\rightarrow B$ transition will require generating a statistically representative ensemble of trajectories of length $T$. Due to the explicit time-dependence of the underlying dynamics, all crossing statistics are to be collected over a two-dimensional region $\mathcal{R}=[\lambda_{\min},\lambda_{\max}]\times[t_{\min},t_{\max}]$. In NS-FFS, milestones are defined along either of the two coordinates, and each milestone is then further divided into bins along the other coordinate.  Similar to conventional FFS, the first stage of NS-FFS involves generating a set of starting configurations at the first milestone through exhaustive sampling of $A$. If milestones are staged along $\lambda$, however, the crossing times also need to be stored for such configurations.  Each such configuration can be the starting point for subsequent trial trajectories, which are terminated after time $T$, or upon crossing a milestone or a bin. If the latter, the trajectory is either branched or pruned with a probability $b(n)$. Here, $n=0,1,\cdots,n_{\max}$ is the number of branched trajectories, with $n=0$ corresponding to the trajectory being pruned. 
 In order for the trajectories to be properly weighed, it is necessary for the average trajectory weight to be conserved across all interface bins.  More precisely, if the weights of the original and branched trajectories are given by $w$ and $w'=wr(n)$, respectively, $r(n)$ can be determined from $\langle nr\rangle = \sum_{n=0}^{n_{\max}}b(n)nr(n)=1$. The weighted visiting statistics at each interface bin can then be used for estimating time-dependent rate constants, and densities of states.  Depending on whether original interfaces are placed along $\lambda$ or $t$, the corresponding NS-FFS variant is called  $\lambda$-based or time-based, respectively. Becker et al. tested their method to probe the kinetics of overcoming a linear barrier by a Brownian particle, and in the genetic switch system.  

\subsubsection{Temporally Coarse-grained FFS}
\label{section:FFS:temporally-cg}
\noindent
FFS is a first passage method in the sense that trial trajectories are terminated when they cross a pre-specified set of milestones for the first time. One might therefore expect that discretizing a dynamical trajectory, which is almost universally practiced in molecular simulations, might result in the underestimation of rate due to missing some intermittent crossing events. It therefore seems plausible that increasing the accuracy of FFS should, in principle, require monitoring the trial trajectories as frequently as possible, i.e.,~at every MD or MC step. This might indeed be the case in systems with fast structural relaxation, although the sensitivity of rate to sampling time (the time between successive calculations of the OP) is not expected to be too large. This assertion might, however, not be true in slowly relaxing systems, particularly those in which different  modes of structural relaxation proceed over widely separated timescales. Under such circumstances, temporal fluctuations of a structural OP-- i.e.,~an OP computed from instantaneous positions and orientations of individual molecules-- might occur at frequencies distributed over several orders of magnitude. Monitoring dynamical trajectories at times commensurate to the higher frequency  fluctuations might hamper the convergence of milestone-based techniques such as FFS by causing a preponderance of false crossing events. Detecting true crossings will therefore involve some level of coarse-graining of the OP time series, in order to filter out undesirable high-frequency fluctuations. The simplest-- and potentially most efficient-- way of doing this is to use a larger sampling time, i.e.,~to compute OPs less frequently, although more sophisticated approaches such as obtaining window averages or removing high frequency fluctuations in the Fourier space might also be utilized. Such coarse-graining might not only be necessary to assure the accuracy of the computed rate, but even for the mere convergence of an FFS calculation.

\begin{figure}
\centering
\includegraphics[width=.45\textwidth]{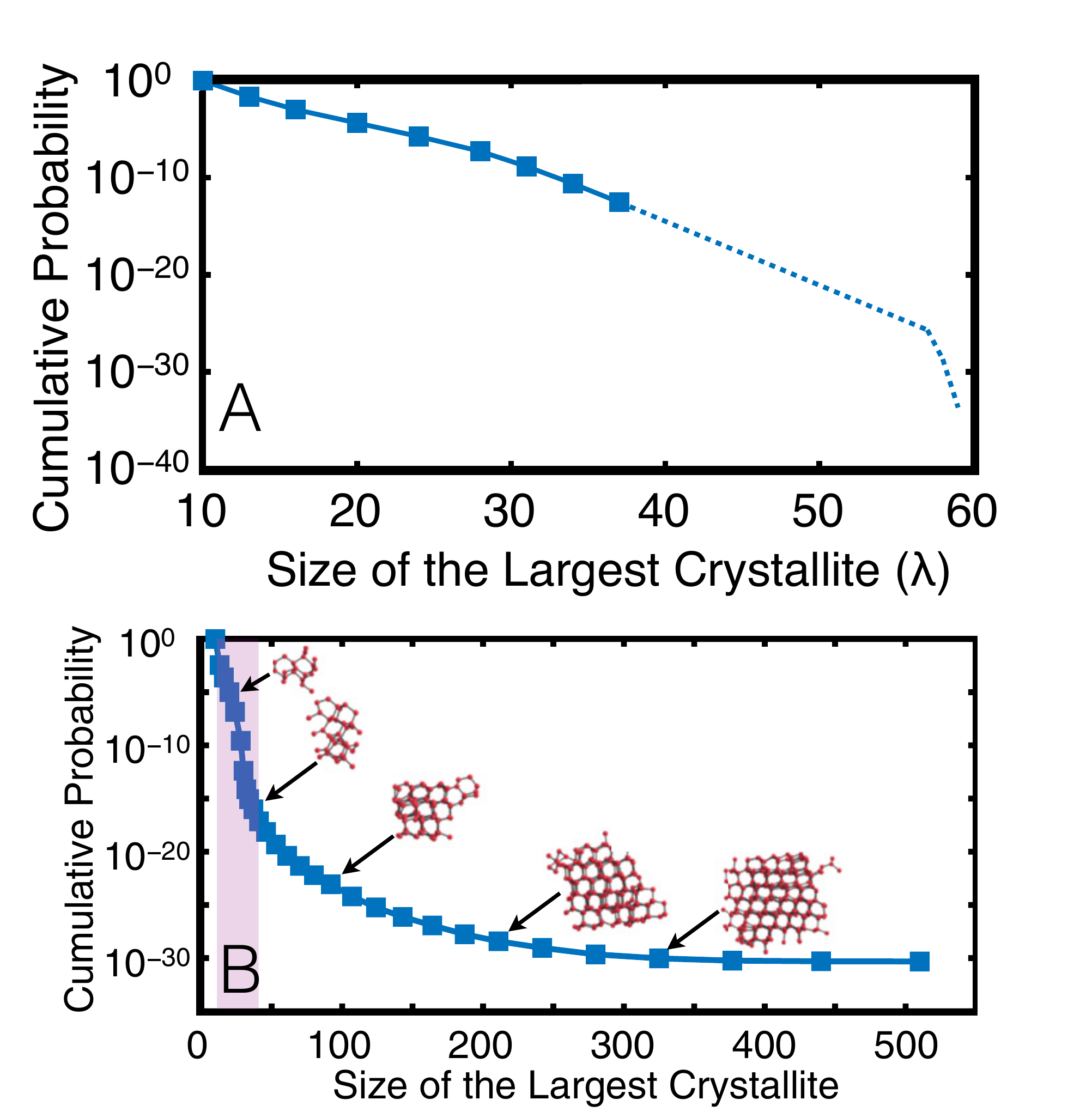}
\caption{(Reproduced from Ref.~\citenum{HajiAkbariPNAS2015}) (A) Cumulative probability curve for homogeneous ice nucleation in the TIP4P/Ice\cite{VegaTIP4PiceJCP2005} system at 230~K and 1~bar with a sampling time of 2~fs. This calculation does not converge due to high-frequency fluctuations in the order parameter. (B) Conducting the same calculation using temporally coarse-grained FFS with a sampling window of 1~ps results in convergence.\label{fig:temporally-cg}}
\end{figure}

Haji-Akbari and Debenedetti~\cite{HajiAkbariPNAS2015} were the first to demonstrate this  in their calculation of homogeneous ice nucleation rate in the TIP4P/Ice system~\cite{VegaTIP4PiceJCP2005}, which is a molecular model of water. Supercooled water is a system in which the characteristic timescale of librational motion is several orders of magnitude smaller than that of diffusion. This results in unphysical high-frequency fluctuations in their utilized OP, i.e.,~the size of the largest crystalline nucleus in the system. The physical insights obtained from their work will be discussed in Section~\ref{section:nucleation:crystal:water}, but as far as FFS is concerned, they considered two sampling times. For a sampling time of 2~fs, which is approximately 6 orders of magnitude smaller than the characteristic diffusion timescales (such as cage escape time or hydrogen bond de-correlation time), the FFS calculation did not converge (Fig.~\ref{fig:temporally-cg}A). They only achieved convergence when they increased the sampling time to 1~ps-- therefore cutting the separation between the sampling and diffusion times to three orders of magnitude (Fig.~\ref{fig:temporally-cg}B).  Since then, temporally coarse-grained FFS has been used by multiple researchers to compute rates in several other systems~\cite{SossoJPhysChemLett2016, HajiAkbariPNAS2017, SossoChemSci2018, JiangJChemPhys2018, JiangJChemPhys2018p, Malmir2019Arxiv}.

\subsubsection{Jumpy FFS (jFFS)}
\label{section:FFS:jFFS}
\begin{figure}
	\centering
	\includegraphics[width=.5\textwidth]{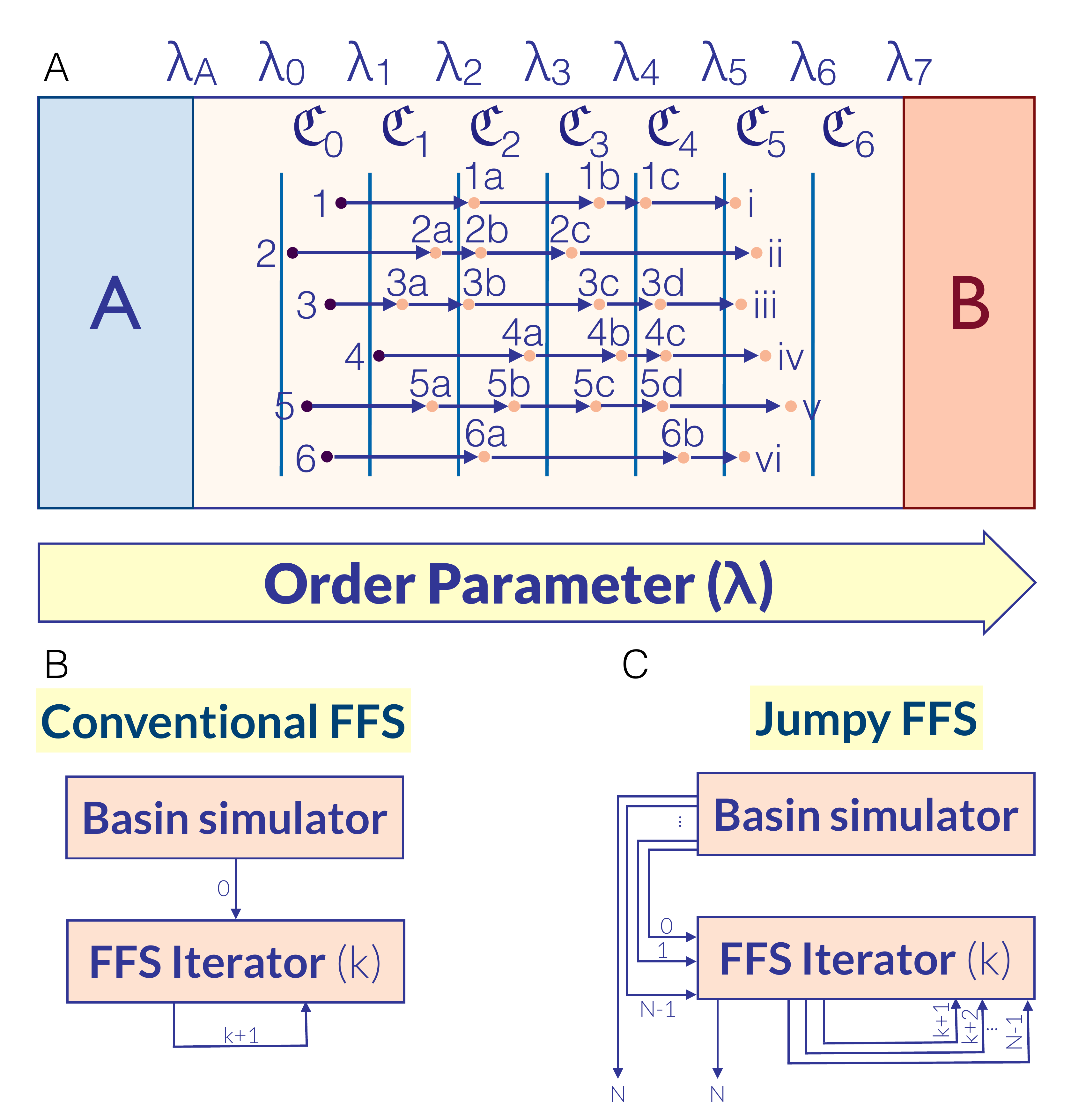}
	\caption {\label{fig:jFFS}(Reproduced from Ref.~\citenum{AmirHajiAkbariJCP2018}) (A) Schematic representation of different possibilities for multi-milestone jumps when the OP is jumpy. Purple configurations arise from sampling the basin but have differing distances from $\lambda_0$. Each arrow corresponds to a successful trial trajectory starting from a particular configuration and aimed at crossing the next milestone.  Such crossing events, however, can result in multi-milestone jumps (such as 1$\rightarrow$1a or 6a$\rightarrow$6b) (B-C) Comparison between conventional FFS (B) in which FFS iterators are called sequentially, and jFFS (C) in which each iterator can, in principle, call any higher order iterator.}
\end{figure}

\noindent
A key assumption in all FFS variants discussed so far is that milestones are crossed \emph{sequentially}, i.e.,~that a trajectory cannot cross a milestone before crossing the previous one at an earlier time. This will only be possible if the underlying order parameter is smooth, i.e.,~if it does not undergo high-amplitude temporal fluctuations. The smoothness criterion cannot, however, be satisfied for a wide variety of OPs, including those that describe aggregation phenomena, or those that are temporally coarse-grained.  
Lack of smoothness can lead to multi-milestone jumps in the sense that the first crossing of $\lambda_k$ can result in a configuration in $\mathfrak{C}_{k+1}=\{\textbf{x}:\lambda(\textbf{x})\in[\lambda_{k+1},\lambda_{k+2})\}$ (e.g.,~crossings of $\lambda_1$ by trajectories started from (1) and (6) in Fig.~\ref{fig:jFFS}A). Moreover, even if such a crossing yields a configuration in $\mathfrak{C}_k$, that configuration might be closer to $\lambda_{k+1}$ than $\lambda_k$ (e.g.,~2a and 5a in Fig.~\ref{fig:jFFS}A that are both far from $\lambda_1$). None of these pathological scenarios have rigorous remedies in conventional FFS. 

Jumpy FFS (jFFS)~\cite{AmirHajiAkbariJCP2018} is a generalized FFS algorithm for which the smoothness criterion is no longer required. In principle, a reactive trajectory can be broken into $1\le q\le N$ sub-trajectories corresponding to $q$ crossings of the $N$ milestones, with the sequence of such crossings called the \emph{jump history}. For instance, the jump history for the top trajectory in Fig.~\ref{fig:jFFS}A is $[-1,0,2,3,4,5]$. For a smooth order parameter, the jump history is always $j=[0,1,2,\cdots,N-1]$, while for a jumpy order parameter, a maximum of $2^N$ distinct jump histories might be possible. Consequently, the transition rate can be expressed as the sum of rates for trajectories with shared jump histories. $k_{AB}^j$, the rate constant for trajectories with jump history $j=[s_1,s_2,\cdots,s_{q_j}]$ will be given by:
\begin{eqnarray}
k_{AB}^j &=& \frac{\langle U_{-1,s_1}\rangle_{\mathcal{E}_A}}{\langle T_B\rangle_{\mathcal{E}_A}}\langle U_{s_1,s_2}\rangle_{[s_1]}\cdots \langle U_{s_{q_j},N}\rangle_{[s_1,s_2,\cdots,s_{q_j}]}\notag\\
\end{eqnarray}
Here $\mathcal{E}_A$ is the ensemble of trajectories initiated in $A$, and $T_B[X]$ and $U_{i,j}[X]$'s are stoppage times and success indicators for the time-invariant Markovian trajectory $X\equiv (x_0,x_1,\cdots)$ (e.g.,~generated via MC or MD):
\begin{eqnarray}
T_{B}[X] &:=& \min_{q \ge L[X]}\{ x_q \in A \cup B\} \\
T_i[X] &:=& \min_{q \geq L[X]} \{x_q \not\in \cup_{i=0}^{i} \mathfrak{C}_i \}\\
U_{ij}[X] &:=& \left\{ 
\begin{array}{ll}
\theta_i(x_{T_i[X]})\theta_j(x_{T_{i+1}[X]})& i\geq0 \\
\phi_0(x_{L[X]})\theta_j(x_{T_0[X]}) & i = -1
     \end{array} 
\right.
\end{eqnarray}
With $L[X] := \min_{q > 0}\{x_q \not\in A \}$ the time that $X$ leaves $A$ for the first time. $\theta_i(x)$ is an indicator function that is unity when $x\in\mathfrak{C}_i$ and zero otherwise, with $\mathfrak{C}_{-1}=\{x\in\mathcal{Q}:\lambda_A\le\lambda(x)<\lambda_0\}$ and $\phi_i(x)=\sum_{j=0}^i\theta_{j-1}(x)$. Here $T_i[X]$ is the earliest time that a trajectory crosses $\lambda_i$ or returns to $A$ after $L[X]$, while $U_{i,j}(j>i)$ is unity only if a trajectory that has landed in $\mathfrak{C}_i$ after crossing $\lambda_i$ for the first time, lands in $\mathfrak{C}_j$ after crossing $\lambda_{i+1}$ for the first time.  The quantity $\langle U_{-1,s_1}\rangle_{\mathcal{E}_A}/\langle T_B\rangle_{\mathcal{E}_A}$ is called the \emph{immediate flux} and is denoted by $\Psi_{A\rightarrow s_1}$ while $\langle U_{s_i,s_{i+1}}\rangle_{[s_1,\cdots,s_j]}$ are jump history-dependent generalized transition probabilities. $k_{AB}$ will thus be given by:
\begin{eqnarray}
k_{AB} &=& \Psi_{A\rightarrow N}+\sum_{q=1}^N\sum_{0\le s_1<\cdots<s_q\le N}\Psi_{A\rightarrow s_1}p_{AB}^{[s_1,s_2,\cdots,s_q]}\notag\\\label{eq:kAB-jFFS}
\end{eqnarray}
Here, $p_{AB}^{[s_1,s_2,\cdots,s_q]}=\langle U_{s_1,s_2}\rangle_{[s_1]}\cdots\langle U_{s_q,N}\rangle_{[s_1,\cdots,s_q]}$. For a smooth order parameter, Eq.~(\ref{eq:kAB-jFFS}) reduces to Eq.~(\ref{eq:FFS}) with $\Psi_{A\rightarrow0}=\Phi_{A,0}$ and $\langle U_{k,k+1}\rangle= P(\lambda_{k+1}|\lambda_k)$.

From an implementation perspective, jFFS has the same operational ingredients of conventional FFS (Fig.~\ref{fig:jFFS}B-C). Similar to conventional FFS, the $A$ basin is exhaustively sampled, and first crossings of $\lambda_0$ are monitored, and the corresponding configurations are stored. Each such configuration, however, is sorted based on its landing index $s(x)$ defined as:
\begin{eqnarray}
s(x) := i,~~~x\in\mathfrak{C}_i
\end{eqnarray}
The basin simulator results in $s_0$ configurations with landing index zero, $s_1$ configurations with landing index $1$, etc. The immediate fluxes are then computed as $\Psi_{A\rightarrow k}:=s_k/T$ with $T$ the total duration of the basin trajectory.  The $s_k$ configurations with landing index $k$ are then passed along to an FFS iterator aimed at crossing $\lambda_{k+1}$. This results in $\tilde{s}_{k+1}, \tilde{s}_{k+2},\cdots, \tilde{s}_N$ configurations in $\mathfrak{C}_{k+1}, \mathfrak{C}_{k+2}, \cdots, \mathfrak{C}_{N}$, respectively, which are then passed along to iterators aimed at crossing the next respective milestones. Therefore, a maximum of $2^N-1$ iterations might be necessary in jFFS, unlike conventional FFS in which $N$ FFS iterations are required. The generalized transition probabilities are then computed as the ratio of each $\tilde{s}_l$ over the total number of trial trajectories. 

In order to decrease the number of necessary iterations, Haji-Akbari~\cite{AmirHajiAkbariJCP2018} proposed a scheme in which the next target milestone is chosen after the current iteration is complete. In sampling the $A$ basin, for instance, $\lambda_{0,\max}$, the largest value of the order parameter is computed for configurations corresponding to first crossings of $\lambda_0$ and $\lambda_1$ is chosen to be larger than $\lambda_{0,\max}$. The same is done for every ensuing FFS iteration. With this scheme, jFFS differs from conventional FFS in that all the configurations collected as a result of crossing $\lambda_k$-- and not just the ones that are close to $\lambda_k$-- are used for initiating trajectories aimed at crossing the next milestone.

Since the main motivation behind deriving the jFFS method is the inherent jumpiness of OPs describing crystal nucleation, Haji-Akbari~\cite{AmirHajiAkbariJCP2018} tested it numerically by computing homogeneous crystal nucleation rates in three different systems, and found that conventional FFS can underestimate nucleation rates by as much as four order of magnitude when OP jumpiness is not taken into consideration. Since then jFFS has been utilized for studying nucleation of NaCl crystals~\cite{JiangJChemPhys2018p} and ion transport through semipermeable membranes~\cite{Malmir2019Arxiv}.

\subsubsection{Forward Flux Sampling in Time (FFST)}
\noindent
A key assumption that is used in the derivation of both conventional FFS and jFFS is that $P(\lambda_B|\lambda_0)$ is vanishingly small, and therefore $\langle T_B\rangle_{\mathcal{E}_A}$ is dominated by failing trajectories. This assumption, however, might not be valid when $P(\lambda_B|\lambda_0)$ is not astronomically small. Adams~\emph{et al.}~\cite{AdamsJCP2010} developed an extension of FFS, which they called FFS in time (FFST), in which the average transition time, $T_{\text{tot}}$, is estimated from transition probabilities and the average lengths of successful and failing trial trajectories. They argued that $T_{\text{tot}}$, the mean first passage time from $A$ to $B$, will be given by:
\begin{eqnarray}
T_{\text{tot}} &=& \left(\frac{1}{p}-1\right)\left(T_{\text{int}}+T_{\text{ext}}\right) + T_f
\label{eq:FFST}
\end{eqnarray}
Here $p$ is the probability of reaching $B$ from $\lambda_A$, $T_{\text{int}}$ is the average time needed for leaving $A$ starting from a configuration just at the surface of $A$, and $T_{\text{ext}}$ is the time it takes to return to $A$ starting from a configuration right outside the surface of $A$. The authors provide expressions for $T_f$ and $T_{\text{ext}}$, which can be computed from transition probabilities, and the average length of successful and failing trajectories initiated at different milestones.  $T_{\text{int}}$, however, can be readily computed from exhaustive sampling of $A$. Note that for $p\ll1$, $1/p-1\approx 1/p$, and Eq.~(\ref{eq:FFST}) will reduce to $T_{\text{tot}} = 1/\Phi_{A,0}P(B|\lambda_0)$. Adams~\emph{et al.} utilized the FFST method to compute the nucleation rate in the Ziff-Gulari-Barshad (ZGB) system~\cite{ZiffPRL1986}, an on-lattice model for carbon monoxide (CO) oxidation on Pt.   In a second paper~\cite{AdamsJCP2010p}, the same authors developed an FFS-like method called the \emph{barrier method} in which $T_{\text{tot}}$ is exclusively expressed in terms of the lengths of surviving and failing trajectories. The barrier method, however, involves a backtracking step that can only be conducted for simple potential energy surfaces, and not for the multi-dimensional spaces commonly considered in molecular simulations. 

\subsubsection{Multistate Forward Flux Sampling}
\label{section:FFS:multistate}

\begin{figure}
\centering
\includegraphics[width=.45\textwidth]{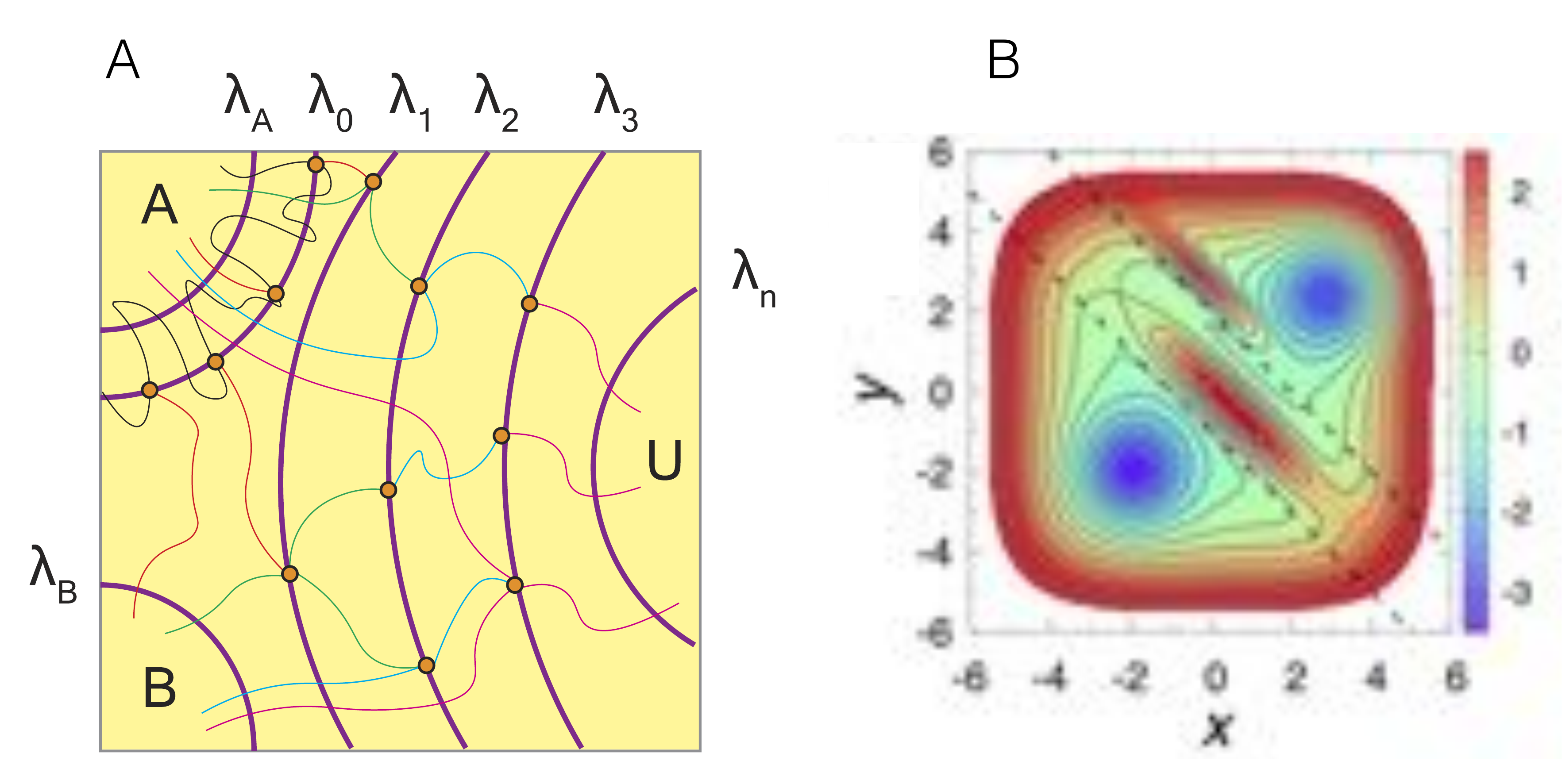}
\caption{\label{fig:multistate-cffs} (A) (Reproduced from Ref.~\citenum{VijaykumarJCP2018}) The basin landscape considered in the development of the multistate FFS method (Section~\ref{section:FFS:multistate}). (B) (Reproduced from Ref.~\citenum{DeFeverJChemPhys2019}) A potential energy surface that cannot be readily probed using a one-dimensional OP. The rate estimates obtained from conventional FFS using $x$ as an OP can be more than an order of magnitude smaller from the rate computed from cFFS (Section~\ref{section:FFS:cFFS}). }
\end{figure}

\noindent
All the FFS variants discussed thus far are formulated based on the assumption that only one basin can be accessed by the trajectories originating in $A$. In many systems, however, additional metastable basins might be accessible to such trajectories. Under such circumstances, some of the trajectories reaching $B$ will have partially proceeded towards-- or might have even passed through-- those intermediates at an earlier time. Capturing the kinetics and mechanism of the $A\rightarrow B$ transition will therefore require accurately accounting for such "off-ramp`` trajectories. Recently, Vijaykumar~\emph{et al.}~\cite{VijaykumarJCP2018} developed an extension of FFS to address this issue. They formulated their multi-state FFS algorithm for a three-state system, but their formalism can be easily extended to transitions that pass through multiple intermediates. The particular basin layout that they considered is comprised of two stable basins $A$ and $B$ that can both access a metastable basin $U$ (Fig.~\ref{fig:multistate-cffs}A). Such a scenario can, for instance, occur for  binding of an enzyme to two distinct binding sites of a ligand wherein $A$ and $B$ will correspond to the bounded states, while $U$ will correspond to the unbounded state.  Following the distance between the substrate and the enzyme as order parameter will map out the $A\rightarrow U$ transition. The rate of the $A\rightarrow B$ transition, however, can be estimated as:
\begin{eqnarray}
R_{A\rightarrow B} &=& \Phi_{A\rightarrow\lambda_0}\sum_{i=0}^{N}P(B|\lambda_i)\prod_{q=0}^{i-1}P(\lambda_{q+1}|\lambda_q) \label{eq:FFS-multi}
\end{eqnarray}
with $P(B|\lambda_i)$ the probability that a trajectory initiated at $\lambda_i$ will reach $B$ before returning to $A$ or crossing $\lambda_{i+1}$. Similar to two-state FFS, the first step of multistate FFS is to exhaustively sample the starting basin with the aim of collecting a large number of configurations at (or beyond) $\lambda_0$.  In subsequent iterations, however, trial trajectories are started from $\lambda_i$ and terminated when they reach $B$ in addition to when they cross $\lambda_{i+1}$ or when they return to $A$. After each iteration, $P(B|\lambda_i)$ is enumerated in addition to the conventional transition probability $P(\lambda_{i+1}|\lambda_i)$.  Vijaykumar~\emph{et al.} tested this method by applying it to a simple enzyme-substrate model.  

\subsubsection{Contour Forward Flux Sampling}
\label{section:FFS:cFFS}

\noindent 
As mentioned earlier, the accuracy of an FFS calculation does not typically rely on having an optimal OP, while its efficiency can be adversely affected if a non-optimal OP is employed. Lack of optimality can be  particularly problematic when the transition rate is astronomically small, or when the transition can proceed through multiple distinct  pathways with transition states corresponding to different values of the suboptimal OP.  Unfortunately, it is not always easy to \emph{a priori} identify an optimal one-dimensional OP for a rare event.  It is, however, generally  easier to invoke physical intuition to identify a handful of suboptimal OPs collectively expected to  describe the transition of interest. DeFever and Sarupria~\cite{DeFeverJChemPhys2019} developed the contour FFS (cFFS) algorithm that allows for conducting FFS over a multi-dimensional order parameter space. They introduced a recip\'{e} for placing the FFS milestones on-the-fly, by collecting the visiting statistics of trial trajectories over a collective variable grid, and placing milestones within that grid with the requirement of a uniform flux across each milestone. Consistent with multi-state FFS, they computed the overall rate using Eq.~(\ref{eq:FFS-multi}) in order to account for the possibility that a trial trajectory reaches $B$ before crossing the next target milestone. They validated cFFS by applying it to several simple model systems, such as several model two-dimensional potential energy surfaces, as well as conformational rearrangements of alanine dipeptides. For the two-dimensional OPs considered therein, cFFS was found to be more efficient than conventional FFS. For certain potential energy surfaces (such as the one in Fig.~\ref{fig:multistate-cffs}B), the rate computed from conventional FFS is more than an order of magnitude smaller than the true rate, which is correctly predicted by cFFS.  The efficiency of cFFS, however, is expected to diminish when more than three collective variables are utilized, as a considerably larger number of trajectories will be needed to enforce the constants flux criterion across a multi-dimensional manifold.  Moreover, the requirement that a trajectory should rarely skip several grid points over a single sampling window can limit its applicability when some of the utilized OPs are jumpy.

\subsection{Optimizing Order Parameters}
\label{section:variants:new:op}

\noindent
The natural reaction coordinate of any transition is the committor probability. For $x\in\mathcal{Q}$, $p_B(x)$ is the probability that a trajectory initiated from $x$ reaches $B$ before $A$. Note that $p_B(x)\approx0$ and $\approx1$ for $x\in A$ and $x\in B$, respectively. In this context, a good OP is a collective variable for which $p_B(x|\lambda(x)=\lambda)$ is narrowly distributed around $p_B(\lambda)=\langle p_B(x)\rangle_{\lambda}$. In FFS-like schemes, it is in principle possible to compute the committor probability of each configuration by tracing forward the trajectories  that connect it to $B$. Motivated by the maximum-likelihood approach of Peters and Trout~\cite{PetersJCP2006}, Borrero and Escobedo~\cite{BorreroJCP2007} proposed a least square estimation approach for \emph{a posteriori} construction of an optimal reaction coordinates from FFS trajectories. Their proposed approach, which they call FFS-LSE, works as follows. Let $x_1,\cdots,x_n\in\mathcal{Q}$ be a collection of configurations in the transition region collected from FFS. Assuming that $p_B(x_i)$ is known for every $i\le n$, a linear regression model is constructed as:
\begin{eqnarray}
p_B(\textbf{q}) = \pmb\beta^T\textbf{q}+\textbf{q}^T\textbf{A}\textbf{q} +\beta_0+\epsilon
\label{eq:ffs-lse-model}
\end{eqnarray}
Here, $\textbf{q}(x) \equiv (q_1(x),q_2(x),\cdots,q_m(x))$ is a column vector of $m$ candidate reaction coordinates selected based on physical intuition. $\pmb\beta\equiv(\beta_1,\beta_2,\cdots,\beta_m)$ corresponds to the relative contribution of each coordinate to $p_B$ while $A$ is an $m$-by-$m$ matrix that quantifies correlations between the $m$ coordinates. $\beta_0$ is a constant that allows to set the committor probability of the transition state to $\frac12$ and $\epsilon$ is the deviation of each configuration from the model. The goal is to find $\pmb\beta$ and $\textbf{A}$ so that:
\begin{eqnarray}
\mathscr{L}[\pmb\beta,\textbf{A},\beta_0] = \sum_{i=1}^n \left[\pmb\beta^T\textbf{q}(x_i)+\textbf{q}^T(x)\textbf{A}\textbf{q}(x) +\beta_0 - p_B(x_i)\right]^2\notag
\end{eqnarray}  
is minimized. With the optimal $\pmb\beta^*$ and $\textbf{A}^*$ at hand, it is possible to use statistical tests such as ANOVA to assess the statistical significance of the entire model, as well as the importance of each coordinate in the model. Note that Eq.~(\ref{eq:ffs-lse-model}) describes a linear regression model, and will therefore only be accurate if the transition pathway is flat enough to be approximated by a hyperplane within the collective variable landscape. It is, however, fairly straightforward to generalize FFS-LSE to non-linear models when this flatness criterion is violated. The authors validated their model by applying it to several simple model systems, including the folding of a lattice protein. Since its development, this approach has been used for constructing optimal OPs for many different processes~\cite{VelezJCP2009, VelezVJCP2010, BorreroBioPhy2010, ShahrazLangmuir2014, DeFeverJChemPhys2017, DeFeverJChemPhys2019}.

\subsection{Free Energy Landscapes from FFS}
\label{section:variants:new:free-energy}
\noindent
As mentioned in Section~\ref{section:intro}, a rare event is not only characterized by its rate, but also by its equilibrium free energy landscape. It is, however, necessary to emphasize that the notion of a free energy  might not even be well-defined for the out-of-equilibrium and driven processes that cannot be mappable  onto a proper thermodynamic ensemble.  As to how non-equilibrium notions of free energy can be defined in such systems is beyond the scope of this section and is discussed elsewhere~\cite{TakatoriPRE2015, MandalPRL2017, delJuncoPNAS2018}. We will instead focus on processes for which a well-defined notion of  free energy can be constructed, which only constitute a subset of all rare events whose kinetics can be probed using FFS. 

Consider a rare event that occurs within a system with a well-defined notion of free energy and let $\textbf{q}(\textbf{x}) \equiv [q_1(\textbf{x}),\cdots,q_l(\textbf{x})]$ be a set of collective variables that might or might not coincide with $\lambda(\textbf{x})$. It is, in principle, possible to define a generalized Landau free energy (in terms of $\textbf{q}$) by enumerating a constrained partition function. In canonical ensemble, for instance, the generalized Helmholtz free energy $F(\textbf{q})$ will be given by:
$$
F(\textbf{q}) := -\beta^{-1}\ln\int_{\mathscr{Q}} e^{-\beta U(\textbf{x})}\delta\left[\textbf{q}(\textbf{x})-\textbf{q}\right]d\textbf{x}
$$
As discussed earlier, free energy profiles can be readily computed using bias-based techniques, while path sampling techniques can only provide qualitative estimates of free energy.  In the case of FFS, in particular, the cumulative transition probability is an indirect and inaccurate measure of free energy, i.e.,~$F(\lambda)\sim-\beta^{-1}\ln P(\lambda|\lambda_0)$. The lack of accuracy arises from the fact that trial trajectories in FFS are terminated when they cross the target milestone, and thus cannot  sample the pre-target region with the correct statistical weight. Developing numerical algorithms that properly reweigh such trial trajectories, and thus allow for a simultaneous calculation of $k_{AB}$ and $F(\textbf{q})$ from FFS has been an active area of exploration in recent years, as employing such algorithms will  save the added computational cost of a separate free energy calculation using another method.  Since the original work of Allen~\emph{et al.},  several algorithms have been developed for extracting equilibrium free energy profiles from FFS. Most of these algorithms, however, rely on analyzing \emph{full} trial trajectories, and require more extensive storage of trajectories (or at the minimum the order parameter or collective variable time series) than what is needed for calculating rates. Moreover, the accuracy of some of these algorithms depends on how the collective variables evolve over time along a dynamical trajectory, and whether their evolution can be described using the Smoluchowksi equation.  
In this section, we will discuss the theoretical bases and implementation details of such approaches.

\subsubsection{Free energy profiles from forward and backward FFS calculations:}
\label{section:method:forward-backward-ffs}
\noindent The first algorithm for extracting free energy profiles from FFS was developed by Valeriani~\emph{et al.}~\cite{ValerianiJChemPhsy2007}, and involves reconstructing the stationary distribution from two FFS calculations conducted in opposite directions, namely a forward and a backward calculation probing the $A\rightarrow B$ and $B\rightarrow A$ transitions, respectively. The overall stationary density of states $\rho(\textbf{q})=e^{-\beta F(\textbf{q})}$ is then estimated from:
\begin{align}
\rho(\textbf{q}) = \Psi_A(\textbf{q}) + \Psi_B(\textbf{q})
\label{eq:fbFFS}
\end{align}
Here, $\Psi_A(\textbf{q})$ and $\Psi_B(\textbf{q})$ correspond to contributions to $\rho(\textbf{q})$ from the trajectories originating in $A$ and $B$, respectively. It can be shown that $\Psi_A(\textbf{q})$ and $\Psi_B(\textbf{q})$ can be expressed as:
\begin{eqnarray}
\Psi_A(\textbf{q}) &=& p_A\Phi_{A,0}\tau_{+}(\textbf{q};\lambda_0)\label{eq:2FFS-Psi-A}\\
\Psi_B(\textbf{q}) &=& p_B\Phi_{B,0}\tau_{-}(\textbf{q};\lambda_N)\label{eq:2FFS-Psi-B}
\end{eqnarray} 
Here, $p_A$ and $p_B$ are the probabilities of the system being in the $A$ and $B$ basins, $\Phi_{A,0}$ and $\Phi_{B,0}$ are the initial fluxes for the forward and backward calculations, and $\tau_{+}(\textbf{q};\lambda_0)$ and $\tau_{-}(\textbf{q};\lambda_N)$ correspond to the average time spent at $\textbf{q}$ by a trajectory originating in $\lambda_0$ and $\lambda_N$, respectively. These quantities can be directly estimated from the forward and backward FFS runs, and, when combined with brute force simulations in the basins and Eq.~(\ref{eq:fbFFS}), can be used for estimating the full stationary distribution. For  forward FFS, $\tau_{+}(\textbf{q};\lambda_0)$ can be related to $\pi_{+}(\textbf{q};\lambda_i)$, or the average time spent at $\textbf{q}$ by a trajectory originating at $\lambda_i$ and ending either at $\lambda_{i+1}$ or $\lambda_0$:
\begin{eqnarray}
\tau_{+}(\textbf{q};\lambda_0) &=& \pi_{+}(\textbf{q};\lambda_0) \notag\\
&& + \sum_{i=1}^{N-1}\pi_{+}(\textbf{q};\lambda_i)\prod_{k=0}^{i-1}P_{\text{forward}}(\lambda_{k+1}|\lambda_{k})\notag\\\label{eq:2FFS-tau-plus}
\end{eqnarray}
Similarly, $\tau_{-}(\textbf{q};\lambda_N)$ is given by:
\begin{eqnarray}
\tau_{-}(\textbf{q};\lambda_N) &=& \pi_{-}(\textbf{q};\lambda_N)\notag\\&& + \sum_{i=N-1}^{1}\pi_{-}(\textbf{q};\lambda_i)\prod_{k=N}^{i+1}P_{\text{backward}}(\lambda_{k-1}|\lambda_{k})\notag\\
\label{eq:2FFS-tau-minus}
\end{eqnarray}
with $\pi_{-}(\textbf{q};\lambda_i)$, the average time spent at $\textbf{q}$ by a trajectory initiating at $\lambda_i$ and ending at $\lambda_{i-1}$ or $\lambda_N$. Note that in computing $\tau_+$ and $\tau_-$, $\pi_+(\textbf{q};\lambda_i)$ and $\pi_-(\textbf{q};\lambda_i)$ are reweighed by cumulative probabilities $P_{\text{forward}}(\lambda_i|\lambda_0)$ and $P_{\text{backward}}(\lambda_i|\lambda_N)$ computed from forward and backward FFS, respectively. If the underlying system has two accessible (meta)stable states, and is in steady-state, $p_A$ will be given by:
\begin{eqnarray}
p_A &=& \frac{k_{BA}/k_{AB}}{1 + k_{BA}/k_{AB}}\label{eq:2FFS-pA}
\end{eqnarray}
and $p_B=1-p_A$. Here, $k_{AB}$ and $k_{BA}$ are the rate constants computed from forward and backward FFS, respectively. All the quantities in Eq.~(\ref{eq:2FFS-Psi-A}-\ref{eq:2FFS-pA}), except for $\pi_{+}$'s and $\pi_-$'s,  can be accurately and efficiently computed from forward and backward FFS. Computing $\pi_+$'s and $\pi_-$'s, however, involves analyzing full trial trajectories initiated from different milestones. More precisely, $\pi_{\pm}(\textbf{q};\lambda_i)$ are given by:
\begin{eqnarray}
\pi_{\pm}(\textbf{q};\lambda_i) &=& \frac{N_\textbf{q}}{M_i\Delta^l\textbf{q}}
\end{eqnarray}
Here, $M_i$ is the number of trial trajectories initiated from $\lambda_i$, while $N_{\textbf{q}}$ is the number of sampled configurations along such trajectories in which the collective variable is within a bin of sides $(\Delta{q}_1,\Delta{q}_2,\cdots,\Delta{q}_l)$ centered at $\textbf{q}$.  $\Delta^l\textbf{q}$ is the bin volume within the $l$-dimensional collective variable space. 

\begin{figure}
\centering
\includegraphics[width=.45\textwidth]{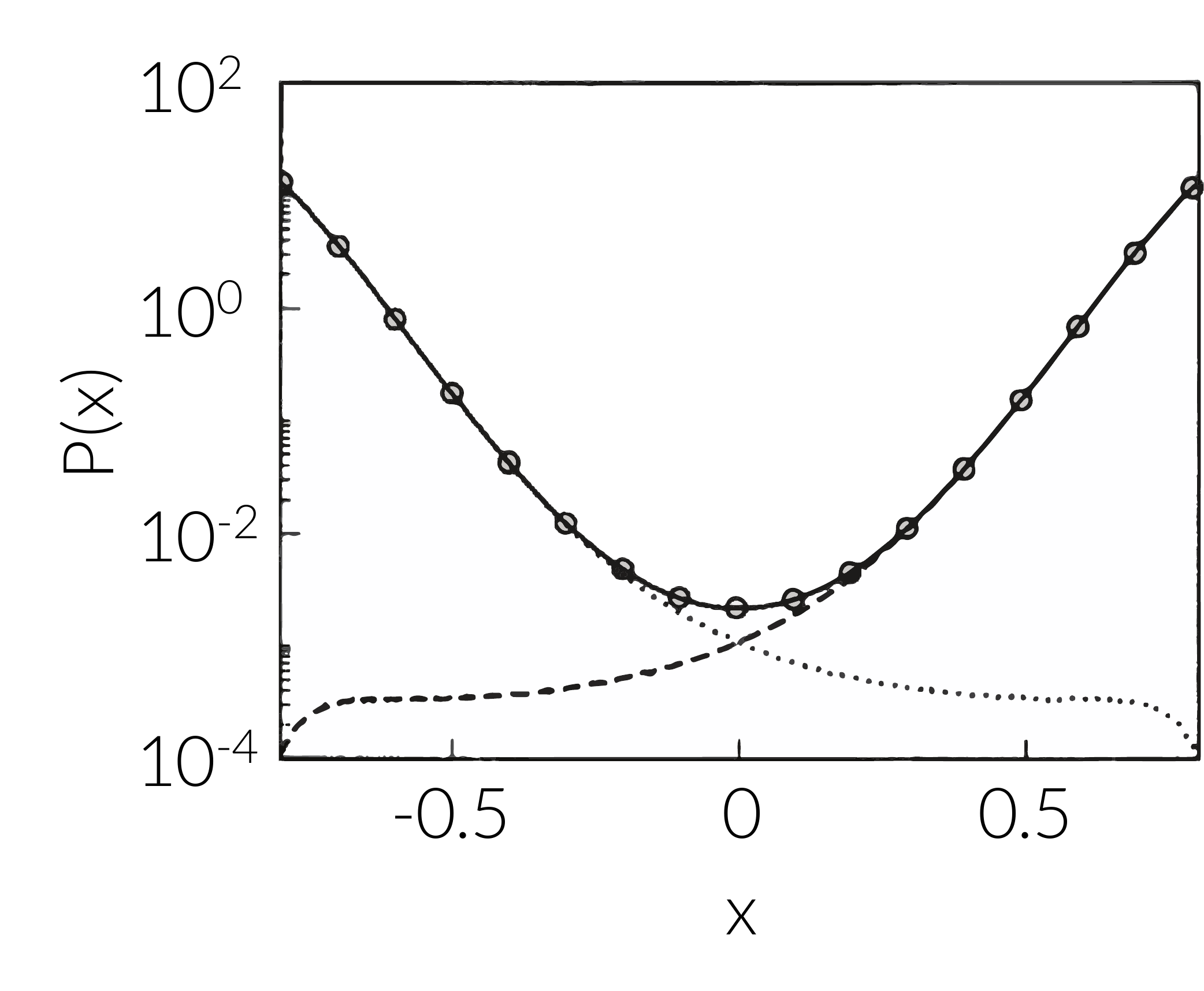}
\caption{(Adapted from Ref.~\citenum{ValerianiJChemPhsy2007} and relabeled) Stationary distribution for the Maier Stein system computed using forward and backward FFS. The dotted lines correspond to $\Psi_A(x)$ and $\Psi_B(x)$ computed from forward and reverse FFS, respectively. }
\label{fig:Meir-Stein}
\end{figure}

Valeriani~\emph{et al.} verified this approach by applying it to several model systems, including the  one-dimensional double-well potential, and the two-dimensional Meir Stein system~\cite{MaierPhysRevLett1992}, and found excellent agreement with the expected stationary distributions. As can be seen in the stationary distribution computed for the Meir Stein system (Fig.~\ref{fig:Meir-Stein}), the reweighing of Eq.~(\ref{eq:fbFFS}) is only necessary in the transition region as $p(x)$ is dominated by $\Psi_A(x)$ and $\Psi_B(x)$ close to the $A$ and $B$ basins. They also utilized their approach to study two more complex processes, i.e.,~switching events in a bistable genetic switch, and nucleation in the 2D Ising model under an external magnetic field. As for the Ising model, which will be discussed in detail in Section~\ref{section:nucleation:Ising}, the transition of interest occurs between the metastable down-spin basin, $A$, and the stable up-spin basin, $B$, with $S$, the number of up spins as the FFS order parameter.  Under such circumstances, the backward  $B\rightarrow A$ calculation will be far more costly computationally than the forward $A\rightarrow B$ calculation. The authors resolved this issue by artificially constructing a modified stable state $B'$ by placing a reflecting wall at a value of $S$ sufficiently far from the transition state $S^*$. Since the main quantity of interest is the free energy barrier for transitioning from $A$ to $B$, replacing $B$ with  $B'$ will not change the shape of the free energy profile prior to and through the transition region as long as $S_{B'}\gg S^*$. The barriers computed from forward and backward FFS matched perfectly with those obtained from umbrella sampling. For the bistable genetic switch, the profiles obtained from FFS and brute force KMC were in excellent agreement. Since its development, this approach has been utilized for computing free energy profiles in multiple systems~\cite{MorelliBiophysJ2008, MorelliJChemPhys2008, ZhangCJChemPhys2012, SharmaJPhysChemB2012, ShahrazLangmuir2014, KratzerSoftMatter2015, LuikenJPhysChemB2015, BiJChemPhys2016, AltabetPNAS2017, AltabetJChemPhys2017}.

\subsubsection{Forward-Flux Sampling/Mean First Passage Time Method (FFS-MFPT)} 
\label{section:method:ffs-mfpt}

\noindent The FFS-MFPT method was developed by Thapar and Escobedo~\cite{ThaparJCP2015}, and unlike the method of Valeriani~\emph{et al}.~\cite{ValerianiJChemPhsy2007}, can reconstruct free energy profiles from a single FFS calculation. The method is based on the theoretical description of Wedekind and Reguera~\cite{WedekindJPCB2008} who provide a rigorous approach to compute free energy profiles from order parameter histograms and mean first passage time distributions.  According to Wedekind and Reguera, $G(\lambda)$, the Gibbs free energy as a function of an order parameter $\lambda$ can be expressed as~\cite{WedekindJPCB2008}:
\begin{eqnarray}
\beta G(\lambda) &=& \ln{[B(\lambda)]} - \int \frac{d\lambda'}{B(\lambda')} + C 
\label{eq:MFPT}
\end{eqnarray}
with $B(\lambda)$ given by:
\begin{eqnarray}
B(\lambda) &=& \frac{1}{P_{st}(\lambda)}\left[\int_{\lambda_A}^{\lambda} P_{st}(\lambda')d\lambda' - \frac{\tau(\lambda;\lambda_A)}{\tau(\lambda_B;\lambda_A)}   \right]\label{eq:MFPT-Blambda}
\end{eqnarray}
Here, $P_{st}(\lambda)$ is the steady-state probability density for the forward trajectories from $\lambda_A$ to $\lambda_B$, and $\tau(\lambda; \lambda_A)$ is the average time that it takes for a trajectory starting at $\lambda_A$ to cross an interface with order parameter $\lambda$. Wedekind and Reguera~\cite{WedekindJPCB2008} derived Eqs.~(\ref{eq:MFPT}) and (\ref{eq:MFPT-Blambda}) by assuming that the temporal evolution of the underlying system along the order parameter space is diffusive in nature, and can be described using the Smoluchowski  equation, and utilized it for accurately computing induction times for rare events occurring during a long-- but computationally tractable-- unbiased MD (or MC) simulation.  Thapar and Escobedo~\cite{ThaparJCP2015} adapted that approach by developing procedures for estimating $P_{\text{st}}(\lambda)$ and $\tau(\lambda;\lambda_A)$ from an FFS calculation using the approach described below.\\

\noindent
\textbf{{Estimation of $P_{st}(\lambda)$}}: $P_{st}(\lambda)$ is the probability that a trajectory originating in $A$ visits a microstate with order parameter $\lambda$, and can be obtained using the following expression:
\begin{eqnarray}
P_{st}(\lambda) &=& \omega(\lambda;\lambda_A) + \pi(\lambda; \lambda_0) \notag\\&&+ \sum_{i=1}^{N-1} \pi(\lambda;\lambda_i) \prod_{j=0}^{i-1} P(\lambda_{j+1}|\lambda_{j})
\label{eq:Nlambda}
\end{eqnarray}
Here, $\pi(\lambda;\lambda_i)$ is the average time spent at $\lambda$ by a trajectory originating in $\lambda_i$. The above equation adds the contributions of trajectories originating in $\lambda_0$ to the re-weighted contribution of those originating from $\lambda_i$ ($i \geq 1$). The additional term included in Eq.~(\ref{eq:Nlambda}) is $\omega(\lambda;\lambda_A)$, which corresponds to the average time spent at $\lambda$ by a trajectory originating in $\lambda_A$ and terminating at $\lambda_0$.
The quantities in Eq.~(\ref{eq:Nlambda}) can be easily estimated from FFS, using the following expressions:
\begin{eqnarray}
\pi(\lambda; \lambda_i) &=& \frac{N_{\lambda}}{M_i\Delta\lambda} \\
\omega(\lambda;\lambda_A) &=& \frac{N_{\lambda}}{N_0\Delta\lambda}
\end{eqnarray}
Here, $N_{\lambda}$ is the number of configurations along a trajectory initiated in $\lambda_i$ (or $\lambda_A$) and terminated at $\lambda_{i+1}$ (or $\lambda_0$) with an order parameter in the range $[\lambda,\lambda+\Delta\lambda)$.  $M_i$ and $N_0$ are the number of  trajectories initiated at $\lambda_i$ and $\lambda_A$, respectively. Note the similarity between  Eq.~(\ref{eq:Nlambda}) and Eq.~(\ref{eq:2FFS-tau-plus}) utilized for computing $\tau_+$ in Valeriani~\emph{et al.}'s forward-backward FFS method~\cite{ValerianiJChemPhsy2007}. \\

\noindent
{\textbf{Estimation of Mean First Passage Times}}: The approach utilized for computing $\tau(\lambda;\lambda_A)$ depends on the value of $\lambda$. For $\lambda_A\le\lambda\le\lambda_0$, $\tau(\lambda;\lambda_A)$ can be computed from conventional (MD or MC) trajectories within $A$. Let $n_0$ be the number of uncorrelated configurations at $\lambda_A$, and let $t_i(\lambda;\lambda_A)$ be the time that it takes for a trajectory initiated from the $i$th such configuration to reach an order parameter value between $\lambda$ and $\lambda + \Delta \lambda$. $\tau(\lambda;\lambda_A)$ will then be given by:
\begin{eqnarray}
\tau(\lambda;\lambda_A) &=& \frac{1}{n_0}\sum_{i=1}^{n_0}t_i(\lambda;\lambda_A),~~~~~ \lambda_A\le\lambda\le\lambda_0\label{eq:tau-lmb0-lmbA}
\end{eqnarray}
For $\lambda_k<\lambda\le\lambda_{k+1}$, $\tau(\lambda;\lambda_A)$ is computed from the following expression:
\begin{widetext}
\begin{eqnarray}
\tau(\lambda;\lambda_A) &=& \frac{M_k\tau(\lambda_k;\lambda_A)+\sum_{i=1}^{S_\lambda}t_i(\lambda;\lambda_k|\lambda_A)+\sum_{i'=1}^{M_k-S_\lambda}t_{i'}(\lambda_A;\lambda_k|\lambda)}{S_\lambda},~~~~\lambda_k<\lambda\le\lambda_{k+1}\notag\\&&\label{eq:tau-lmbk-lmbA}
\end{eqnarray}
\end{widetext}
Here, $\tau(\lambda_k;\lambda_A)$ is the mean first passage time computed from the previous iteration (Eq.~(\ref{eq:tau-lmb0-lmbA}) for $k=0$ and Eq.~(\ref{eq:tau-lmbk-lmbA}) for $k\neq0$). $M_k$ is the number of trial trajectories initiated at $\lambda_k$, while $S_\lambda$ is the number of those that reach $\lambda$ earlier than returning to $A$. $t_i(\lambda;\lambda_k|\lambda_A)$ and $t_{i'}(\lambda_A;\lambda_k|\lambda)$ correspond to the average time needed for a first crossing of $\lambda$ (in the case of "successful`` $S_\lambda$ trajectories), and to return to $\lambda_A$ (in the case of ''failing`` $M_k-S_\lambda$ trajectories), respectively. 

In order to assess the performance of FFS-MFPT, Thapar and Escobedo~\cite{ThaparJCP2015} utilized it for computing free energy profiles in three different systems, switching nucleation in the 2D Ising model (also studied in Ref.~\citenum{ValerianiJChemPhsy2007}), a mean field model system called the EVB potential~\cite{PetersJChemPhys2013}, and homogeneous crystal nucleation in hard polyhedra. They found reasonable agreement between FFS-MFPT and umbrella sampling for the 2D Ising model, and nucleation in  several hard particle systems. For the EVB potential, however, the agreement depended heavily on choosing a good order parameter. Finally, for  hard cuboctahedra, the agreement was not good considering the proximity of FFS milestones and artifacts arising from the jumpiness of the utilized order parameters. Since its development, the FFS-MFPT method has been utilized for computing free energy profiles in several systems~\cite{RichardJCP2018, JiangJChemPhys2018, Malmir2019Arxiv}.

\subsubsection{Probability Splitting Method} 
\label{section:method:probability-splitting}

\noindent This method is due to Richards and Speck~\cite{RichardJCP2018} and is based on expressing the equilibrium distribution $P(\lambda)$ as:
\begin{eqnarray}
P(\lambda) &=& \frac{P_{\text{st}}(\lambda)}{P_A(\lambda)}
\label{eq:splitting}
\end{eqnarray}
Here, $P_A(\lambda)$ is called the splitting probability, and is the likelihood that a configuration with order parameter value $\lambda$ falls back to $A$. In general, $P_A(\lambda)$ can be estimated from:
\begin{eqnarray}
P_A(\lambda) &=& \frac{\displaystyle\int_{\lambda}^{\lambda_B}\dfrac{d\overline{\lambda}}{f(\overline{\lambda})P(\overline{\lambda})}}{\displaystyle\int_{\lambda_A}^{\lambda_B}\dfrac{d\overline{\lambda}}{f(\overline{\lambda})P(\overline{\lambda})}}
\end{eqnarray}
with $f(\lambda)$ the attachment rate. In general,  Eq.~(\ref{eq:splitting}) can be solved self-consistently to obtain $P(\lambda)$, a task that is not trivial. Richards and Speck~\cite{RichardJCP2018}, however, argue that for sufficiently large free energy barriers, $P_A(\lambda)$ can be directly estimated from FFS as $P_A(\lambda_i)=1-P_B(\lambda_i) = 1-\prod_{k=i+1}^{N}P(\lambda_k|\lambda_{k-1})$. They utilize the method to compute the free energy profiles for crystal nucleation in the Weeks-Chandler-Andersen (WCA) system~\cite{WeeksChandlerAndersenPRA1971} and found excellent agreement with FFS-MFPT and umbrella sampling methods.

\subsubsection{Forward-flux/Umbrella Sampling (FFS-US)} 

\noindent
This method, which was developed by Borrero and Escobedo~\cite{BorreroJPCB2009}, is conceptually related to an earlier technique known as partial path transition interface sampling~\cite{MoroniPhysRevE2005}. Due to the insufficient statistical accuracy of the attained free energy profiles and due to possible inaccuracies caused by memory effects, they refine $F^{\text{FFS}}(\lambda)=-\ln h^{\text{FFS}}(\lambda)$, the free energy profile estimate obtained from FFS via  auxiliary umbrella sampling simulations conducted within windows separating consecutive FFS milestones. For $\lambda\in [\lambda_i,\lambda_{i+1}]$, $h^{\text{FFS}}_i(\lambda)$ is given by:
\begin{eqnarray}
h^{\text{FFS}}_i (\lambda) &=& f_{i\rightarrow i+1}(\lambda) + W_{i+1}l_i(\lambda)  \notag\\
&& + W_i\left[b_{i+1\rightarrow i}(\lambda) + l_{i+1}(\lambda)\right]
\label{eq:h-FFS-US}
\end{eqnarray} 
Here, $f_{i\rightarrow i+1}(\lambda)$ and $l_i(\lambda)$ are histograms populated from trajectories initiated at $\lambda_i$, and correspond to parts of such trajectories that connect $\lambda_i$ and $\lambda_{i+1}$ without any intermittent crossings of $\lambda_i$, and those that loop around $\lambda_i$, respectively. $b_{i+1\rightarrow i}(\lambda)$ is, however, computed from parts of trajectories initiated at $\lambda_{i+1}$ that directly connect $\lambda_{i+1}$ and $\lambda_i$ without any intermittent crossing of $\lambda_{i+1}$. $W_i$'s are included to properly reweigh forward and backward trajectories to construct the equilibrium free energy profiles. More precisely, the flux of forward and backward trajectories that cross each milestone need to be equal under steady-state, a condition that is not satisfied in FFS, and is mitigated by reweighing backward trajectories by the factor $W_i=n_{i+1\rightarrow i}/n_{i\rightarrow i+1}$. Here $n_{i+1\rightarrow i}$ and $n_{i\rightarrow i+1}$ are the number of partial paths originating at $\lambda_i$ and $\lambda_{i+1}$ that meet $\lambda_{i+1}$ and $\lambda_i$ before reaching $\lambda_{i-1}$ and $\lambda_{i+2}$, respectively. Note that $W_0=1$ since a trajectory that crosses $\lambda_0$ is allowed to crawl back to $A$ while sampling the $A$ basin.

The next step is to refine the profile computed from Eq.~(\ref{eq:h-FFS-US}) using umbrella sampling within windows separating consecutive FFS milestones. In particular, the authors prescribe the approach described by Virnau and M\"{u}ller~\cite{VirnauJChemPhys2004} in which the biasing potential for the $i$th window  is given by:
\begin{eqnarray}
u_i^{\text{bias}} (\textbf{r}) &=& \left\{
\begin{array}{ll}
0 & ~~~~~\lambda_i\le \lambda(\textbf{r})\le\lambda_{i+1}\\
+\infty & ~~~~~\text{otherwise}
\end{array}
\right.
\end{eqnarray}
In other words, each sampling is conducted in unbiased fashion within windows with two hard boundaries. $H^{\text{US}}(\lambda)$, the visiting histogram for $\lambda_i\le\lambda\le\lambda_{i+1}$, can be estimated as:
\begin{eqnarray}
H^{\text{US}} (\lambda) &=& H^{\text{US}} (\lambda_0)\frac{H_i^{\text{US}}(\lambda)}{H_i^{\text{US}}(\lambda_i)}\prod_{j=0}^{i-1} \frac{H_j^{\text{US}}(\lambda_{j+1})}{H_j^{\text{US}}(\lambda_j)}
\end{eqnarray}
Here, $H_i$'s are the visiting statistics gathered from $i$th umbrella sampling simulations within the $i$th window. Note that $H^{\text{US}}(\lambda)$ can be computed from a stand-alone umbrella sampling calculation. The role of FFS, however, is twofold. First, the configurations collected as part of FFS are utilized as starting points for US calculations. Secondly, in order to attain a desired level of uncertainty in $H(\lambda) = H^{\text{FFS}}(\lambda)+H^{\text{US}}(\lambda)$, fewer steps are needed with each US window.  Borrero~\emph{et al.}~\cite{BorreroJPCB2009} utilized FFS-US to compute free energy profiles for folding of a lattice protein, as well as a potential energy surface proposed by Chopra~\emph{et al.}~\cite{ChopraJChemPhys2008} and found excellent agreements with pure umbrella sampling, albeit at a lower computational cost. Recently, Qin~\emph{et al.} proposed a similar method based on the idea of partial path reweighing. Their approach~\cite{QinJCP2019} provides a rigorous procedure for assessing the importance of memory effects, and thus the need to conduct auxiliary umbrella sampling calculations. 

\subsection{Heuristics for Accurate and Efficient Implementation}
\label{section:variants:new:efficiency}

\noindent
The main functional goal of an FFS calculation is to efficiently and accurately predict the rate and the mechanism of a rare eventd. Achieving this goal will not only depend on the quality of the utilized OPs, but might also be affected by implementation parameters, such as the number and the location of milestones, and the number of trial trajectories.  In Section~\ref{section:accuracy}, we discuss the basics of assessing the efficiency and statistical accuracy of FFS. One can, in principle, formulate the problem of determining the optimal values of such parameters as a constrained non-linear optimization problem. In recent years, several researchers have employed this approach to derive heuristics for milestone staging, as well as deciding the amount of sampling at each iteration. In order to keep their proposed heuristics simple and universal, however, they have all conducted their analyses assuming that landscape variance does not contribute significantly to the overall error. This section is dedicated to a detailed discussion of such efforts. But it is necessary to emphasize that applying these heuristics should always be done with caution, as overlooking the effect of landscape variance can result in large errors that are not only difficult to quantify, but can also propagate over many FFS iterations.


\subsubsection{Milestone Placement Strategies}

\noindent
A critical aspect of any FFS calculation is to choose the number and locations of milestones, as well as to decide the number of trial trajectories initiated at each milestone. Both these decisions are typically made manually pursuant to certain guidelines discussed in the literature. One such widely utilized heuristic is due to Borrero and Escobedo~\cite{BorreroEEJCP2008, EscobedoJPCM2009}, who minimized the statistical uncertainty  subject to a fixed overall transition probability. They found the optimal staging of milestones to correspond to a constant flux across milestones, namely a fixed $M_ip_i$. This will, for instance, imply maintaining a constant transition probability when the number of trial trajectories initiated at each milestone is also kept constant. By minimizing the statistical error subject to a constant computational cost, however, they obtained the optimal number of trajectories per milestone for a given staging of milestones. They demonstrated that the optimal $M_i$'s need to satisfy  $M_i^2p_i(\lambda_i-\lambda_A)=\text{const}$. As mentioned above, these heuristics are based on the assumption that landscape variance is unimportant. In reality, however, more extensive sampling is required in iterations for which landscape variance is large.

In the approach  proposed by Borrero and Escobedo, the total number of FFS milestones, $N$, is an input parameter. They, however, provide no guidelines on how to choose $N$.   Kratzer~\emph{et al.}~\cite{AllenJCP2013} invoked the constant-probability  heuristic of Borrero and Escobedo~\cite{BorreroEEJCP2008, EscobedoJPCM2009} to conclude that the optimal $N$ is given by:
\begin{eqnarray}
N &=& \frac{\log P(\lambda_B|\lambda_0)}{p}
\end{eqnarray}
Here $p$ is the transition probability between successive milestones, and is chosen to maximize the FFS efficiency defined in (\ref{eq:efficiency}). They estimated the computational cost $\mathcal{C}$ by using Eq.~(\ref{eq:cost-per-conf}) and assuming a fixed number of trial trajectories per milestone:
\begin{eqnarray}
\mathcal{C} &\approx& N_0R + M\sum_{i=0}^{N-1}\mathcal{C}_i \label{eq:cost-simplified}
\end{eqnarray}
They also further assumed that the computational cost of each trajectory is proportional to the number of FFS milestone that it crosses, i.e.,
\begin{eqnarray}
\mathcal{C}_i & \approx&\frac{S}{N} \bigg[p + i(1-p)\bigg]
\end{eqnarray}
with $S$ the cost of conducting a trajectory from $A$ to $B$. Similar to Ref.~\citenum{BorreroEEJCP2008}, they neglected the effect of landscape variance  in Eq.~(\ref{eq:variance-pj}) to conclude:
\begin{eqnarray}
\mathcal{V} &\approx& \frac{N(1-p)}{Mp} \label{eq:variance-simplified}
\end{eqnarray}
In general, $\mathcal{E}(p)$ is not a strong function of $p$, hence the authors prescribe a range of $p\in [0.3,0.7]$ for obtaining reasonable efficiency without making successive milestones correlated. They then proposed two procedures for automatically staging the FFS milestones to maintain a fixed transition probability. In both approaches, the location of $\lambda_{k+1}$ is determined  \emph{after} the transition from $\lambda_{k-1}$ to $\lambda_k$ is already complete. 

In the \emph{trial interface method}, a suitable initial guess  $\lambda_{k+1}^{\text{est}}$ is chosen, and a small number of $M_{\text{trial}}\ll M$ (typically $\approx 15$) exploratory trajectories are initiated at $\lambda_k$ to compute $p_{\text{est}}$, the transition probability between $\lambda_k$ and $\lambda_{k+1}^{\text{est}}$. If $p_{\text{est}}$ lies within the prescribed range $[p_{\min},p_{\max}]$, $\lambda_{k+1}^{\text{est}}$ is accepted as the $(k+1)$-th milestone. Otherwise, a new guess, $\lambda_{k+1}^{\text{est,new}} = \lambda_{k+1}^{\text{est}} + \lambda_{\text{step}}\Delta p$ is chosen until $p_{\text{est}}$ lies within $[p_{\min},p_{\max}]$. Here, $\Delta p = (p_{\text{est}}-p_{\max})$ or $(p_{\text{est}} - p_{\min})$, depending on whether $p_{\text{est}} > p_{\max}$ or $p_{\text{est}}  < p_{\min}$. As described in Ref.~\citenum{AllenJCP2013}, a different correction rule might be adopted based on the energetics of the underlying system. This algorithm also requires specifying a minimum distance $d_{\min}$ between successive interfaces in order to prevent correlations between successive milestones. Note that the trial interface method can be computationally costly if the initial estimate of $\lambda_{k+1}^{\text{est}}$ is poor.

The second algorithm is called the \emph{'Exploring-Scouts` method}, and does not require an initial estimate of $\lambda_{k+1}$. Instead, the $M_{\text{trial}}$ trajectories initiated at $\lambda_k$ are monitored for their largest $\lambda$ value, which are then used to get a distribution of potential $\lambda_{k+1}'s$. Each trajectory is terminated upon reaching $\lambda_A$ or $\lambda_B$, or after $m_{\max}$ steps, and their maximum $\lambda$'s are sorted as $\lambda_{\max}^{(1)}\le \lambda_{\max}^{(2)}\le\cdots\le\lambda_{\max}^{(M_{\text{trial}})}$. It can be easily noted that the tentative next milestone can be chosen as $\lambda_{\max}^{(j)}$ so that $p=(M_{\text{trial}}-j)/M_{\text{trial}}$ lies within the prescribed range. Despite depending on fewer number of user-defined parameters, the success of this approach depends on choosing a reasonable $m_{\max}$ in order to balance the computational cost of conducting a large number of exploratory trajectories, and the risk of having correlated milestones for  $m_{\max}$'s that are too small.

\subsubsection{FFPilot}  \label{section:FFSPilot}

\noindent
This method is due to Klein and Roberts~\cite{KleinBioRxiv2018}, who use a nonlinear optimization approach to identify $M_k$'s that minimize the computational cost of FFS subject to a user-defined level of statistical uncertainty in the mean first passage time, i.e.,~the inverse of the rate constant given by Eq.~(\ref{eq:FFS}). Unlike Borrero and Escobedo~\cite{BorreroEEJCP2008, EscobedoJPCM2009} and Kratzner~\emph{et al.}~\cite{AllenJCP2013}, Klein and Roberts~\cite{KleinBioRxiv2018} do not make any assumptions about how the computational cost of each iteration scales with $\lambda$. Their adopted notation are slightly different from those widely in use in the rest of the literature, and need to be introduced here. More precisely, they define the random variable $\xi_0$ as the waiting time between successive crossings of $\lambda_0$ during basin exploration, while $\xi_{k(>0)}$ is a Bernoulli random variable that is one when a trajectory initiated at $\lambda_{k-1}$ reaches $\lambda_k$ before returning to $A$ and is zero otherwise. The mean first passage time $W$ can then be expressed as:
\begin{eqnarray}
W &=& \frac{w_0}{\prod_{k=1}^Nw_k} \label{eq:FFSpilot-W}
\end{eqnarray}
with $w_k=\langle\xi_k\rangle$.  The statistical estimators of $w_k$'s from FFS are denoted by $\hat{w}_k$ and are given by:
\begin{eqnarray}
\hat{w}_k &=& \left\{
\begin{array}{ll}
\tau_A & k=0\\
P(\lambda_k|\lambda_{k-1}) & k>0
\end{array}
\right.
\end{eqnarray} 
where $\tau_A=1/\Phi_{A,0}$ is the average wait time between successive first crossings of $\lambda_0$. The statistical uncertainty in $\widehat{W}$ is given by:
\begin{eqnarray}
\zeta(\widehat{W};\alpha) &=& \frac{z_\alpha\sqrt{V[\widehat{W}]}}{\left\langle \widehat{W}\right\rangle} = \frac{z_\alpha\sqrt{V[\widehat{W}]}}{W}
\end{eqnarray}
Here, $\alpha$ is the confidence level associated with $\widehat{W}$ and $z_\alpha$ is its associated $z$ score. The authors estimate $V[\widehat{W}]$ by using the multivariate delta method~\cite{WassermanSpringer2013}. Furthermore, since the number of first crossings of $\lambda_0$, as well as the trial trajectories initiated at each $\lambda_k$ are both large, the central limit theorem will imply that $\hat{w}_k$ will converge in distribution to a Gaussian random variable with mean $\langle\xi_k\rangle$ and variance $V[\xi_k]/M_k$. Assuming that $\xi_k$'s are uncorrelated, $V[\widehat{W}]$ will be given by:
\begin{eqnarray}
V[\widehat{W}] &=& W^2\sum_{k=0}^N\frac{V[\xi_k]}{w_k^2n_k}
\end{eqnarray}
Here, $n_k$ is the number of sampled $\xi_k$ replicas, namely the number of first crossings of $\lambda_0$ for $k=0$ and $M_{k-1}$, the number of trajectories initiated at $\lambda_{k-1}$ for $k>0$.  The next step is to minimize $\mathcal{C}=\sum_{k=0}^Nn_kc_k$ with the constraint that $\zeta(\widehat{W};\alpha)$ is equal to a user-defined value. Here $c_k$ is the average computational cost of observing a first crossing of $\lambda_0$ for $k=0$, and the average duration of a trial trajectory for $k>0$. Using Lagrange multipliers and neglecting the effect of landscape variance, it can be shown that $\mathcal{C}$ is minimized by the following $n_k$'s:
\begin{eqnarray}
n_0^* &=& \frac{z_\alpha^2}{\zeta^2} \sqrt{\frac{V[\hat{w}_0]}{w_0^2c_0}}\Bigg[\sqrt{\frac{V[\hat{w}_0]}{w_0^2c_0}} 
\notag\\&& + \sum_{j=1}^N\sqrt{\frac{c_j(1-w_j)}{w_j}}\Bigg]\label{eq:FFSpilot-opt-n0}\\
M_{k-1} = n_{k(>0)}^*  &=& \frac{z_\alpha^2}{\zeta^2} \sqrt{\frac{c_k(1-p_k)}{p_kc_k}}\Bigg[\sqrt{\frac{V[\hat{w}_0]}{w_0^2c_0}}\notag\\&&  + \sum_{j=1}^N\sqrt{\frac{c_j(1-w_j)}{w_j}}\Bigg]\label{eq:FFSpilot-opt-nk}
\end{eqnarray}
Note that optimal $n_k$'s depend on the transition probabilities and computational costs of \emph{all} $N$ iterations. The authors therefore propose a two-step approach in which $V[\xi_0]/w_0^2$, $c_k$'s, and $p_k$'s are all estimated during a pilot simulation, and the optimal $n_k^*$ values obtained from Eqs.~(\ref{eq:FFSpilot-opt-n0}) and (\ref{eq:FFSpilot-opt-nk}) are then utilized to launch a production FFS calculation.  The pilot stage is conducted according to a blind optimization scheme, wherein phase $k$ is terminated only when a certain number of successful crossings are obtained, allowing for obtaining an upper bound on $V[\xi_k]$. For a relative error of 2\%, for instance, $n_{\text{pilot}}$ needs to be $\approx10^4$, but smaller $n_{\text{pilot}}$'s can be used if higher relative errors are permitted. The authors also argued that an optimal placement of $\lambda_0$ will assure that $V[\xi_0]\approx\tau_A$, as $\xi_0$ is expected to be a Poisson random variable with mean $\tau_A$. 

The authors tested the FFPilot method in three systems, a toy model called the rare event model (REM), a self-regulatory gene model (SRG), and a genetic toggle switch (GTS) model. The algorithm was found to efficiently control the level of uncertainty for REM and SRG. For GTS, however, deviations from the specified level of uncertainty were larger, due to the importance of landscape variance. This problem can only be remedied through an across-the-board oversampling since such errors are correlated.

A fact that severely limits the applicability of FFPilot to real molecular systems is its  large computational cost, since the number of successful crossings needed even in its pilot stage is much larger than what is practically possible in the applications of FFS discussed in Section~\ref{Impact of FFS}. Moreover, the existence of landscape variance in almost all real systems could make the error estimates obtained from FFPilot unreliable.


\section{FFS Implementations\label{section:implementation}}
\noindent The FFS variants discussed in Sections~\ref{section:classical:variants} and~\ref{section:variants:new} have been utilized for studying a wide variety of rare events in many systems, some of which will be discussed in Section~\ref{Impact of FFS}.  Expanding the applicability of FFS to more systems and/or rare events can, however, face the following implementation bottlenecks.  On one hand, it is necessary to develop computer programs that interface with a suitable MD or MC engine, and efficiently compute all the necessary collective variables.  Furthermore, workflows need to be devised to ensure proper system set-up, interface placement, monitoring of trial trajectories, and keeping track of crossing events and order parameter time series. Achieving all these objectives via a low-level system- and process-specific implementation can prove prohibitive to potential new users.
Developing more scalable and generalizable software packages that allow for  a simplified, high-level application of FFS and other advanced sampling techniques has therefore become an intense focus of activity in recent years. Several software packages with FFS capability, such as PLUMED~\cite{BonomiCompPhyComm2009, TribelloCompPhyComm2014}, flexible rare event sampling harness system (FRESHS)~\cite{KratzerCompPhyCommu2014} and SSAGES~\cite{SidkyJCP2018}, already offer open-source functional implementations, while others, such as scalable automated FFS for illuminating rare events (SAFFIRE)~\cite{XuanIEEE2013, HangerScaFFS2015, DeFeverPEARC2019}, parallel forward flux sampling (PFFS)~\cite{AllenPFFS2013}, and AdvSamp~\cite{HajiAkbariPNAS2017, AmirHajiAkbariJCP2018}, are yet to become publicly available. In addition, several python libraries, such as PyRETIS~\cite{LervikPyretis2017} and Open paths sampling~\cite{openpathsamp}, have been developed for conducting committor analysis as well as several path sampling techniques such as TPS, TIS and replica exchange TIS (RETIS). Due to their built-in capabilities, these libraries can be modified for conducting FFS calculations.

One of the first open-source packages suitable for advanced sampling calculations is the PLUgin for MolEcular Dynamics (PLUMED)~\cite{BonomiCompPhyComm2009, TribelloCompPhyComm2014} package, which is a plugin-based software that provides patching procedures for several common MD engines, such as GROMACS~\cite{PronkBioinformatics2013}, LAMMPS~\cite{PimptonLAMMPS1995}, NAMD~\cite{PhillipsJCompChem2005} and {\sc Quantum ESPRESSO}~\cite{GiannozziJPCM2009, GiannozziJPCM2017}. The original PLUMED package was capable of conducting several advanced sampling techniques such as metadynamics, umbrella sampling, thermodynamic integration and replica exchange MD using a wide variety of collective variables. In the newer PLUMED 2.0~\cite{TribelloCompPhyComm2014} version, which is parallelized using MPI, it is possible to define new free energy and path sampling methods and collective variables without editing the core functionalities of the software.

The software package FRESHS~\cite{KratzerCompPhyCommu2014} has been developed for parallel implementation of splitting methods such as FFS and stochastic process rare event sampling (S-PRES)~\cite{BerrymanJCP2010}. Similar to PLUMED, FRESHS provides plugins to interface with GROMACS, LAMMPS and ESPResSO~\cite{LimbachCompPhyComm2006, ArnoldMeshfree2013, WeikEuroPhyJ2019}. It works on a server-client framework wherein the client side has an MD engine attached to it while the server side implements sampling-algorithm in a modular fashion and communicates with the client via a socket layer. This allows FRESHS to have efficient parallelization with low communication and high modularity for both the sampling module and the MD engine. 

Another recent package called Software Suite for Advanced General Ensemble Simulations (SSAGES)~\cite{SidkyJCP2018} allows for the application of several sampling techniques, such as umbrella sampling~\cite{TorrieJCompPhys1977}, FFS, adaptive biasing force~\cite{DarveJCP2001}, nudged elastic band (NEB)~\cite{BohnerJCP2014}, metadynamics~\cite{LaioPNAS2002, BarducciCompMolSci2011,  SuttoCompMolSci2012} and artificial neural network (ANN) sampling~\cite{SidkyJCP2018p}, with the ability to interface with most of common MD engines in an engine-agnostic fashion. The latter is achieved by interfacing with an MD engine using an adapter called a "hook``, which allows the user to employ any engine for which an appropriate hook is already developed. Moreover, the concrete steps needed for adding new methods and collective variables to the package are properly outlined in the documentation. 

In addition to these publicly available packages, several other packages with FFS capability, such as SAFFIRE~\cite{XuanIEEE2013, HangerScaFFS2015, DeFeverPEARC2019}, PFFS\cite{AllenPFFS2013} and AdvSamp~\cite{HajiAkbariPNAS2017, AmirHajiAkbariJCP2018}, are yet to become publicly available.
Among them, SAFFIRE is a software framework specifically designed to handle the large throughput of data generated from millions of simulation tasks needed for applying FFS to complex systems. SAFFIRE uses the HADOOP~\cite{ShvachkoIEEE2010} open-source data management infrastructure to provide an efficient and fault-tolerant framework for implementing large-scale FFS jobs. Thus far, SAFFIRE has been tested with common MD engines like LAMMPS~\cite{PimptonLAMMPS1995} and GROMACS~\cite{PronkBioinformatics2013}. The platform has been used to perform FFS for studying nucleation in clathrate hydrates~\cite{DeFeverJChemPhys2017} and its developers are working on releasing it to the public soon.  Another package is PFFS\cite{AllenPFFS2013}, which was originally developed as a message passing interface (MPI) implementation of FFS. Its development, however, seems to be still incomplete. Finally,  AdvSamp~\cite{HajiAkbariPNAS2017, AmirHajiAkbariJCP2018} provides a modular implementation of path sampling methods by providing independent base modules for evolving the intrinsic dynamics, computing order parameters, and conducting path sampling simulations. This modularity makes extending the package fairly straightforward. Moreover, the modules that are responsible for evolving the intrinsic dynamics maintain memory-level communications with the underlying MD engine. This makes FFS calculations more efficient by minimizing the overheads associated with reading and  writing trajectories. Similar to SAFFIRE, the developers of AdvSamp are working on releasing it to the public in the near future.

  These publicly available softwares have made it much easier for users to implement FFS, in some cases in its optimized forms (e.g.,~with automatic interface placement) without worrying about the underlying workflow and data management. Consequently, FFS has been made available to a wider community of researchers, which has, in turn, resulted in its applications to a broader class of problems discussed in Section~\ref{Impact of FFS}.

\section{FFS Applications}\label{Impact of FFS}

\noindent 
In the past decade, FFS has been used to study the kinetics and mechanisms of a wide variety of rare events in many different systems. We dedicate this section to overview these diverse applications with a particular emphasis on the operational aspects of such applications, such as the employed FFS variants and the utilized order parameters. Moreover, we  discuss the physical insight provided from such FFS calculations. The process that has been most widely studied using FFS is nucleation, a topic that will be thoroughly discussed  in Section~\ref{section:nucleation}. In addition, we will overview  applications of FFS to processes such as conformational rearrangements in biomolecules (Section~\ref{section:conformation-bio}), structural relaxation in polymer melts and solutions (Section~\ref{section:polymer-relaxation}), solute transport in membranes (Section~\ref{section:filtration}), and rare switching events (Section~\ref{section:rare-switching}). 

\subsection{Nucleation}\label{section:nucleation}

\begin{figure}
	\centering
	\includegraphics[width=.5\textwidth]{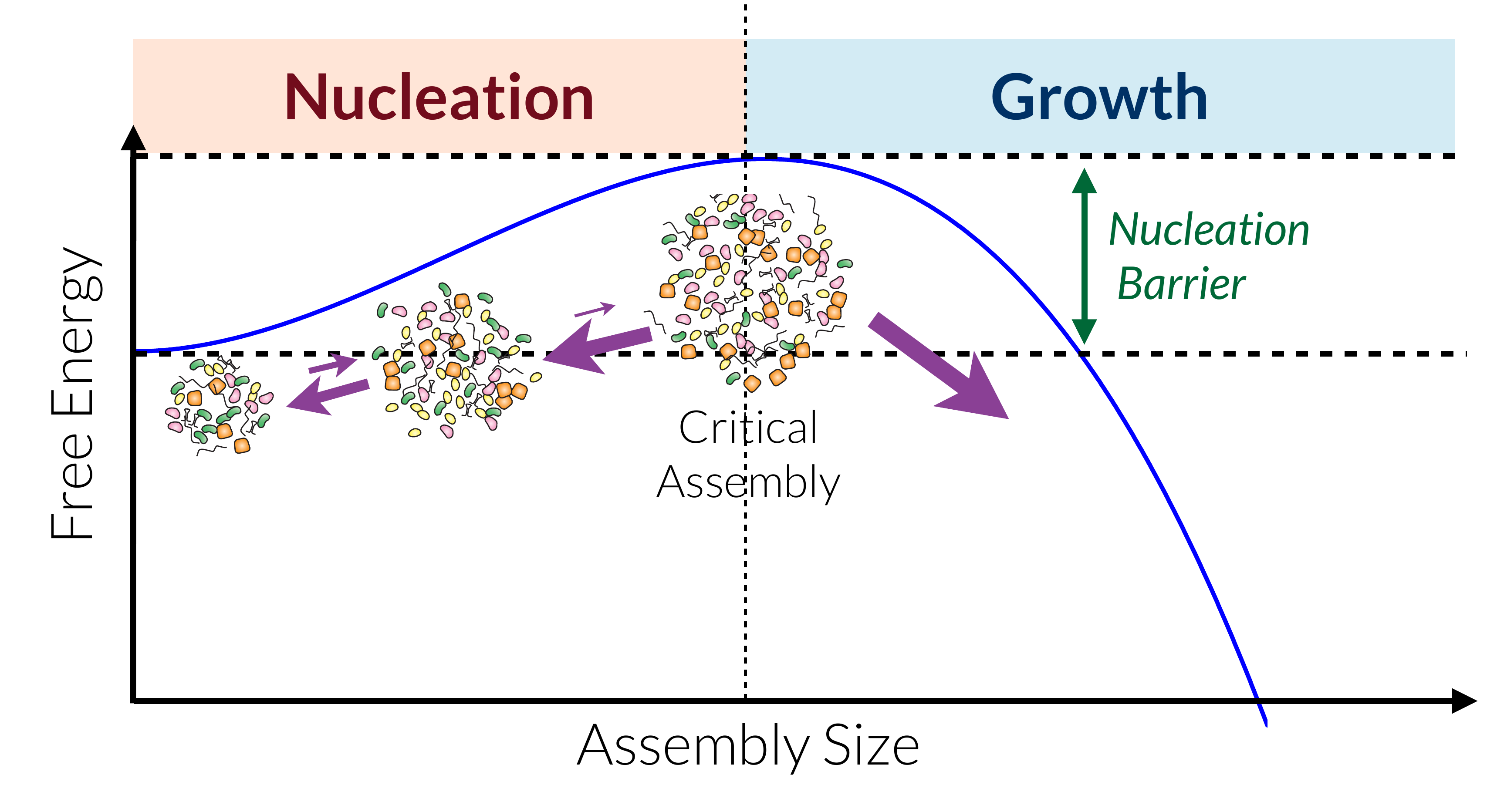}
	\caption{\label{fig:nucleation}(Reproduced from Ref.~\citenum{FalahatiSoftMatter2019}) Schematic representation of the nucleation and growth process.}
\end{figure}

\noindent
Nucleation and growth is a process through which first-order phase transitions proceed under conditions at which the thermodynamic driving force is not too large (Fig.~\ref{fig:nucleation}). During nucleation, a sufficiently large nucleus of the new phase emerges within the old metastable phase, while during growth, this critical nucleus  grows until thermodynamic equilibrium is achieved. At smaller thermodynamic driving forces, nucleation becomes a fluctuation-driven activated process, and involves crossing a free energy barrier typically known as \emph{nucleation barrier} and denoted by $\Delta{G}_{\text{nuc}}$. The existence of a barrier  arises from the dominance of surface effects in small nuclei and the energetic penalty associated with forming a two-phase interface. This makes nucleation a rare event occurring at a rate proportional to $\exp[-\Delta{G}_{\text{nuc}}/kT]$. 

Nucleation is the prevailing mechanism-- and the primary rata-limiting step-- for a wide variety of phase transitions from crystallization and liquid-liquid phase separation, to evaporation and magnetization~\cite{FalahatiSoftMatter2019}. As mentioned earlier, FFS is fairly robust to the selection of a subpar oder parameter. Yet, probing the kinetics of nucleation using FFS still requires devising order parameters that are \emph{local}, i.e.,~that can spatiotemporally resolve the formation and evolution of the emerging nucleus. This usually requires devising criteria for distinguishing the molecules that belong to different phases. For transitions occurring between two disordered phases, quantities such as local density or coordination number can usually be used for that purpose. When one  phase is liquid-crystalline, crystalline or quasicrystalline, however, a more sophisticated approach is warranted, and particular symmetries of the ordered phase need to be taken into consideration. A thorough discussion of how translational and rotational order is quantified can be found elsewhere~\cite{SteinhardtPRB1983, KeysAnnRevCondMatPhys2011, KeysJCompPhys2011, HajiAkbariJPhysA2015}. The most widely used approach, however, is to map out the local orientational signature of each molecule's coordination shell using a set of metrics known as \emph{Steinhardt order parameters}~\cite{SteinhardtPRB1983}, which are scalar invariants of complex-valued functions computed from the vectors that connect that molecule to a select set of its nearest neighbors. 
Screening for solid- and liquid-like molecules is then conducted by identifying invariants that adopt non-overlapping distributions in the two phases, and labeling each molecule as ordered or disordered accordingly. After identifying the local phase of each molecule, the neighboring molecules belonging to the target phase are clustered, and the number of molecules within the largest cluster is chosen as the FFS order parameter. This two-step framework results in a spatially resolved local order parameter, and is conceptually distinct from another approach-- commonly used in bias-based methods-- in which Steinhardt order parameters for different molecules are spatially averaged to obtain a global order parameter. The local (per-particle) and global Steinhardt order parameters are typically denoted by $q_l$ and $Q_l$ respectively, with $l$ the order of the Legendre polynomial utilized in their calculation. A more detailed discussion of Steinhardt order parameters can be found in Ref.~\citenum{SteinhardtPRB1983}.

\subsubsection{Nucleation in On-Lattice Spin Models}
\label{section:nucleation:Ising}

\begin{figure}
	\centering
	\includegraphics[width=.45\textwidth]{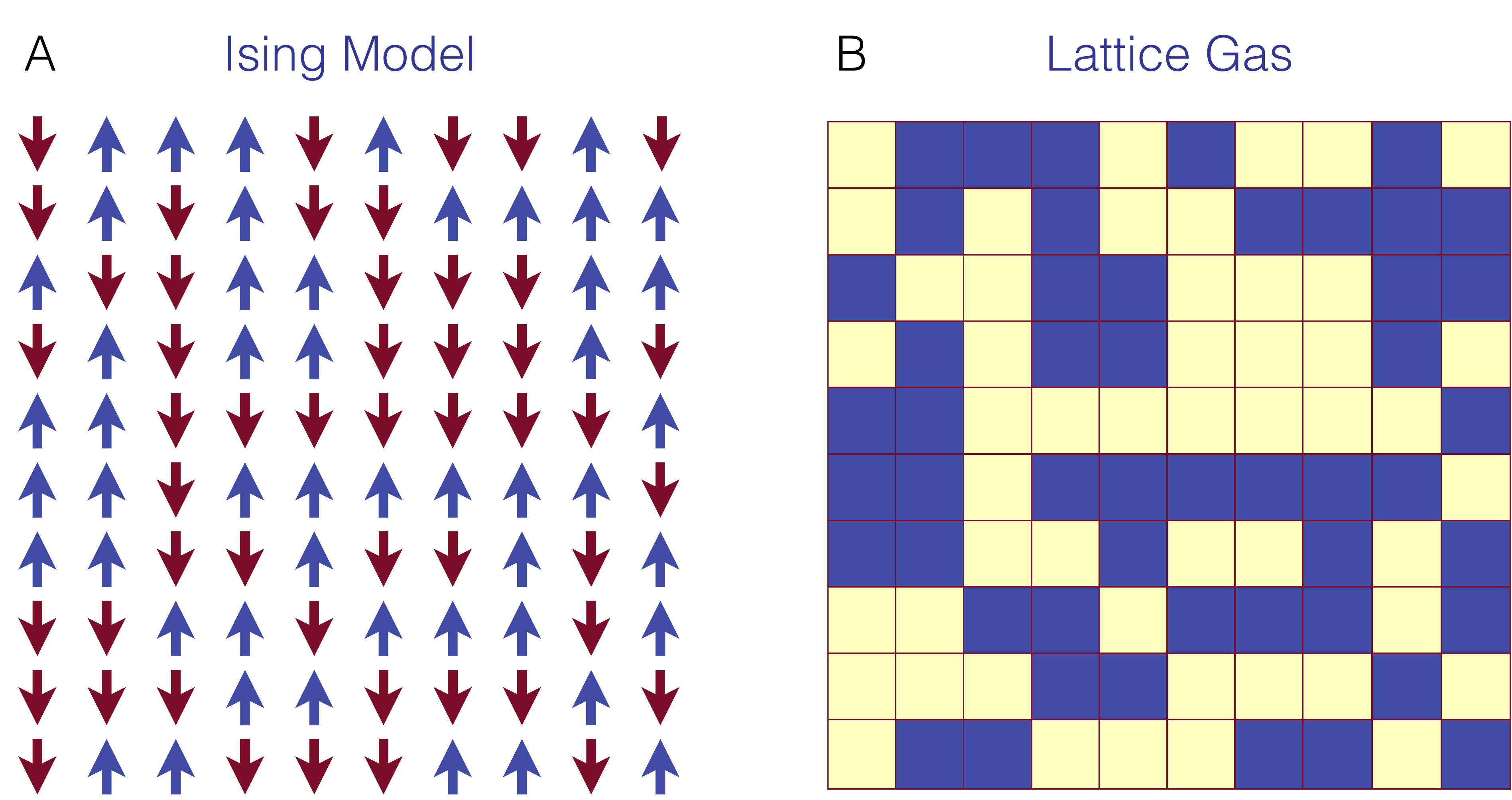}
	\caption{\label{fig:ising}Schematic representation of the (A) Ising and (B) lattice gas models. Occupied and empty sites in (B) are depicted in dark blue and yellow, respectively. The Ising model configuration in (A) is mapped onto the lattice gas configuration in (B). }
\end{figure}

\noindent
Spin models constitute minimal models of ferromagnetic materials. In discrete spin models, magnetic spins $\pmb\sigma\equiv(\sigma_1,\sigma_2,\cdots)$ are assigned to the vertices of a graph, and each spin interacts with its neighboring spins (i.e.,~spins connected via edges)  and an external magnetic field. Spin models are typically formulated on a lattice, i.e.,~a graph with a periodic (or self-repeating) arrangement of vertices and edges.   Due to their simplicity and on-lattice nature,  spin models have been extensively studied to deduce generic features of phase transitions and critical phenomena~\cite{LeePhysRev1952, StanleyPhaseTransitions1971, FerrenbergPhysRevB1991}. The oldest and the most famous spin model is the \emph{Ising model}~\cite{IsingZPhys1925} in which only two types of spins (up and down) are permitted (Fig.~\ref{fig:ising}A), and the Hamiltonian of the system is given by:
\begin{eqnarray}
H(\pmb\sigma) &=& -J\sum_{\langle i,j\rangle} \sigma_i\sigma_j - h\sum_j\sigma_j
\end{eqnarray}
Here, $\langle i,j\rangle$ refers to a summation over  nearest neighbors, i.e.,~spins connected via an edge in the graph, and $J$ and $h$ are the coupling parameter, and the external field, respectively. One- and two-dimensional Ising models are among a handful of systems with nontrivial Hamiltonians for which the partition function can be calculated analytically, including the \emph{tour de force} solution of Onsager for the Ising model on a square lattice~\cite{OnsagerPhysRev1944}. For lattices in higher dimensions as well as more complex networks,  Metropolis Monte Carlo has been extensively used for probing the thermodynamics of the Ising model~\cite{FerrenbergPhysRevB1991}. It can be shown that for dimensions $d\ge2$, a transition from a spin-disordered into a spin-ordered phase occurs below a critical temperature. Such a transition will be first-and second-order for $h\neq0$ and $h=0$, respectively. It is also necessary to note that the Ising model can be easily mapped onto another popular model, called a \emph{lattice gas} (Fig.~\ref{fig:ising}B), by defining $n_i:=(1+\sigma_i)/2$. In a lattice gas, each vertex will be either  occupied $(n_i=1)$ or vacant $(n_i=0)$. 

In recent years, FFS has been extensively utilized for studying the sensitivity of the nucleation kinetics in the Ising and lattice gas models to  the existence and energetic properties of impurities~\cite{SearJPhysChemB2006, SearEPL2008, SearJPhysCondensMatter2012}, mechanical deformations~\cite{AllenJChemPhys2008, AllenProgTheorPhys2008}, internal and external interfaces~\cite{PagePRL2006, SearEPL2008, SearJPhysCondensMatter2012, HedgesSoftMatter2012, HedgesSoftMatter2013}, and the topological properties of the underlying network~\cite{ChenPhysRevE2011, ChenPhysRevE2011p, ShenChaos2013, ShenEPL2015}. Since nucleation will only occur if the underlying transition is first-order, all these simulations were conducted in the presence of an external magnetic field or chemical potential, and the starting basin $A$ is always set to the respective metastable phase in each system. Due to the simplicity of the Ising model, it is possible to conduct FFS using both global and local order parameters, namely the total number of up spins (or occupied sites), and the size of the largest cluster of the new phase, respectively. 

The first use of FFS to study nucleation in the Ising model was conducted by Sear~\cite{SearJPhysChemB2006}, who computed heterogeneous nucleation rates in the presence of lines of fixed spins in order to assess the effect of impurities.  He observed that even the existence of one fixed up spin enhances the nucleation of the up spin phase by four orders of magnitude. The rate then increases by a factor of five for each additional fixed spin. In a more recent paper~\cite{SearJPhysCondensMatter2012}, he utilized diffusive-limited aggregation~\cite{WittenPRL1981} to generate quenched impurities (i.e.,~arbitrarily shaped clusters of up spins with a given mean size). He then employed FFS to compute the rate of nucleation in the presence of each quenched impurity, and observed that  rates can differ by as much as ten orders of magnitude. These findings underscore the perils of simple averaging of nucleation rates in different experimental realizations of a system with a known concentration of impurities with unknown size and shape distributions.

The Sear Group have also utilized FFS to study other aspects of heterogeneous nucleation. For instance, Page and Sear~\cite{PagePRL2006} used FFS to investigate heterogeneous nucleation in the presence of inert lines (i.e.,~lines with fixed spins but no interactions with the lattice), and rectangular pores. They found such walls to enhance nucleation considerably despite their inertness.  In the case of rectangular pores, nucleation involved a two-step process, i.e.,~filling of the pore with the up-spin phase, and the nucleation of the up-spin phase in the bulk from the filled pore. The kinetics of these two processes scale differently with the pore width $W$, resulting in a non-monotonic dependence of the overall rate on $W$. In a later paper, Sear utilized the lattice gas model to investigate the impact of nanoparticle solubility on the rate of condensation~\cite{SearEPL2008}, which was found to increase considerably upon increasing solubility.

The question of heterogeneous nucleation in the lattice gas models have been investigated by other groups as well. For instance, Hedges and Whitelam\cite{HedgesSoftMatter2012, HedgesSoftMatter2013} studied condensation in the presence of pores.  In their 2012 paper~\cite{HedgesSoftMatter2012}, they considered nucleation on  surfaces regularly etched with rectangular pores in two dimensions and cuboidal pores in three dimensions, and demonstrated that only the pores with proper sizes and aspect ratios enhanced nucleation (in comparison to flat surfaces). They also confirmed that FFS is robust to lack of locality in the OP, as they obtained identical rates with global and local OPs, namely the total number of filled sites, and the size of the largest dense cluster, respectively. In a second paper~\cite{HedgesSoftMatter2013}, they demonstrated that the characteristic length of a pore that yields the largest rate at each state point is almost identical to the characteristic size of the critical cluster in the bulk under the same thermodynamic conditions. 

FFS has also been utilized for studying homogeneous nucleation under mechanical deformation. For instance, Allen~\emph{et al.}~\cite{AllenJChemPhys2008} used FFS to probe nucleation in the sheared Ising model. They introduced shearing using the algorithm of Cirillo~\emph{et al.}~\cite{CirilloPhysRevE2005} in which each shearing trial move corresponds to randomly selecting a row and shifting all sites above it along a fixed pre-specified direction. They found the nucleation rate to depend non-monotonically on the shear rate, i.e.,~the average number of shear moves per MC sweep. This observation was attributed to the fact that lower shear rates facilitate the coalescence of precritical clusters, while higher shear rates induce their breakup. They further confirmed this explanation by conducting rattle shearing simulations in which the direction of shearing was randomly chosen at each move. By doing so, the nucleation rate increased monotonically with  shear rate, as the stochasticity in shear direction suppressed  cluster breakup. In a follow-up study~\cite{AllenProgTheorPhys2008}, the same authors used FFS to probe the effect of external magnetic field on nucleation under shear, and found the non-monotonicity to disappear in the presence of sufficiently strong external fields.

In addition to regular lattices, several studies have focused on homogeneous and heterogeneous nucleation in non-periodic networks, such as random graphs. For instance Chen and Hou~\cite{ChenPhysRevE2011} used FFS to study homogeneous nucleation in a graph comprised of two modules with differing intra- and inter-modular connectivities. They found that the rate exhibits a non-monotonic dependence on modularity, due to a transition from  two-step  to one-step nucleation. In other words, nucleation within a network with intermediate modularity was faster than that in a uniform network. 

In three subsequent papers~\cite{ChenPhysRevE2011p, ShenChaos2013, ShenEPL2015} from the same group, nucleation was studied in Barab\'{a}si-Albert (BA)~\cite{BarabasiScience1999} scale-free random networks. For instance, Chen~\emph{et al.}~\cite{ChenPhysRevE2011p} showed that the rate of homogeneous nucleation decays exponentially with the number of nodes in the  BA network, implying that homogeneous nucleation is not possible in the thermodynamic limit (i.e.,~when the number of nodes goes to infinity). Adding a few impurities (i.e.,~fixed up-spin nodes), however, resulted in a significant increase in nucleation rate. They found the enhancement to be more prominent when those impurities are added to high-connectivity nodes. In a later paper, Shen~\emph{et al.}~\cite{ShenChaos2013} compared the mechanism of nucleation in degree-uncorrelated and degree-correlated BA networks. In uncorrelated networks, nucleation was a gradual process. In degree-correlated networks, however,  small nuclei formed in high-connectivity regions, and then rapidly burst in a merging manner. In their most recent paper, Shen~\emph{et al.}~\cite{ShenEPL2015} studied the impact of mobile impurities within a scale-free network on heterogeneous nucleation. Like Ref.~\citenum{PagePRL2006}, impurities were inert and did not interact with either of the spins. The authors demonstrated that increasing the mobility of impurities resulted in faster nucleation. They also observed a non-monotonic dependence of rate on $\alpha$, a control parameter that biased the random walk towards higher- ($\alpha>0$) and lower-degree ($\alpha<0$) nodes, and found an unbiased random walk to result in faster nucleation.

An important extension of the Ising model is the $q$-state \emph{Potts model}~\cite{PottsProcCambPhilSoc1952, WuRevModPhys1982} in which each spin can take $q$ distinct values, and the strength of spin-spin and spin-field interactions depends on the identities of the individual spins.  This can create a potentially rich phase diagram with multiple metastable states. Consequently, the Potts model has been widely studied to understand metastability~\cite{BltePRL1979, KirkpatrickPhysRevB1987}. Nucleation in the Potts model has therefore been primarily studied to understand the consequences of the existence of metastable intermediates. One important consequence of metastability is the Ostwald step rule which states that a first-order phase transition will usually occur in stages passing through such intermediates. For instance, Sanders~\emph{et al.}~\cite{SandersPhysRevB2007} tuned the interaction parameters of the Potts model to generate metastable states with differing stabilities and employed FFS to compute the rates of transitions between them.  They demonstrated that if the competing metastable intermediates are equally stable, one of them will be chosen randomly during the nucleation process. However, the selection probability is not uniform, and the phases that can access more stable intermediates downstream will be easier to nucleate. These findings suggest that  in systems with a complex web of metastable intermediates, the Ostwald step rule will be satisfied in a probabilistic sense, and no unique nucleation pathway might exist.

In recent years,  nucleation in the three-state Potts model has been studied to understand the role of metastable liquid intermediates in crystal nucleation. For instance, Sear ~\cite{SearJPhysCondensMatter2007} employed FFS to probe heterogeneous nucleation of a spin-3 phase from two coexisting spin-1 and spin-2 phases in the vicinity of an inert wall. He found that nucleation is the fastest at the three-phase contact point. Moreover,  he observed that heterogeneous nucleation at the two-phase interface is faster than homogeneous nucleation in the bulk. Sear argued that these findings are relevant in the context of explaining the relative importance of different modes of heterogeneous ice nucleation. In a second paper, Sear~\cite{SearJChemPhys2009} assessed the importance of a metastable intermediate on heterogeneous nucleation by altering the metastability of the intermediate phase via changing the interaction parameters in the three-state Potts model. He observed that nucleation occurs at considerably higher rates when an intermediate exists, and initiates at concave portions of a pore. In another paper, Chen and Shen~\cite{ChenPhysicaA2015} investigated nucleation kinetics and mechanism in a three-state Potts model, and observed different nucleation scenarios depending on the magnitudes of the external field and the the chemical potential.

\subsubsection{Crystal Nucleation}

\noindent
Crystal nucleation is a process that has been extensively studied using FFS, perhaps more so than any other rare event. The interest in crystal nucleation predates FFS, and conventional MD and bias-based techniques  (such as umbrella sampling and Benette-Chandler type methods) have been used for studying the free energy landscape and kinetics of crystal nucleation in a wide variety of systems. Most of these studies employ global order parameters, such as total potential energy or global Steinhardt order parameters, to monitor and/or drive nucleation. With the advent of FFS, it has become possible to compute nucleation rates that span over tens of orders of magnitude, and to obtain unbiased information about the mechanism and free energy landscape of the nucleation process. As a result, FFS has been utilized to study crystal nucleation in a wide variety of systems from simple model systems such as Lennard Jones and hard spheres, to complex crystals such as gas hydrates. This section is dedicated to discussing this large body of work, organized in accordance to the system type.  \\

\begin{figure}
	\centering
	\includegraphics[width=.45\textwidth]{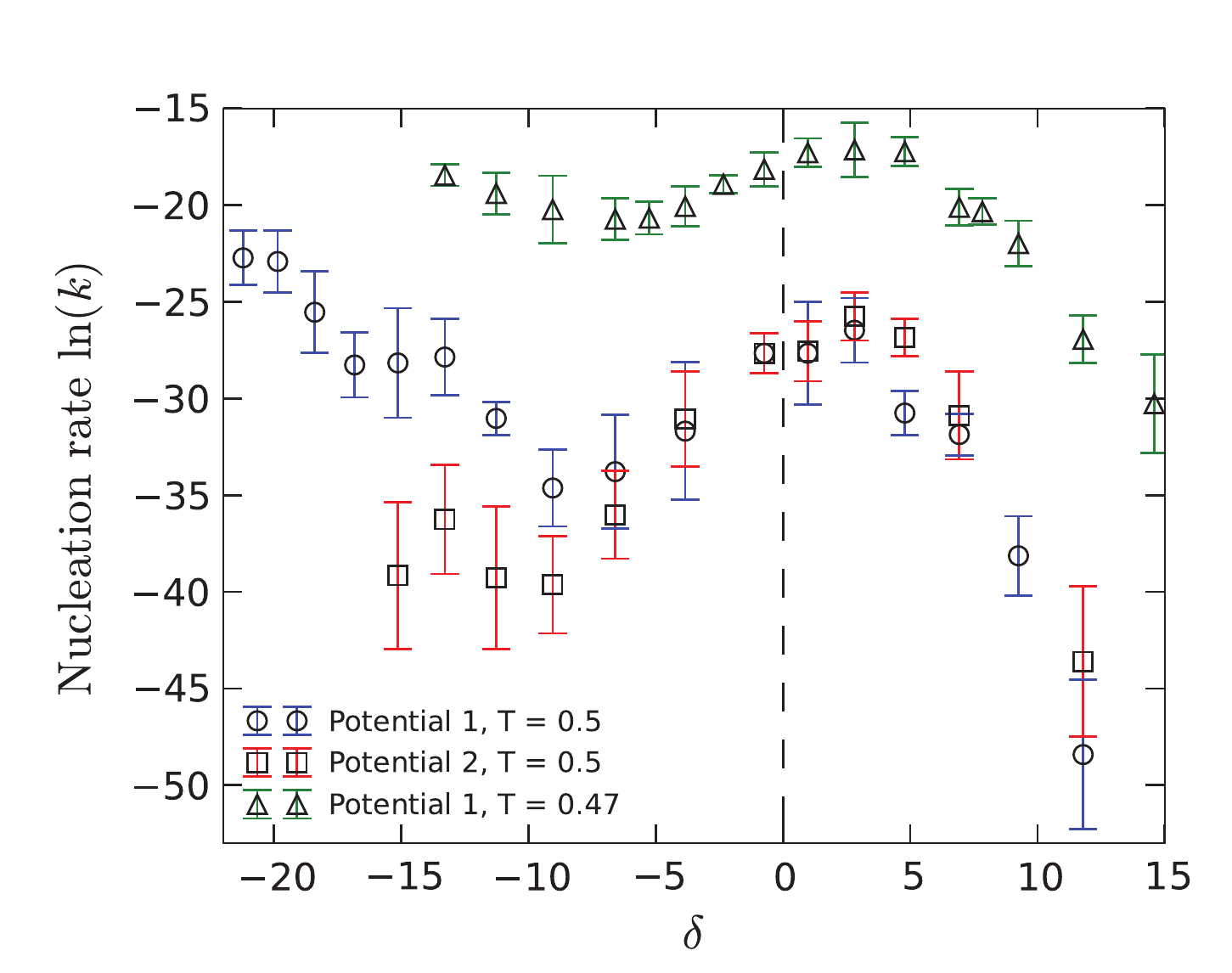}
	\caption{\label{fig:lj-mismatch}(Reproduced from Ref.~\citenum{MithenJChemPhys2014}) Dependence of heterogeneous nucleation rate in the LJ system on the lattice mismatch $\delta$. Note that the maximum rate is observed for $\delta$'s that are very close to, but not equal to, zero. }
\end{figure}

\paragraph{Lennard-Jones:}\label{Lennard-Jones System}
The Lennard-Jones (LJ) potential~\cite{LJProcRSoc1924} is the simplest model for representing interactions between small non-polar molecules (such as argon, neon and methane). Its phase diagram is very simple, and consists of a gas-liquid-solid triple point,  a gas-liquid critical point, and positively sloped melting curve. Historically, the LJ potential has been the model of choice for studying the underlying physics of structural relaxation, and phase transitions in simple liquids~\cite{JnssonRPL1988, KobAndersenPRE1995,  DonatiPhysRevE1999}. The thermodynamically stable crystalline form of the LJ system is the face-centered cubic (FCC) crystal. Computational investigations of crystal nucleation in the LJ system date back to as early as 1976, when Mandell~\emph{et al}. utilized conventional MD to crystallize the LJ system~\cite{MandellJChemPhys1976}. Since then, a large number of studies have been carried out using conventional~\cite{TruduPhysRevLett2006} and bias-based techniques~\cite{FrenkelFarDissLJ1996} to study different aspects of  the thermodynamics and kinetics of crystal nucleation in the LJ system. For instance, ten Wolde~\emph{et al.} utilized umbrella sampling and linear response theory to compute the barriers and rates of homogeneous crystal nucleation~\cite{FrenkelFarDissLJ1996}. FFS investigations of crystal nucleation in the LJ system have therefore been primarily conducted for benchmarking purposes, e.g.,~for assessing the performance and accuracy of different FFS variants as in Refs.~\citenum{HajiAkbariPNAS2015} and~\citenum{AmirHajiAkbariJCP2018}. There have, however, been several studies utilizing FFS to gain insight into crystal nucleation in the LJ system. For instance, van Meel~\emph{et al.}~\cite{vanMeelJChemPhys2008} utilized FFS to compute the rate of crystal nucleation from LJ vapor below the triple point. They found the nucleation to proceed via a two-step process, with the first step involving the nucleation of a liquid droplet from supersaturated vapor, followed by crystal nucleation within the droplet. The second step only occurred when the droplet size exceeded a critical value and its occurrence did not depend on vapor supersaturation. In a later study, Mithen and Sears~\cite{MithenJChemPhys2014} inspected the impact of lattice mismatch between the crystal nucleating surface and the LJ crystal on the rate of heterogeneous nucleation. They computed rates for substrates with different levels of lattice mismatch $\delta$, and concluded that the highest rates are observed for mismatches that are very close to-- but not equal to-- zero (Fig.~\ref{fig:lj-mismatch}). Deviations from this optimal mismatch (in either direction) were shown to result in a considerable decrease in rate, but no change in the overall heterogeneous nature of nucleation. \\

\begin{figure*}[t]
	\centering
	\includegraphics[width=.7\textwidth]{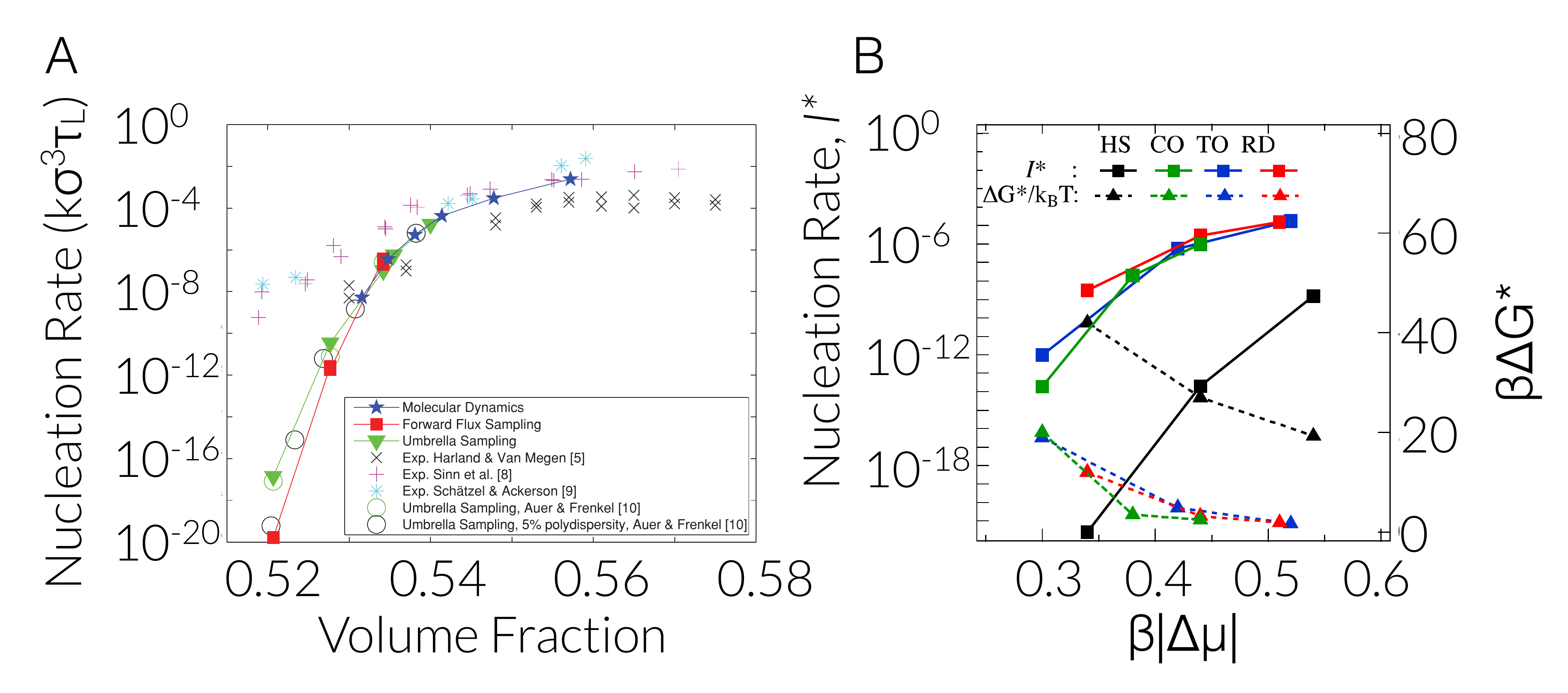}
	\caption{\label{fig:hp-ffs} (A) (Reproduced and re-labeled from Ref.~\citenum{FillionJCP2010}) Crystal nucleation rates in hard spheres computed using different methods. Experimental data are from Harland and van Megen~\cite{HarlandPhysRevE1997}, Sinn~\emph{et al.}~\cite{SinnProgColloidPolymSci2001} and Sch\"{a}tzel and Ackerson~\cite{SchatzelPRE1993}, while the umbrella sampling calculations of Auer and Frenkel are from Ref.~\citenum{AuerFrenkelNature2001}.  (B) (Reproduced and re-labeled from Ref.~\citenum{ThaparPRL2014}) Nucleation rates and barriers for hard spheres, cubotahedra, truncated octahedra and rhombic dodecahedra computed from FFS as a function of the degree of supersaturation $\beta|\Delta\mu|$.}
\end{figure*}

\paragraph{Hard Particles:} Hard particles interact via excluded volume interactions in the sense that the potential energy of the system is zero for a non-overlapping configuration, and is infinite if an overlap exists. For monodispersed hard spheres of diameter $d$, for instance, the binary inter-particle hard potential is given by: 
\begin{eqnarray}
u(r) &=& \left\{
\begin{array}{ll}
0 &~~~~~~~~~r\ge d\\
+\infty &~~~~~~~~~r<d
\end{array}
\right.
\end{eqnarray}
Hard potential is a good approximation for short-range repulsive interactions. Due to the prevalence of such repulsions in dense phases of small molecules, hard spheres have, for decades, been used in theoretical and computational studies of condensed states of matter. In a hard particle system, all permissible configurations have the same potential energy, and therefore its thermodynamics is exclusively determined by entropy. At sufficiently large packing fractions, hard particles can form crystalline structures as the entropy of the crystal exceeds that of the disordered fluid at the same density. This results in \emph{entropy-driven disorder-order transitions}~\cite{FrenkelPhysA1999}, which were first predicted in pioneering works of Onsager~\cite{Onsager1949} and Kirkwood~\cite{Kirkwood1951}, and were later confirmed in Monte Carlo simulations of hard spheres~\cite{AlderWainwright1957, WoodJacobson1957} and several other shapes~\cite{EppengaFrenkel1984, FrenkelPRL1984, MulderFrenkelMolPhys1985, VeermanFrenkelPRA1990, VeermanFrenkel1992, FrenkelNature1993, BolhuisFrenkel1997, BatesFrenkel1998, JohnEscobedoJPCB2005, JohnEscobedo2008, HajiAkbariEtAl2009, EscobedoNatureMaterials2011, HajiAkbaricondmat2011, HajiAkbariDQC2011, PabloScience2012, PabloACSNanot2012, HajiAkbariPRE2013}. Hard spheres crystallize into a combination of face-centered cubic (FCC) and hexagonally closed packed (HCP) crystals. These two polymorphs have identical packing fractions, constitute the densest known packings of monodispersed spheres~\cite{HalesAnnalsMath2005}, and have almost identical free energies at all packing fractions~\cite{FrenkelLaddJCP1984}.

Similar to the Lennard-Jones system, crystal nucleation in the hard sphere system had been extensively studied prior to the development of FFS. In particular, Auer and Frenkel~\cite{AuerFrenkelNature2001, AuerNature2001p} used umbrella sampling and kinetic Monte Carlo~\cite{CichockiPhysicaA1990} to compute rates of homogeneous nucleation in the hard sphere fluid, reporting rates that are several orders of magnitude smaller than those observed in experimental studies of colloidal spheres. The first investigation of hard particle nucleation using FFS was conducted by Filion et al.~\cite{FillionJCP2010}, who found reasonable agreement among the rates computed from FFS (Monte Carlo), umbrella sampling, and direct MD (Fig.~\ref{fig:hp-ffs}A). The critical nuclei obtained from different methods were also structurally similar, and were all comprised of a large fraction of FCC stacks. The problem of large temporal fluctuations of the order parameter was observed in this early study, but similar to later rate calculations conducted in the LJ system~\cite{AmirHajiAkbariJCP2018}, did not result in significant errors in rate estimates.

In recent years, FFS has also been utilized for exploring crystal nucleation in systems of hard polyhedra. In particular,   Thapar and Escobedo~\cite{ThaparPRL2014} used FFS to study the nucleation of a metastable rotator phase in systems of hard cuboctahedra (CO), truncated  octahedra (TO), and rhombic dodecahedra (RD). These rotator phases-- also known as plastic crystals-- form in systems of anisotropic building blocks with low asphericities. In a plastic phase, the building blocks occupy the sites of the lattice otherwise occupied by the isotropic (i.e.,~spherical) building blocks, without adopting any preferred orientations. Therefore, the rotator phases formed by hard COs, TOs and RDs are similar to hard sphere packings in terms of their translational order. Despite this similarity, Thapar and Escobedo~\cite{ThaparPRL2014} demonstrated that in all three systems, the rotator phases nucleate at rates considerably higher than hard spheres (Fig.~\ref{fig:hp-ffs}B). They also computed the nucleation barriers via umbrella sampling and found them to be considerably smaller than those for hard sphere nucleation. This difference lies in the structure of the critical nuclei, which, unlike the hard sphere system, are primarily comprised of HCP-like stacks.  This difference in critical nucleus structure was partially linked to the lower free-energy barrier, along with the presence of a positive coupling between translation and rotational degrees of freedom in the disordered phases of the corresponding polyhedra.

An alternative potential that is widely used for representing excluded volume interactions in colloidal systems is the Weeks-Chandler-Andersen (WCA) potential~\cite{WeeksChandlerAndersenPRA1971}:
\begin{eqnarray}
u_{\text{WCA}}(r) &=& \left\{
\begin{array}{ll}
4\epsilon\left[\left(\dfrac{\sigma}{r}\right)^{12} - \left(\dfrac{\sigma}{r}\right)^{6} + \dfrac{1}{4}\right]&~~~~\dfrac{r}{\sigma} \le 2^{1/6}\\
0 &~~~~\dfrac{r}{\sigma} > 2^{1/6}
\end{array}
\right.\notag\\&&
\label{eqn:WCA}
\end{eqnarray}
The main advantage of the WCA potential over the hard potential is its continuity and piecewise differentiability, which makes it convenient for use with methods such as Brownian dynamics (BD)~\cite{VanGunsterenMolPhys1982} that are widely utilized for studying colloidal systems. In general, the numerical implementation of hard potentials in BD-based schemes, such as the overdamped Langevin dynamics, is not straightforward, and even though attempts have been made to develop event-driven BD algorithms~\cite{ScalaJCP2007}, the WCA potential-- and its generalizations for non-spherical particles-- is the potential of choice in BD simulations. As such, nucleation rate estimates for WCA colloids have been made via direct BD~\cite{FilionJCP2011}, MD~\cite{RichardJCP2018,RichardIIJCP2018}, umbrella sampling~\cite{FilionJCP2011}, the seeding method~\cite{RichardIIJCP2018}, and the persistent embryo method~\cite{RenSoftMatter2018}. FFS was first used by Filion~\emph{et al.}~\cite{FilionJCP2011}, along with direct BD and umbrella sampling, to calculate nucleation rates for the WCA model with $\beta\epsilon = 40$. The rate estimates obtained from all three methods agreed well with each other and the dependence of the obtained rate on  the effective packing fraction, $\phi_{\text{eff}}$, was similar to that of hard spheres. 

More recently, Richard and Speck~\emph{et al.}~\cite{RichardJCP2018} used FFS-MFPT (Section~\ref{section:method:ffs-mfpt}) and the probability splitting method (Section~\ref{section:method:probability-splitting}) to estimate free-energy landscape of crystal nucleation in the monodispersed WCA system. The computed free energy barriers follow the scaling predicted by CNT. Quantitative agreement with CNT was, however, achieved with a an effective solid-fluid surface tension that is $\approx35\%$ larger than the bulk value.  In their follow-up paper~\cite{RichardIIJCP2018}, they combine the nucleation works computed from FFS and several other methods, and pressure differences between solid nuclei and the surrounding melt, to conclude that surface tension in small nuclei is indeed larger than the bulk value. \\

\begin{figure}
\centering
\includegraphics[width=.5\textwidth]{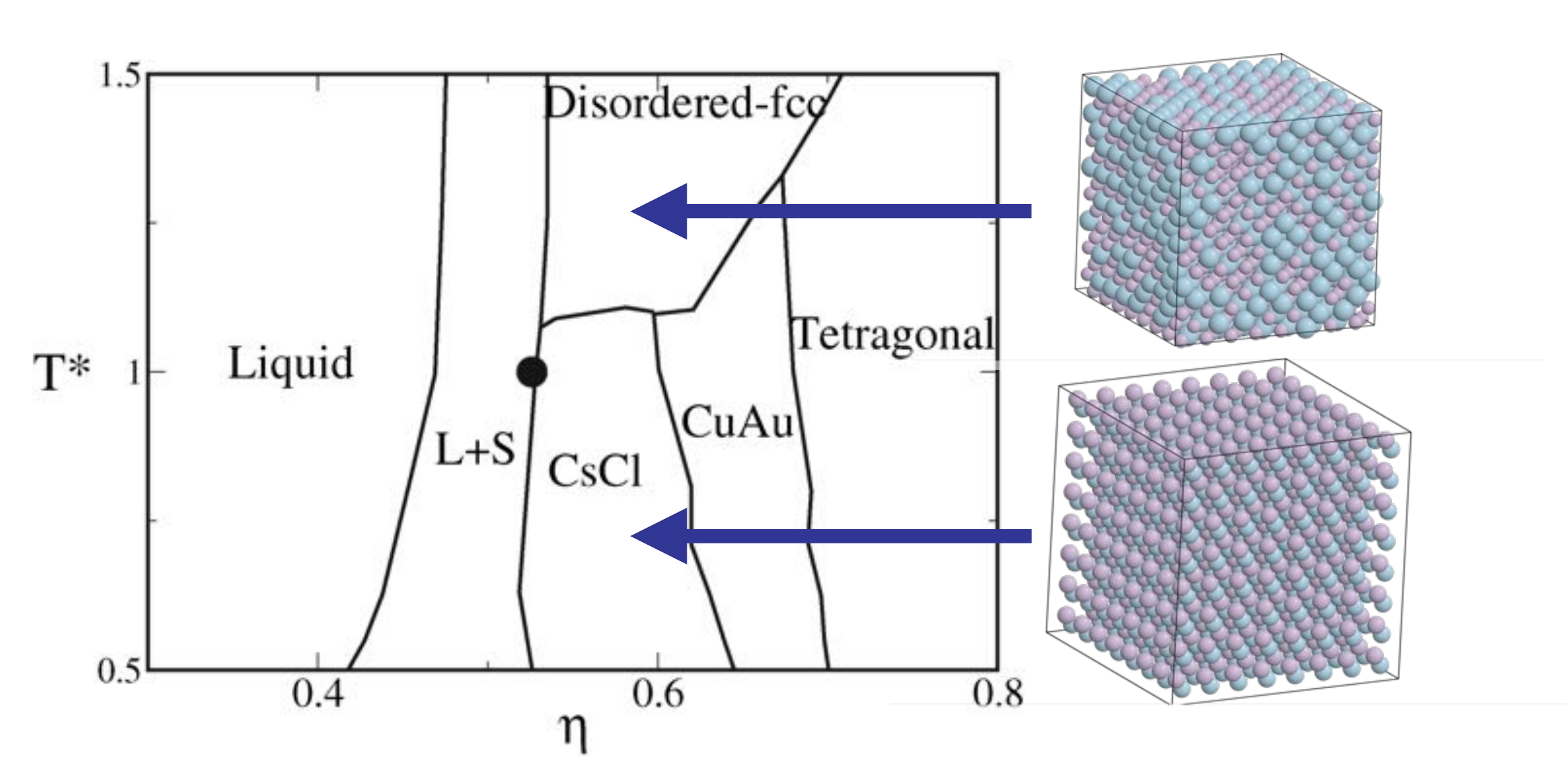}
\caption{\label{fig:charged-colloid-ffs} (Reproduced from Ref.~\citenum{SanzPRL2007} and re-designed) Phase diagram of the 50:50 binary Yukawa system reported by Hynninen~\emph{et al.}~\cite{HynninenPhysRevLett2006}. $\eta=\pi\sigma^3N/6V$ is the packing fraction. Representative renderings of the disordered FCC and CsCl structures are also depicted. }
\end{figure}

\paragraph{Charged Colloids:} One of the earliest studies of nucleation using FFS was conducted for probing polymorph selection in a system of oppositely charged spherical colloids~\cite{SanzPRL2007, SanzJPhysCOndensMatter2008}, represented using the Yukawa potential. This system has a very rich phase diagram, with multiple crystalline phases~\cite{HynninenPhysRevLett2006}. In particular, an FCC lattice with non-periodic distribution of cations and anions, and a CsCl crystal can coexist with the liquid at high and low temperatures, respectively (Fig.~\ref{fig:charged-colloid-ffs}). Exploring crystal nucleation is particularly interesting  in this system due to its polymorphism, and the fact that these competing crystals cannot be converted into one another through simple geometric rearrangements (also known as diffusionless transitions). Sanz~\emph{et al.}~\cite{SanzPRL2007} analyzed crystal nucleation at the state point $P^* = 15$ and $T^* = 1$, a condition under which the CsCl polymorph is thermodynamically preferred. They, however, found the transition state to be predominantly comprised of disordered FCC. By incorporating MC swap moves into the underlying Monte Carlo algorithm, they were able to assemble the CsCl crystal. They also found the nucleation barrier to CsCl assembly to be smaller than that of the disordered FCC. Their work provides evidence for an out-of-equilibrium mechanism in cases where stable crystal formation is kinetically inhibited due to "self-poisoning`` of the precritical nuclei, i.e.,~the predominance of the wrong crystalline motifs at early stages of nucleation. A later study revealed that this non-equilibrium effect can be attributed to anisotropic diffusion~\cite{PetersJChemPhys2009}. 

A more recent study by Kratzer and Arnold~\cite{KratzerSoftMatter2015} used FFS to study crystal nucleation in a system of charged colloids. (The same authors had discussed technical details of implementing FFS in the same system in an earlier paper~\cite{RoehmHighPerfComput2013}.)  Unlike Ref.~\citenum{SanzPRL2007}, they considered a single-component system in which all colloids had the same  charge, resulting in a purely repulsive interaction potential. Moreover, they represented excluded volume repulsions using the WCA potential.   For sufficiently small screening lengths, this system is known to form a metastable body-centered cubic (BCC) crystal, and a thermodynamically stable FCC crystal~\cite{AzharJChemPhys2000, HynninenPhysRevE2003}.  The authors studied crystal nucleation for two different contact energies, with contact energy defined as the ratio of energy scales of the Yukawa and WCA potentials. In both cases, the nucleation of the FCC phase from the fluid was a two-step process, starting with the nucleation of a BCC nucleus followed by  heterogeneous nucleation of FCC at the BCC-fluid interface. The second step was almost spontaneous for the smaller contact energy. For the larger contact energy, however, the second step  involved crossing a large nucleation barrier to the extent that FCC-rich clusters never emerged throughout the first FFS calculation. In order to probe the kinetics and mechanism of the BCC-FCC transition, the authors conducted a second FFS calculation with the size of the largest FCC-like cluster as the order parameter, unlike their first FFS calculation in which no distinction was made between FCC- and BCC-like particles. They found the FCC-rich nuclei to form from the BCC-rich clusters that were post-critical,  i.e.,~that had a low probability of returning to the disordered fluid basin. They also used the forward-backward FFS method of Ref.~\citenum{ValerianiJChemPhsy2007} (Section~\ref{section:method:forward-backward-ffs}) to compute the free energy landscape of the fluid-BCC transition and found reasonable agreement with CNT. By conducting pedigree analysis, the authors did not find any density fluctuations preceding the nucleation of BCC-rich nuclei from the fluid, unlike what has been suggested for other systems~\cite{SchillingPhysRevLett2010}. 
\\

\paragraph{Water:}\label{section:nucleation:crystal:water}
 Water has a very complex phase diagram, with 18 experimentally observed crystalline phases, and fifteen triple points~\cite{HajiAkbariIce2019, MillotNature2019}. Most such crystals, however, form under very high pressures and/or very low temperatures, or are thermodynamically metastable. The only crystalline form of water that is thermodynamically stable under ambient conditions is \emph{ice I}, which has two stacking variants known as \emph{hexagonal} and \emph{cubic} ice. Understanding the kinetics and molecular mechanism of how ice I forms is critical to fields such as cryobiology and atmospheric physics. The nucleation of ice I has therefore been extensively studied in molecular simulations prior to the development of FFS using unbiased~\cite{KusalikPRL1994, KusalikJACS1996, Matsumoto2002} and biased~\cite{TroutJACS2003} techniques. Since its development, FFS has been utilized extensively to study different aspects of homogeneous~\cite{GalliPCCP2011,  GalliNatComm2013, HajiAkbariFilmMolinero2014, HajiAkbariPNAS2015, HajiAkbariPNAS2017} and heterogeneous~\cite{CabrioluPRE2015, SossoJPhysChemLett2016, BiJPhysChemC2016, SossoChemSci2018} ice nucleation.  In general, we can classify these studies into three categories: (i) homogeneous nucleation in the bulk, (ii) role of vapor-liquid interfaces in homogeneous nucleation, (iii) physics of heterogeneous nucleation. \\

\begin{figure}
\centering
\includegraphics[width=0.25\textwidth]{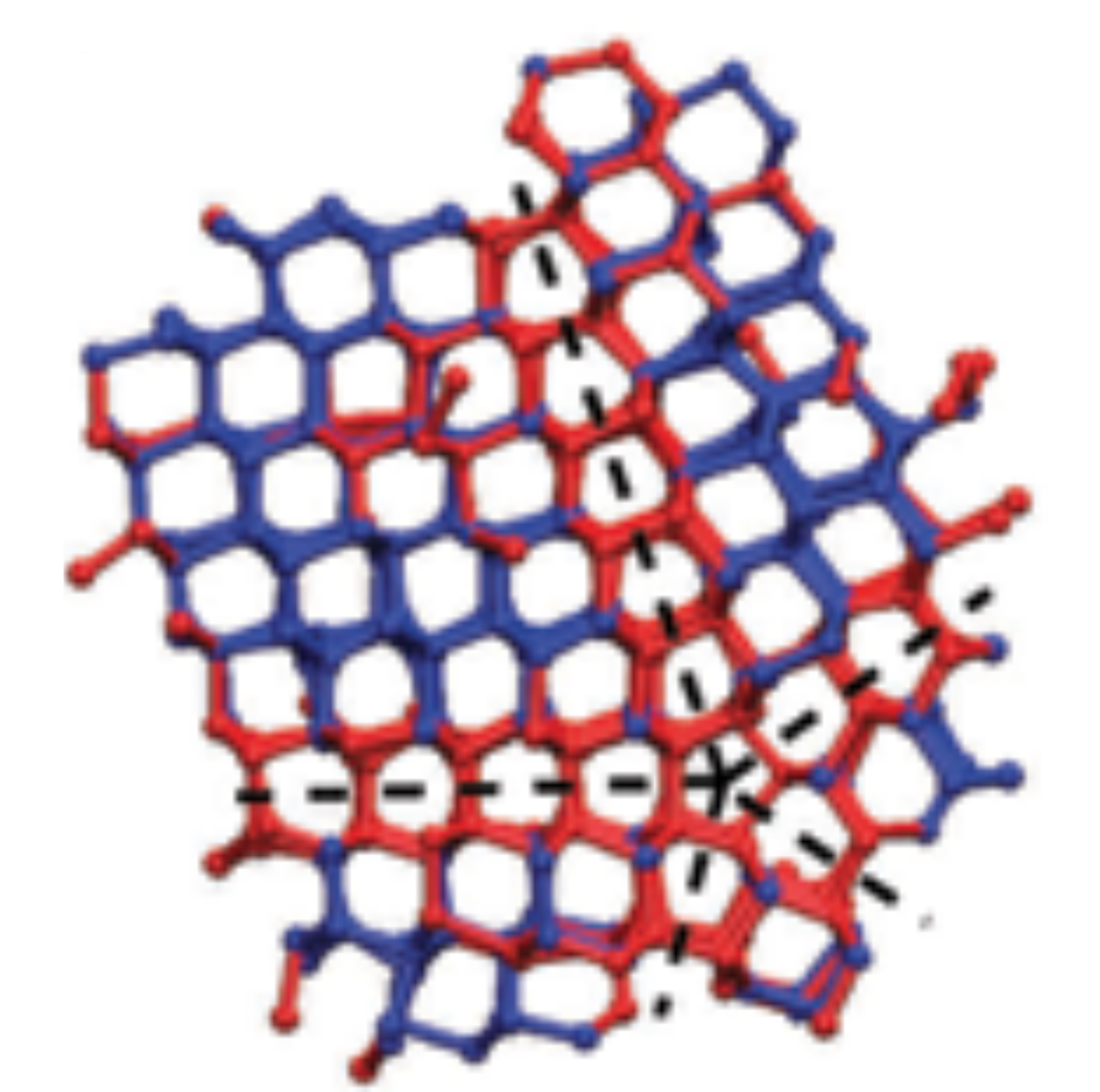} 
\caption{\label{fig:twin-boundaries-mw}(Reproduced from Ref.~\citenum{GalliPCCP2011}) A representative nucleus formed during homogeneous ice nucleation in the mW system at 235~K. Cubic and hexagonal stacks are depicted in blue and red, respectively. This nucleus has a fivefold twin boundary defect.}
\label{fig:Nucleation_3}
\end{figure}

\noindent\textbf{Homogeneous Nucleation in the Bulk:} The first direct calculation of homogeneous ice nucleation rates using FFS was conducted using the the  coarse-grained monoatomic water (mW) potential~\cite{MolineroJPCB2009} in which each water molecule is represented by a single interaction site and hydrogen bonding is mimicked by introducing a three-body term that favors the formation of tetrahedral angles. The mW potential constitutes a re-parameterization of the Stillinger-Weber (SW) potential~\cite{StillingerPRB1985} originally developed for modeling  Group IV elements such as silicon and germanium. Li~\emph{et al.}~\cite{GalliPCCP2011} used FFS to compute homogeneous ice nucleation rates in the mW system in the temperature range 220--240~K, and found the mW potential to underestimate the rate by $5-8$ orders of magnitude. They attributed this discrepancy to possible overestimation of $\gamma_{ls}$, the solid-liquid surface tension by the mW model.  Considering the cubic dependence of $\Delta{G}_{\text{nuc}}$ on $\gamma_{ls}$, even a small deviation in $\gamma_{ls}$ can result in considerable discrepancies in rate. 
By analyzing the transition states, they found the precritical and critical nuclei to be comprised of a mixture of cubic and hexagonal stacks. They also observed  a preponderance of fivefold twin boundary defects (Fig.~\ref{fig:twin-boundaries-mw}), which are thought to play an important role in the nucleation of other tetrahedral crystals (e.g.,~Si), but have not been observed in ice nucleation studies utilizing other water models.

The task of directly computing homogeneous nucleation rates for more realistic molecular-- also known as \emph{atomistic}-- models of water is, however, more challenging~\cite{BrukhnoJPhysCondMat2008}, and was not achieved until recently when Haji-Akbari and Debenedetti~\cite{HajiAkbariPNAS2015} used temporally coarse-grained FFS (Section~\ref{section:FFS:temporally-cg}) to probe homogeneous ice nucleation kinetics in the TIP4P/Ice model~\cite{VegaTIP4PiceJCP2005}, which is one of the best existing non-polarizable molecular models of water. As discussed in Section~\ref{section:FFS:temporally-cg}, their FFS calculation only converged after temporal coarse-graining, i.e.,~computing OP every 500 MD steps.  Similar to Li~\emph{et al.}~\cite{GalliPCCP2011}, their computed rate was several orders of magnitude smaller than experimental estimates, which they attributed to the fact that the TIP4P/Ice model underestimates the chemical potential difference between the liquid and hexagonal ice by almost 20\%. Their estimate of rate was later revised up by four orders of magnitude after Haji-Akbari~\cite{AmirHajiAkbariJCP2018} partially repeated the calculation using jFFS.

\begin{figure}
\centering
\includegraphics[width=.45\textwidth]{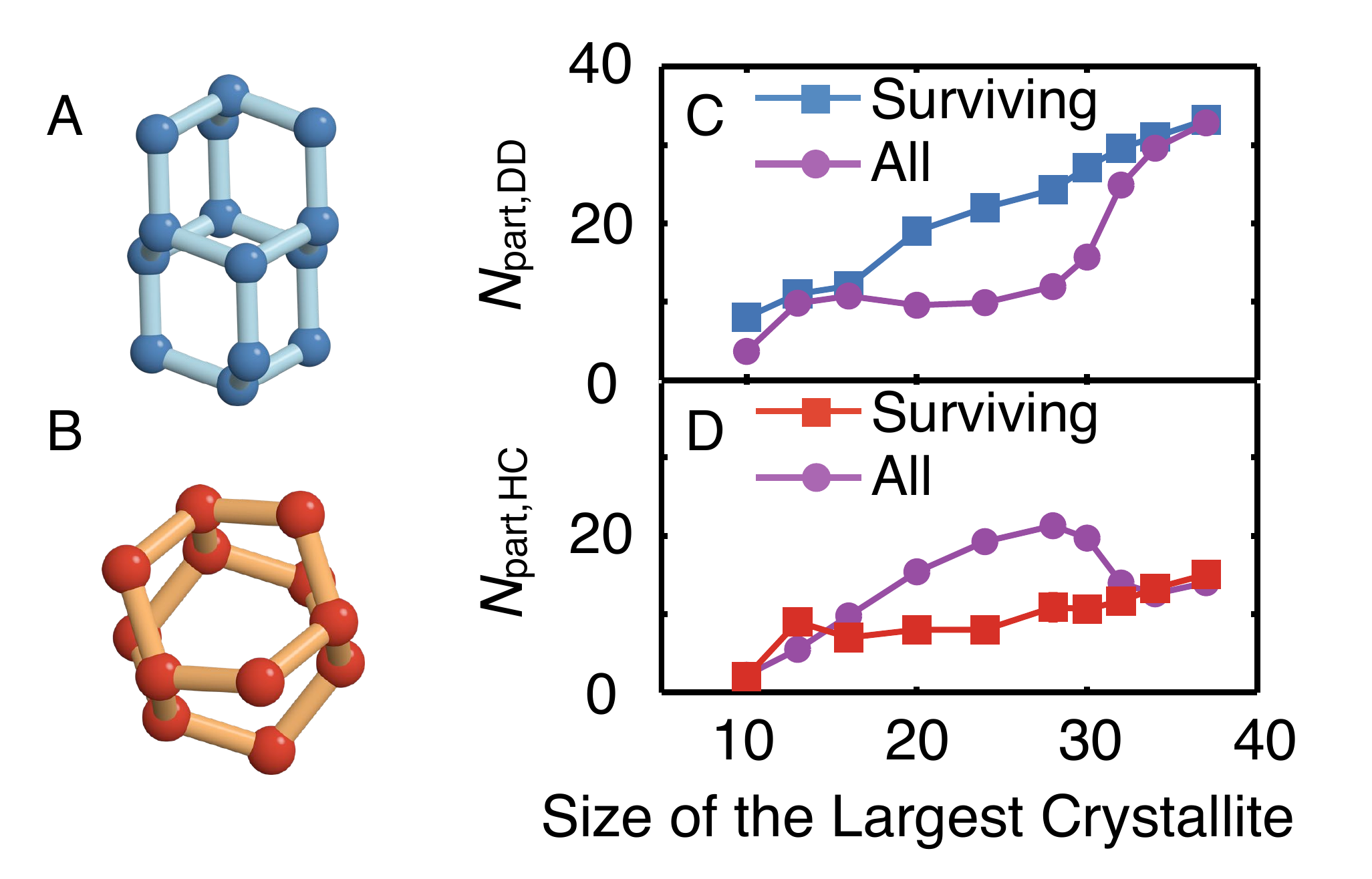} 
\caption{(Reproduced from Ref.~\citenum{HajiAkbariPNAS2015}) (A) DDCs and (B) HCs. Number of water molecules in the largest cluster that are part of a DDC (C) and HC (D). Surviving configurations have progeny beyond the inflection region of Fig.~\ref{fig:temporally-cg}B.} 
\label{fig:DDC_HC_mech}
\end{figure}

In order to probe the nucleation mechanism, Haji-Akbari and Debenedetti~\cite{HajiAkbariPNAS2015} zoomed in on the pronounced inflection of the cumulative probability  for cluster sizes $\sim30$ (Fig.~\ref{fig:temporally-cg}B), and identified the configurations at earlier milestones that survive this inflection, i.e.,~ that have offspring right after the inflection (e.g.,~at $\lambda=41$). They found the 'surviving` (Fig.~\ref{fig:DDC_HC_mech}C) and 'vanishing` (Fig.~\ref{fig:DDC_HC_mech}D) configurations to have an abundance of double-diamond cages (DDCs) (Fig.~\ref{fig:DDC_HC_mech}A) and hexagonal cages (HCs) (Fig.~\ref{fig:DDC_HC_mech}B), respectively. This difference was attributed to the fact that DDCs are more symmetric and can grow more uniformly, while HCs are anisotropic and their growth along their prismatic faces results in the formation of highly aspherical nuclei. This was the first molecular explanation for the experimental observation that the ice nucleating at deep supercoolings was a mixture of cubic and hexagonal stacks, as DDCs and HCs constitute the topological building blocks for cubic and hexagonal ice, respectively. An alternative thermodynamic explanation was recently provided by Lupi~\emph{et al.} based on the importance of the entropy of cubic and hexagonal stacks in small nuclei\cite{LupiNature2017}.

\noindent\textbf{Role of Vapor-Liquid Interfaces in Homogeneous Nucleation:} \label{Role of Vapor-Liquid Interfaces in Homogeneous Nucleation} With these rate calculations at hand, several researchers turned their attention to exploring a confounding-- and controversial-- phenomenon in atmospheric physics known as \emph{surface freezing}~\cite{HajiAkbariJCP2017}, or enhancement of homogeneous ice nucleation close to a vapor-liquid interface.  Surface freezing was first hypothesized by Tabazadeh~\emph{et al.}~\cite{TabazadehPNAS2002} to explain apparent discrepancies among rates measured for microdroplets of different sizes, a hypothesis that was later tested by several researchers~\cite{DuftACPD2004, LuApplPhysLett2005, EarleAtmosChemPhys2010,  KuhnAtmosChemPhys2011, RzesankePCCP2012, RiechersPCCP2013}. The existing evidence is, however, still inconclusive since surface freezing is predicted to become dominant for sub-micron droplets, and it is extremely difficult to generate mono-dispersed droplets of such sizes in experiments. It has indeed been demonstrated that overlooking the effect of polydispersity might result in inaccurate characterization of bulk-dominated freezing as surface-dominated nucleation~\cite{SignorellPhysRevE2008}.

 The first FFS investigation of surface freezing in water was conducted by  Li~\emph{et al.}~\cite{GalliNatComm2013}, who used FFS to probe nucleation kinetics in nano-droplets of mW water. They demonstrated that nucleation in nanodroplets was several orders of magnitude slower than the bulk, with suppression becoming stronger at higher temperatures and for smaller droplets. These findings were in contrast to the surface freezing hypothesis, and were attributed to the presence of larger Laplace pressures and the subsequent drop in thermodynamic driving force. Later FFS calculations of nucleation rates in freestanding thin films of mW by Haji-Akbari~\emph{et al.}, however, concluded that nucleation is suppressed even in the absence of Laplace pressure~\cite{HajiAkbariFilmMolinero2014}. Haji-Akbari~\emph{et al.} also conducted umbrella sampling simulations to demonstrate that crystalline nuclei are less stable at the vapor-liquid interface, presumably due to their high asphericity. They, therefore, unequivocally concluded that the mW potential does not undergo the type of surface freezing proposed by Tabazadeh~\emph{et al.}

\begin{figure}
	\centering
	\includegraphics[width=.46\textwidth]{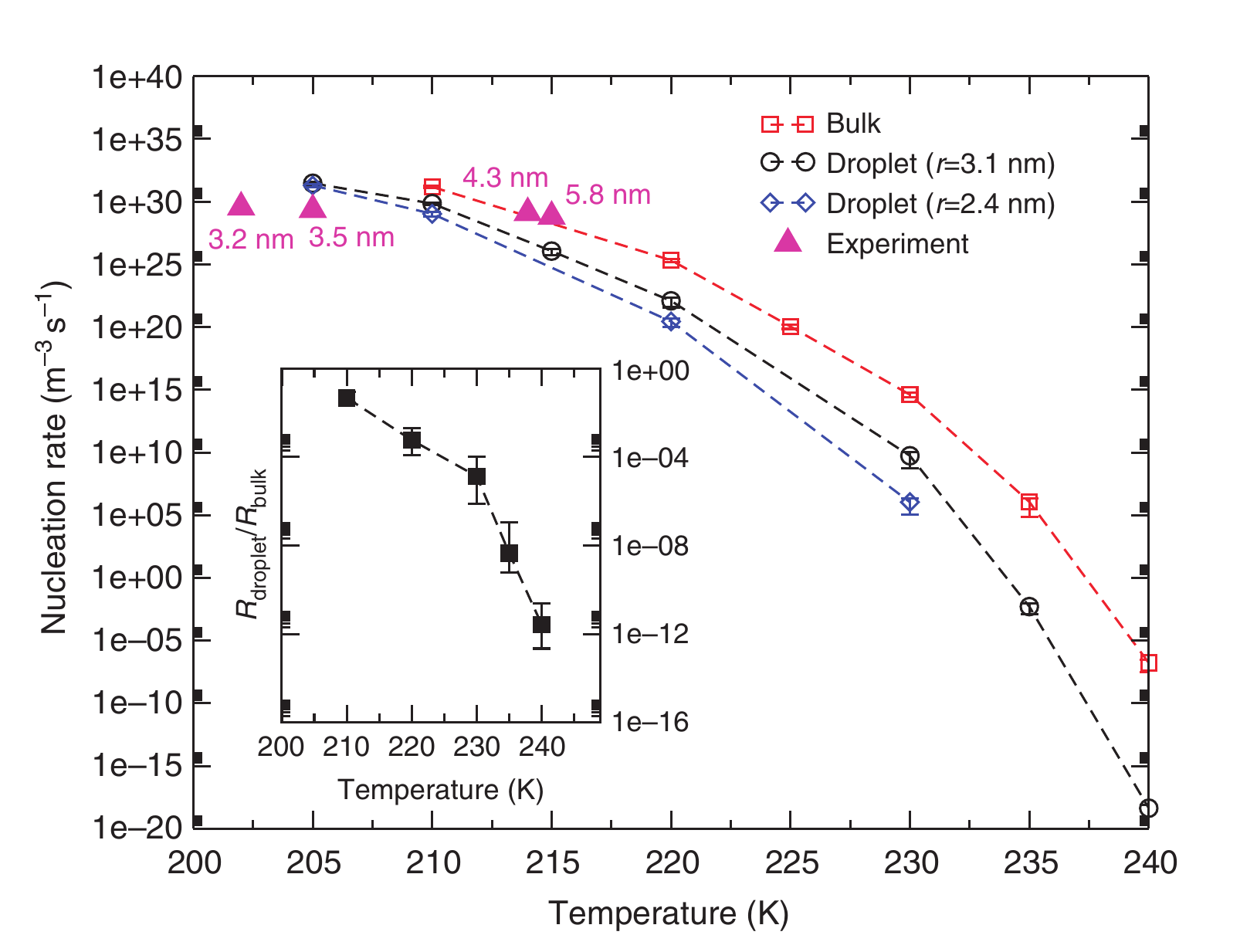}
	\caption{\label{fig:nanodroplet}(Reproduced from Ref.~\citenum{GalliNatComm2013}) Homogeneous ice nucleation rates in the bulk and in nanodroplets of mW water computed using FFS. The inset corresponds to the extent by which nucleation is suppressed in nanodroplets with respect to the bulk. }
\end{figure}

In order to further inspect the sensitivity of nucleation kinetics and surface freezing propensity to the particulars of the employed force-field, Gianetti~\emph{et al.}~\cite{GianettiPCCP2016} utilized FFS to compute homogeneous nucleation rates in the bulk and in 5-nm-thick freestanding liquid films of a family of mW-like potentials with different tetrahedralities. The tetrahedrality parameter controls the energetic penalty for deviating from the tetrahedral angle, and is 23.15 for the mW potential. Gianetti~\emph{et al.} considered three other tetrahedralities, namely 21, 22 and 24, and in order to account for changes in melting temperature as a result of changing tetrahedrality, they computed the rates not at a fixed supercooling, $\Delta T=T_m-T$, but at a fixed supercooling ratio $\xi=T/T_m$. Their computed homogeneous nucleation rates spanned over 48 orders of magnitude. By using CNT, the authors fully attribute this spread to changes in the thermodynamic driving force for nucleation. As for surface freezing propensity, they observed surface-enhanced nucleation only at one tetrahedrality, namely at 21, which seems to be related to enhanced structuring of the free interface at lower tetrahedralities. 

The fact that surface freezing does not take place in the mW system does not rule out its occurrence in real water, and other water models.  Indeed, important qualitative differences have been observed between coarse-grained and molecular models of water, with one notable example being the existence of a liquid-liquid critical point~\cite{LimmerJChemPhys2011, PalmerNature2014, PalmerJChemPhys2018}. As for surface freezing, an old-- but not very accurate-- six-site molecular model of water~\cite{NadaJCP2003} had been shown to exhibit surface freezing propensity~\cite{JungwithJPCB2006}. Motivated by these discrepancies and utilizing temporally coarse-grained FFS (Section~\ref{section:FFS:temporally-cg}), Haji-Akbari and Debenedetti~\cite{HajiAkbariPNAS2017} computed the rate of homogeneous ice nucleation in 4-nm thick freestanding films of TIP4P/Ice at 230~K, and found a considerable increase in nucleation rate in comparison to bulk at the same temperature. They also observed that the nucleation starts within a region of the film that is more conducive to the formation of DDCs, the topological building blocks of cubic ice, which, as shown in their prior study~\cite{HajiAkbariPNAS2015}, are more likely to grow due to their higher symmetry. \\

\noindent\textbf{Heterogeneous Ice Nucleation:} Heterogeneous nucleation is a process in which an extrinsic surface facilitates freezing by decreasing the nucleation barriers. Heterogeneous  nucleation is the primary mechanism~\cite{MurrayChemSocRev2012} for ice formation on earth, and yet its underlying physics is far from fully understood.  In recent years, FFS has been employed to study heterogeneous ice nucleation both using coarse-grained and molecular models. Similar to homogeneous nucleation, earlier FFS studies of heterogeneous nucleation were conducted using the mW model. For instance, Cabriolu~\emph{et al.}~\cite{CabrioluPRE2015} used FFS to compute the rates of heterogeneous ice nucleation on graphene surfaces, and compared them with the homogeneous nucleation rates computed in their earlier study~\cite{GalliPCCP2011} to assess the quantitative predictiveness of classical nucleation theory (CNT) for heterogeneous nucleation~\cite{TurnballJCP1950}. According to CNT, $R_{\text{her}}$, the rate of heterogeneous nucleation is given by $R_{\text{her}}=A\exp[-f_c(\theta_c)\Delta{G}_{\text{hom}}/kT]$, with $\theta_c$ the solid-liquid-substrate contact angle, $f_c(\theta_c)=\frac14(1-\cos\theta_c)^2(2+\cos\theta_c)$, the potency factor, and $\Delta{G}_{\text{hom}}$ the barrier for homogeneous nucleation. They computed $R_{\text{het}}/R_{\text{hom}}$ as a function of temperature and found a behavior commensurate with a fixed potency factor, i.e.,~an $f_c(\theta_c)$ that is independent of temperature. Moreover, the  $f_c(\theta_c)$ computed from nucleation rates was equal to the ratio of the sizes of the critical nuclei in heterogeneous and homogeneous nucleation at each temperature, as predicted by CNT. This level of agreement with CNT, however, is expected only if the contact angle is considerably larger than zero. Under conditions at which $\theta_c\approx0$,  properties such as line tension also need to be taken into account as suggested in some earlier studies~\cite{AuerPhysRevLett2003, WinterPhysRevLett2009}. 

In another work by the same group, Bi~\emph{et al.}~\cite{BiJPhysChemC2016} used FFS to investigate the effect of surface hydrophilicity and crystallinity on ice nucleating capability, and found crystalline surfaces to nucleate ice at higher rates than their amorphous counterparts. They also demonstrated that increasing hydrophilicity beyond a certain value will slow down nucleation on both amorphous and crystalline surfaces, an observation that they attributed to strong templating of water molecules into unfavorable arrangements on super-hydrophilic surfaces. The dependence of rate on hydrophilicity was qualitatively different for crystalline and amorphous surfaces. For crystalline surfaces, the rate decreased very abruptly with hydrophilicity, while for amorphous surfaces, the decline was more gradual and started at lower hydrophilicities. In their FFS simulations, the authors utilizes an approach known as \emph{mirroring} in which the supercooled liquid is sandwiched between two crystal nucleating surfaces that are periodic images of one another. This approach results in faster nucleation, and despite being unphysical, is sometimes used for increasing computational efficiency. 

In recent years, heterogeneous ice nucleation has been studied using molecular models of water as well. For instance, Sosso~\emph{et al.}~\cite{SossoJPhysChemLett2016} utilized temporally coarse-grained FFS (Section~\ref{section:FFS:temporally-cg}) and the TIP4P/Ice force-field\cite{VegaTIP4PiceJCP2005} to calculate the rate of heterogeneous ice nucleation on the (001) plane of a Kaolinite (Al$_2$Si$_2$O$_5$(OH)$_4$) surface, and obtained a rate $\approx16$ orders of magnitude larger than the homogeneous nucleation rate at the same temperature~\cite{HajiAkbariPNAS2015, AmirHajiAkbariJCP2018}. They also demonstrated that unlike homogeneous nucleation in which a mixture of cubic and hexagonal ice emerges during nucleation, the (001) plane of  Kaolinite preferentially nucleates hexagonal ice. Despite being structurally similar to the basal plane of hexagonal ice, the (001) plane of Kaolinite facilitates the formation of the prismatic plane, and its impact on nucleation can therefore not be explained by templating.

Unlike inorganic surfaces whose propensity to nucleate ice is of primary interest to atmospheric sciences, ice nucleation on organic surfaces are also of interest to fields such as cryobiology in which the ability to prevent ice formation is critical for maintaining the longevity of biological cells and organs. Moreover, organic surfaces are  more interesting systems from a fundamental perspective due to their complexity and tunability. After probing heterogeneous ice nucleation on the kaolinite surface, Sosso \emph{et al.}~\cite{SossoChemSci2018} used FFS to investigate heterogeneous ice nucleation on the (001) plane of cholesterol monohydrate (CHLM), with CHLM lattice parameters taken from experiments. The computed rate is almost 16 orders of magnitude larger than the homogeneous nucleation rate at the same temperature. Due to the small surface dimensions considered in this work, however, the rates are likely impacted strongly by finite size effects. The authors also find crystallization to be anisotropic in nature, similar to their earlier findings on kaolinite surfaces. What is different though is the propensity to form both cubic and hexagonal ice at the surface. This qualitative difference is attributed to the diversity of nucleation sites on CHLM surfaces, which enables the formation of both polymorphs at early stages of nucleation. \\

\paragraph{Gas Hydrates:} Gas hydrates form in water-gas mixtures and are crystalline solids in which water molecules form polyhedral cages that encompass small guest molecules, such as methane, CO$_2$, and argon. Despite their numerous potential applications (e.g.,~for energy storage and carbon sequestration), hydrates are notorious for the problems that they cause in gas transition pipelines~\cite{PerrinChemSocRev2013}. There is therefore considerable interest in the petroleum industry to understand the molecular mechanisms that culminate in their formation~\cite{EnglishHydrates2014}. Consequently, computational studies of hydrate nucleation date back to late 1990's~\cite{Hirai1997}, and since then hydrate nucleation has been studied using  conventional MD~\cite{WalshScience2009, JacobsonJACS2010, SarupriaJPCL2012, KusalikJPCB2013}, bias-based techniques~\cite{TroutJCP2002} and the seeding method~\cite{MolineroJACS2012}. Similar to ice nucleation, hydrate nucleation has also been studied using FFS. The order parameters utilized in such calculations, however, are more sophisticated, and require additional clustering and topological analysis to identify the constituent polyhedral cages of the hydrate nucleus. So far, all FFS studies of hydrate nucleation~\cite{LiHydrateJPCB2014, BiJChemPhys2016, DeFeverJChemPhys2017} have utilized the mW potential~\cite{MolineroJPCB2009}, and its parameterizations for water-organic mixtures~\cite{JacobsonJPhysChemB2010}. 

\begin{figure}
	\centering
	\includegraphics[width=.47\textwidth]{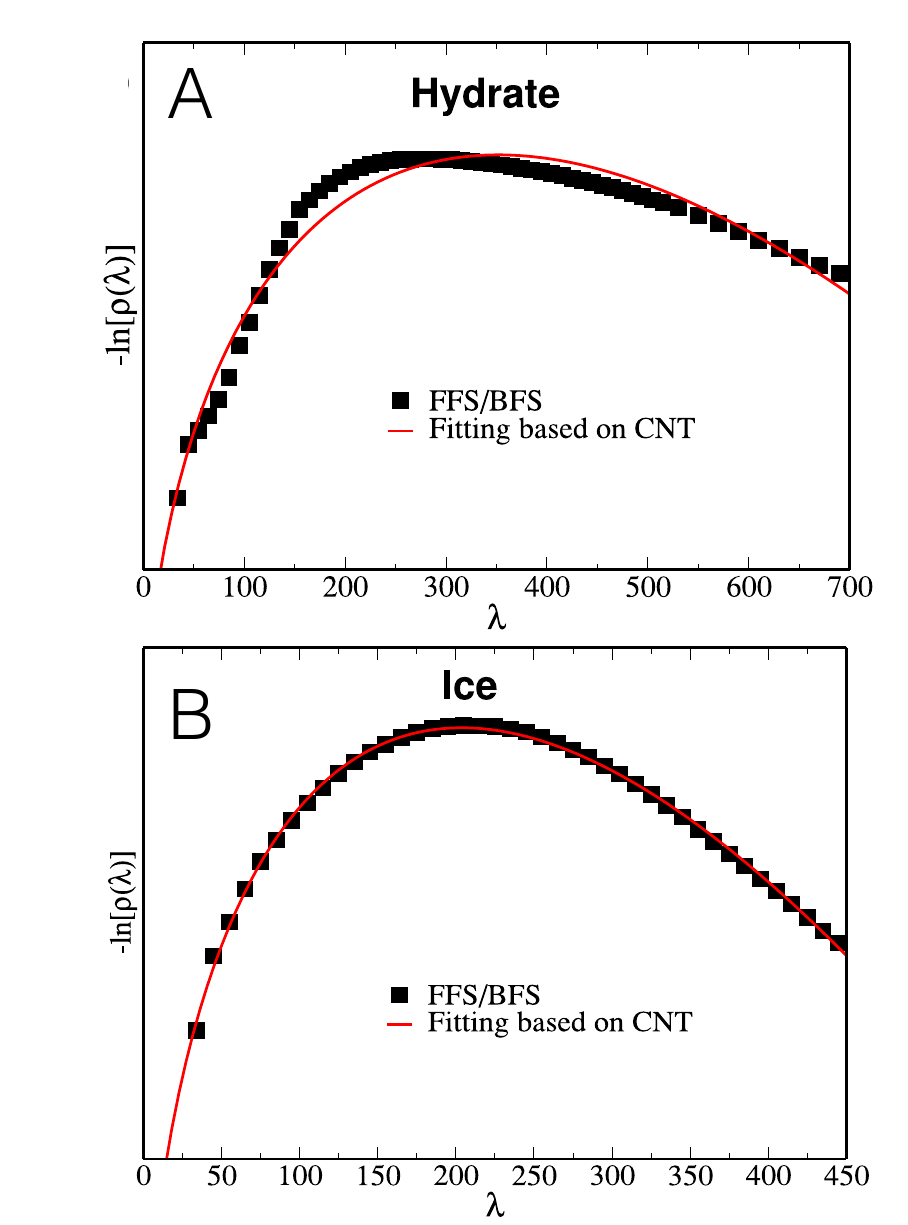}
	\caption{\label{fig:cnt-hydrate-ice}(Reproduced from Ref.~\citenum{BiJChemPhys2016}) The free energy profiles computed from forward-backward FFS and predicted from CNT for (A) hydrate and (B) ice nucleation. }
\end{figure}

Bi~\emph{et al.} were the first to utilize FFS to study methane hydrate nucleation~\cite{LiHydrateJPCB2014}.  They observed that  nucleation starts at the gas-liquid interface, and  proceeds through a two-step process as previously suggested~\cite{MolineroJACS2012}. This work was also important from a methodological perspective, as it was the first systematic effort to assess the sensitivity of the computed rate on the total duration of sampling the starting basin. The authors demonstrated that inadequate sampling of the basin can result in an underestimation of rate, by as much ten orders of magnitude. In their later paper, Bi~\emph{et al.}~\cite{BiJChemPhys2016} investigated hydrate nucleation for oxetane, a hydrophilic guest molecule. Similar to methane hydrates, oxetane hydrates tend to form through a two-step process in which amorphous precursors emerge as  intermediates prior to crystallization. Their careful analysis of the TPE, however, revealed considerable mechanistic variability. While the majority of pathways were consistent with the "two-step`` paradigm, some  did not involve the formation of amorphous intermediates.  They also used forward and backward FFS (Section~\ref{section:method:forward-backward-ffs}) to map out the free energy landscape of nucleation, and observed fair agreement with CNT predictions (Fig.~\ref{fig:cnt-hydrate-ice}A), though the agreement was not as good as for ice nucleation (Fig.~\ref{fig:cnt-hydrate-ice}B) due to the non-classical nature of hydrate formation. A more recent study of hydrate nucleation was conducted by DeFever and Sarupria~\cite{DeFeverJChemPhys2017}, who used FFS, conventional MD and committor analysis to explore hydrate formation by a hydrophilic guest molecule. They analyzed a large number of distinct order parameters using the FFS-LSE method (Section~\ref{section:variants:new:op}) and found those based on water structuring to be better reaction coordinates than those based on guest ordering.   \\

\paragraph{Silicon:} Similar to water, silicon (Si) is a tetrahedral liquid that exhibits several anomalies~\cite{RussoPNAS2018} and can  form multiple polymorphs~\cite{JonesPhysRevB2017}. The ambient-pressure polymorph of silicon, which has the diamond cubic structure and is therefore isostructural to cubic ice, is a semiconductor at room temperature. Due to the abundance of silicon on earth, and the ability of its crystal to function at high temperatures,   crystalline Si is the primary component in computer chips and solar cells. Currently, single-crystalline Si is primarily produced using the Czochralski process~\cite{DupretCzochralski1994}, which is based on controlled growth from a single-crystalline seed. Understanding how silicon crystallizes, and how its nucleation is impacted by different processing factors is therefore  important both from fundamental and technological standpoints, and can result in the development of  cheaper and easier alternatives to the Czochralski process. 
Considering the scientific and technological importance of silicon crystallization, its kinetics and mechanism has been extensively studied in molecular simulations~\cite{SuiCrystEngComm2018}, using methods such as conventional MD~\cite{NakhmansonJPhysCondMater2002, BeaucagePhysRevB2005}, umbrella sampling~\cite{DesgrangesJACS2011}, and metadynamics~\cite{BonatiPhysRevLett2018}. Most such studies were, however, conducted under relatively large thermodynamic driving forces. Using advanced sampling techniques such as FFS has enabled probing the kinetics and mechanism of nucleation closer to coexistence. 

\begin{figure}
\centering
\includegraphics[width=0.48\textwidth]{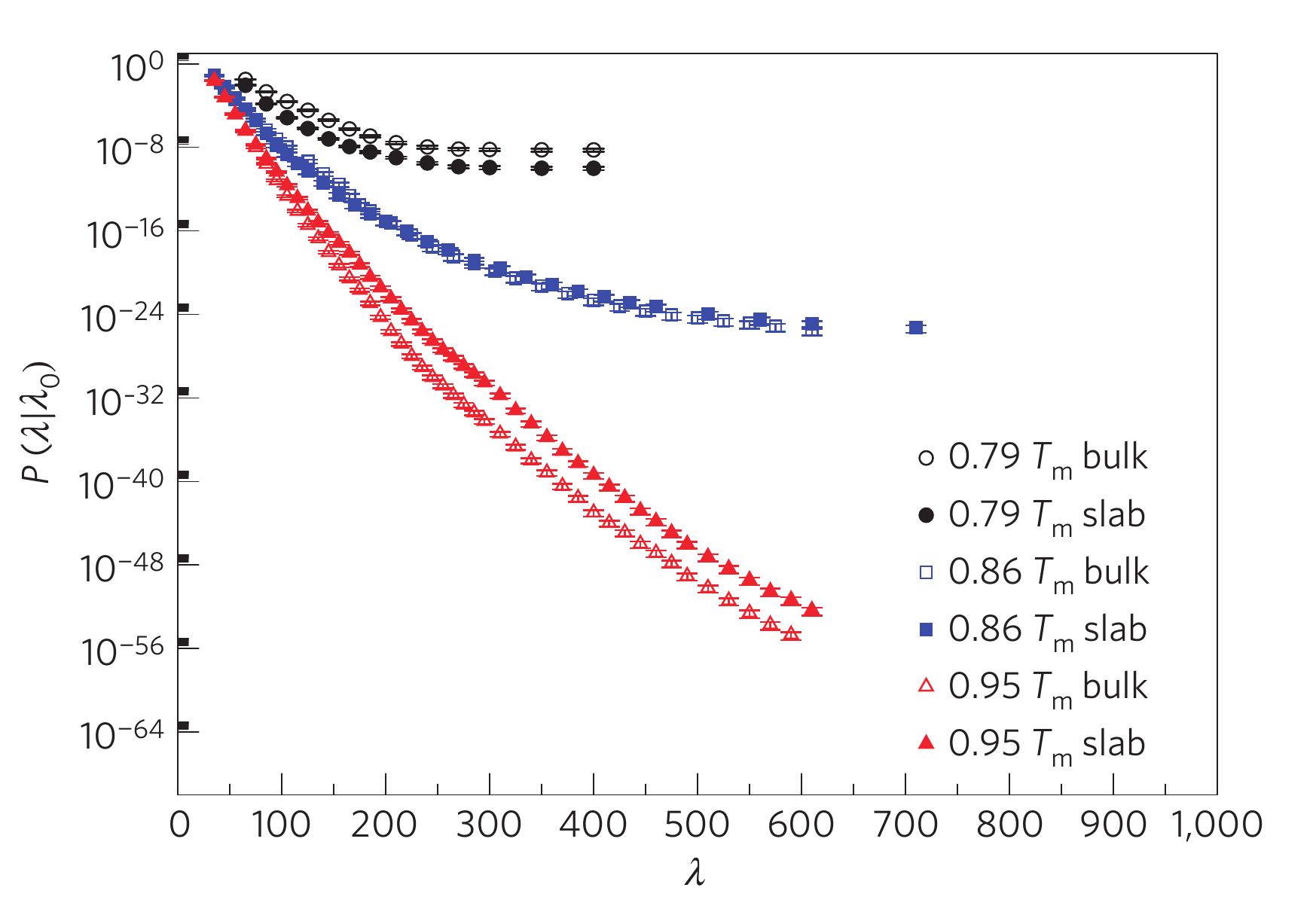} 
\caption{(Reproduced from Ref.~\citenum{LiNatMater2009}) Cumulative transition probability as a function of the OP for crystal nucleation in the bulk and in freestanding thin films of silicon. \label{fig:Si-surface-freezing}
}
\end{figure}

The first computational investigation of homogeneous crystal nucleation in supercooled silicon was conducted by  Li \textit{et al.}~\cite{LiJChemPhys2009}, who represented Si using the  Tersoff potential~\cite{TersoffPRL1988}. They found the nucleation rate to be a strong function of temperature, and to increase by as much as 17 orders of magnitude when temperature is decreased from 0.86$T_m$ to 0.79$T_m$. Their computed rate at 0.79$T_m$ was 18 orders of magnitude larger than the CNT estimate obtained from the coexistence $\gamma_{ls}$, suggesting that  surface tension at 0.79$T_m$ should be at least 28\% smaller than the coexistence value.  They also analyzed the transition path ensemble and observed twin boundary defects of the type depicted in Fig.~\ref{fig:twin-boundaries-mw} within the critical nucleus at 0.79$T_m$, but not at 0.86$T_m$. The fact that such defects are only observed at 0.79$T_m$ was attributed to rapid nucleation at lower temperatures, which, in turn, results in the formation of a large number of precritical nuclei that can then merge to form such defective nuclei. At higher temperatures, not only such nuclei form less frequently, but the arising defects get annealed more efficiently.  The authors suggested that such a dependence of defect density on temperature can be used for engineering and controlling defects within nucleated crystals.

After investigating crystal nucleation in the bulk, the same authors used FFS to study homogeneous nucleation in freestanding thin films of supercooled silicon~\cite{LiNatMater2009} with the aim of understanding whether free interfaces  enhance or suppress crystal nucleation. (We discuss the importance of this question in the context of surface freezing in water in Section~\ref{Role of Vapor-Liquid Interfaces in Homogeneous Nucleation}.) In their work, they used two different force-fields, namely the Tersoff and Stillinger-Weber potentials. In the case of the Tersoff potential, their rate calculations demonstrated a clear transition from bulk-dominated nucleation to surface-induced nucleation upon increasing temperature (Fig.~\ref{fig:Si-surface-freezing}), while for the Stillinger-Weber potential,  nucleation in the film was always faster. The authors attributed the enhancement of nucleation at the surface to the negative slope of the melting curve in silicon, and that density fluctuations needed for nucleation will be better accommodated at the free interface. They even hypothesized that this behavior will be universal for all materials for which the supercooled liquid is denser than the crystal. This  hypothesis, however, was later disproven as discussed in Section~\ref{Role of Vapor-Liquid Interfaces in Homogeneous Nucleation}.\\

\paragraph{NaCl:} Table salt, or NaCl, is the most abundant salt on earth, including in seawater. It is therefore of immense fundamental and practical importance to understand the formation of NaCl crystals, both within supercooled melts and aqueous solutions. The thermodynamically stable form of the NaCl crystal is a charge-ordered FCC lattice, with the space group $Fm\overline{3}m$~\cite{BraggProcRSoc1913}. Studies of the nucleation of NaCl crystals date back to 1990's, and have been conducted using a plethora of techniques~\cite{OhtakiPureApplChem1991, MuchaJPhysChemB2003, ZahnPhysRevLett2004, KawskaJChemPhys2006, AlejandrePhysRevE2007, MendozaJMolLiq2013, ChakrabortyJPhysChemLett2013, ZimmermannJACS2015}. In order to observe nucleation over computationally tractable timescales, however, most such studies were conducted in small systems and/or under larger supersaturations. Understanding crystal nucleation under experimentally relevant conditions requires utilizing techniques that allow for rate calculations at lower supersaturations. Consequently, the nucleation of NaCl crystals from supercooled melts~\cite{ValerianiJCP2005} and supersaturated solution~\cite{JiangJChemPhys2018, JiangJChemPhys2018p} has been extensively studied using FFS. Valeriani~\emph{et al.}~\cite{ValerianiJCP2005} utilized FFS to explore crystal nucleation from a supercooled melt of NaCl modeled using the Tosi Fumi force-field~\cite{FumiJPhysChemSolids1964}. They found the critical nuclei to share the structure of bulk NaCl crystal, and to be cubical in shape. They computed nucleation rates using umbrella sampling and diffusive barrier crossing, FFS, and TIS. Even though the rates computed from different methods were in reasonable agreement with one another, they all were  five orders of magnitude smaller than experimental estimates, which the authors attributed to possible inadequacies of CNT and the utilized force-field. 

\begin{figure}
	\centering
	\includegraphics[width=.31\textwidth]{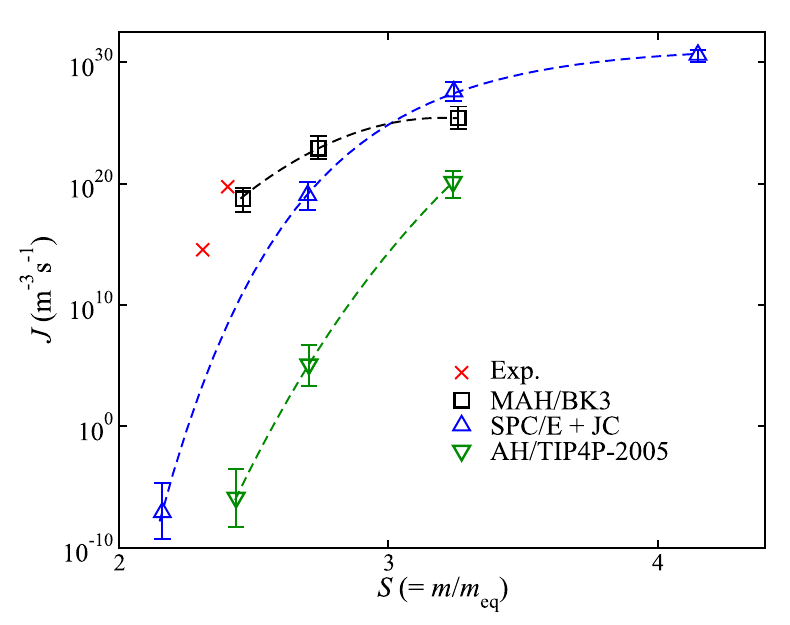}
	\caption{\label{fig:NaCl-pol}(Reproduced from Ref.~\citenum{JiangJChemPhys2018p}) Homogeneous NaCl nucleation rate as a function of supersaturation computed for different force-fields.  The experimental data are from Refs.~\citenum{NaJCrystGrowth1994} and~\citenum{GaoJPhysChemA2007}, while the rates for the SPCE/JC force-field are from Ref.~\citenum{JiangJChemPhys2018}.}
\end{figure}

More recent applications of FFS are concerned with NaCl nucleation from aqueous solutions. Jiang~\emph{et al.}~\cite{JiangJChemPhys2018} utilized FFS to study homogeneous crystal nucleation from supersaturated aqueous NaCl solutions of varying concentrations using the extended simple point charge (SPC/E)~\cite{BerendsenJPhysChem1987} and the Joung-Cheatham (JC)~\cite{JoungJPhysChemB2009} force-fields for water and NaCl, respectively.  The computed rates were considerably lower than the experimental estimates, and the nucleation pathways were satisfactorily described using CNT, unlike predictions from earlier studies that had observed a two-step nucleation process with amorphous precursors~\cite{ChakrabortyJPhysChemLett2013}. They attributed this discrepancy to large supersaturations in those earlier studies, and the possibility that distinct mechanisms might be in play at small and large supersaturations. They also observed a positive correlation between the crystallinity of a nucleus and its likelihood of becoming post-critical at later milestones. In their more recent paper,  Jiang~\emph{et al.}~\cite{JiangJChemPhys2018p} utilized jFFS (Section~\ref{section:FFS:jFFS}) to compute rates for both non-polarizable and polarizable force fields. Similar to their earlier work, they observed considerable disagreement with experiments when they utilized the  AH/TIP4P-2005 non-polarizable model~\cite{BenavidesJChemPhys2017}. Upon utilizing the polarizable  MAH/BK3~\cite{KolafaJChemPhys2016} force-field, however, better agreement with experiments was achieved (Fig.~\ref{fig:NaCl-pol}),  demonstrating the potentially important role of polarizability in accurately predicting the kinetics of nucleation in ionic systems.\\

\subsubsection{Evaporation} \label{Evaporation} 
\noindent
Evaporation is a process in which a liquid transforms into a gas, and is usually barrierless when a macroscopic vapor-liquid interface exists in the system. In the absence of free interfaces, however,  the formation of a thermodynamically stable vapor phase will require crossing a nucleation barrier, and will involve the emergence of sufficiently large pockets of vapor within the metastable liquid. Similar to crystal nucleation, vapor nucleation can be induced by an external surface (e.g.,~in hydrophobic evaporation), or can occur homogeneously within the liquid (e.g.,~in cavitation under negative pressures). Probing the kinetics and mechanism of evaporation under such circumstances is of considerable interest from a fundamental perspective, and can not only help us assess the effectiveness of CNT in predicting evaporation rates~\cite{OxtobyJCP1988, ZengJCP1991}, but can also provide us with valuable insights into the impact of interfacial properties (such as hydrophobicity) on cavitation and biological self-assembly~\cite{SumitPNAS2012, AltabetPNAS2017}. Furthermore, predicting and controlling the spatiotemporal distribution of evaporation events is critical to efficient operation of heat exchangers~\cite{WangJCPB2009} and electro-spray experiments~\cite{WangIJHMT2014}, as well as self-assembly of nanoparticles and biological entities~\cite{ChengJCP2013}. Similar to crystallization, probing evaporation events at a molecular level is challenging in experiments, and as a result, theoretical and computational studies have emerged as attractive alternative in understanding activated evaporation. In the theoretical realm, several authors, such as  Oxtoby~\emph{et al.}~\cite{OxtobyJCP1988}, Zeng~\emph{et al.}~\cite{ZengJCP1991} and Talanquer~\emph{et al.}~\cite{TalanquerJChenPhys1994} have used mean field density functional approaches to develop generalized nucleation theories for condensation and cavitation. In the realm of molecular simulations, evaporation and cavitation have been extensively studied both in bulk and confined geometries, using a wide variety of techniques such as conventional MD~\cite{KinjoFluid1998, WuMicroThermoEng2003, NovakPRB2007}, conventional~\cite{NeimarkJCP2005} and grand canonical transition-matrix MC~\cite{ShenJCPB2004}, umbrella sampling~\cite{ShenJCP1999, SundeepIndEngChem2002}, TPS~\cite{BolhuisJCP2000} and TIS\cite{MenzlPNAS2016}. Since its development, FFS has also been utilized for studying cavitation~\cite{WangJCPB2009, SumitPNAS2012, MeadleyJCP2012,  AltabetPNAS2017, AltabetJChemPhys2017}. The order parameters utilized in such investigations are much simpler than those employed in crystal nucleation studies, and typically quantify the size of the largest cavity in the system (in the case of homogeneous cavitation) or the number of liquid molecules in a pre-specified region of space (in the case of heterogeneous evaporation). In this section, we discuss applications of FFS to study evaporation in different systems. \\

\paragraph{The Lennard-Jones System:} As mentioned in Section~\ref{Lennard-Jones System}, the LJ potential is widely utilized for studying structural relaxation and phase transitions in simple liquids. Consequently, most theoretical and computational studies of evaporation have employed the LJ potential. One of the earliest such studies was by Zeng~\emph{et al.}~\cite{ZengJCP1991}, who utilized a nonclassical density functional theory approach to study cavitation in the LJ fluid.  Later on, Shen and Debenedetti~\cite{ShenJCP1999} used umbrella sampling to study  vapor nucleation in the superheated LJ fluid, and found the critical nuclei to be aspherical system-spanning voids. Evaporation in the LJ system was further investigated using direct MD ~\cite{KinjoFluid1998,WuMicroThermoEng2003,NovakPRB2007}, direct MC~\cite{NeimarkJCP2005}, umbrella sampling ~\cite{SundeepIndEngChem2002} and grand canonical transition-matrix MC~\cite{ShenJCPB2004}. Using FFS to probe evaporation kinetics is more recent, with the first study conducted by Wang~\emph{et al.}~\cite{WangJCPB2009}, who computed cavitation rates in the LJ fluid under  conditions similar to Ref.~\citenum{ShenJCP1999}. They chose the size of the largest bubble in the system as the order parameter, and computed it by distinguishing liquid- and vapor-like molecules based on the number of molecules within their first solvation shell, and identifying the grid elements within the simulation box that do not have any liquid-like molecules within a certain distance. Unlike Shen and Debenedetti~\cite{ShenJCP1999} who had identified the transition states to be spanning cavities, Wang~\emph{et al.}'s analysis of the TPE revealed that the critical bubbles were compact in shape.  They attributed this difference to the locality of their  order parameter, while Shen and Debenedetti~\cite{ShenJCP1999} had utilized a global order parameter, namely the total system density. However, their computed rates were still considerably larger than those predicted by CNT.

\begin{figure}
\centering
\includegraphics[width=0.45\textwidth]{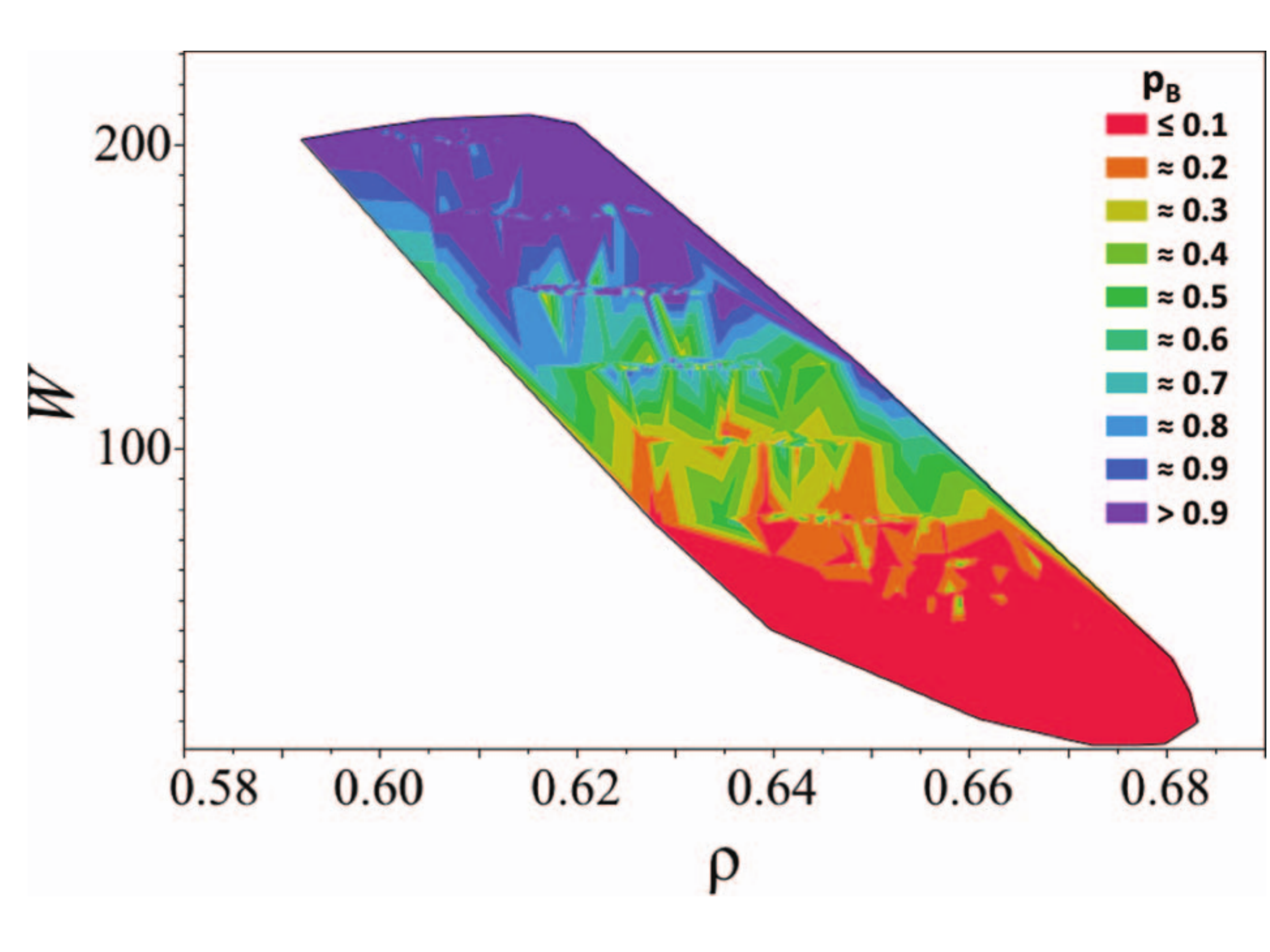} 
\caption{(Reproduced from Ref.~\citenum{MeadleyJCP2012}) Committor probability as a function of two order parameters, the volume of the largest bubble ($W$) and the global density ($\rho$).}
\label{fig:evapo_1}
\end{figure}

Another recent study by Meadley and Escobedo~\cite{MeadleyJCP2012} used FFS along with hybrid umbrella sampling MC and boxed MD~\cite{GlowackiJPhysChemB2009}, to probe homogeneous bubble nucleation in the stretched and superheated regime. They used two different order parameters, i.e.,~the global density ($\rho$) and the size of the largest bubble ($W$), with the barriers estimated from $\rho$ and $W$ differing by several $kT$'s. Using the FFS-LSE algorithm of Borrero and Escobedo~\cite{BorreroJCP2007} (Section~\ref{section:variants:new:op}), they concluded that $W$ is a better order parameter than $\rho$, as the committor probability $p_B$ is peaked around 50\% for $W$, but had a broader distribution for $\rho$ (Fig.~\ref{fig:evapo_1}). The fact that $\rho$ is such a poor order parameter arises from its inability to resolve the local density fluctuations relevant to cavitation. The shapes and sizes of critical bubbles predicted by Meadley and Escobedo were in disagreement with earlier studies that had considered smaller system sizes.  Such finite-size effects can skew the kinetics and mechanism of the underlying rare event in nontrivial ways by artificially altering the growth propensities of the pre-critical nuclei, due to their interactions with their periodic images in small systems.\\

\begin{figure*}
	\centering
	\includegraphics[width=.8\textwidth]{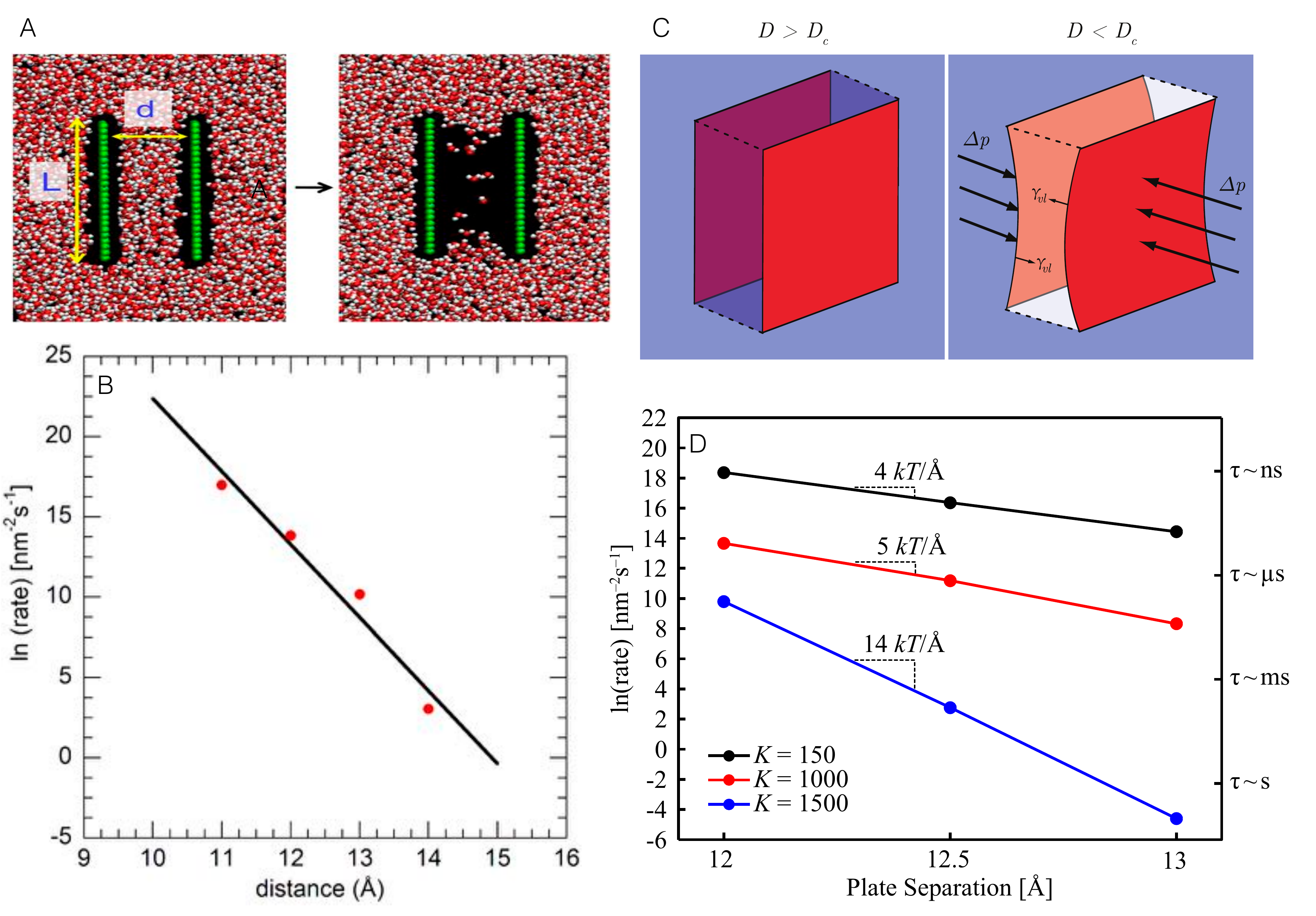}
	\caption{\label{fig:hydrophobic-evaporation} (A-B) (Reproduced from Ref.~\citenum{SumitPNAS2012}) Two hydrophobic surfaces separated by a gap $d$ are immersed within a bath of SPC/E water molecules. Capillary evaporation occurs between the plates at a rate depicted in (B). (C) (Reproduced from Ref.~\citenum{AltabetJCP2014}) Hydrophobic evaporation when the plates are flexible, in which case the critical separation $D_c$ will depend on the Young modulus of the plates. (D) (Reproduced from Ref.~\citenum{AltabetPNAS2017}) Dependence of rate on plate flexibility. $K$ is a measure of plate stiffness, with the sensitivity of rate on separation larger for larger $K$'s. }
\end{figure*}

\paragraph{Hydrophobic Evaporation:} It can be demonstrated~\cite{WallqvistJPC1995, ClaudioJPCL2011, AltabetJCP2014} using simple thermodynamic arguments that liquid water  confined between two hydrophobic surfaces will become metastable with respect to water vapor if the surfaces are sufficiently close (Figs.~\ref{fig:hydrophobic-evaporation}A,C). The process that culminates in the capillary evaporation of the confined liquid is called \emph{hydrophobic evaporation}, which has been linked to long-range hydrophobic interactions~\cite{BerardJCP1993, LumJPCB1999, LuzarJCP2000} in self-assembling systems and is thought to play an important role in biophysical processes such as ligand binding and protein folding~\cite{TanfordJACS1962}. Understanding the thermodynamics and kinetics of hydrophobic evaporation has therefore been the focus of numerous computational studies~\cite{LeungPRL2003, LuzarJPCB2004, LuzarJCP2000, BolhuisJCP2000, XuJPCB2010} utilizing both conventional and advanced sampling techniques. FFS investigations of hydrophobic evaporation are more recent, and mostly employ the SPC/E force-field~\cite{BerendsenJPhysChem1987} of water with a generic LJ-like representation of the hydrophobic plates.  The first such study was conducted by Sharma and Debenedetti~\cite{SumitPNAS2012} to understand how evaporation kinetics is impacted by $A$, the surface areas of the confining hydrophobic plates, and $d$, the separation between them (Fig.~\ref{fig:hydrophobic-evaporation}A). They utilized the total number of water molecules between the plates as the order parameter, which is conceptually simpler than the OPs used for studying homogeneous cavitation, but still satisfies the locality condition discussed earlier. They estimated evaporation barriers from the Arrhenius relationship, and found them to increase linearly with $d$ (Fig.~\ref{fig:hydrophobic-evaporation}B). As expected, the evaporation rates were higher for plates with larger surface areas.  The molecular mechanism of evaporation was also different for small and large plates. For smaller plates, only a few water molecules were present between the plates at the transition state. For larger plates, however,  the region between the  plates was not fully cavitated at the transition state, which was comprised of spanning cylindrical cavities instead. In a follow-up study, the same authors used~\cite{SharmaJPhysChemB2012} forward and reverse FFS (Section~\ref{section:method:forward-backward-ffs})-- alongside umbrella sampling-- to compute evaporation barriers, and observed an approximately linear dependence of the barrier on $d$, consistent with what they had previously concluded from the Arrhenius relationship~\cite{SumitPNAS2012}.

The hydrophobic plates considered by Sharma and Debendetti~\cite{SumitPNAS2012} were rigid in the sense that they were not thermalized throughout the simulation. In a subsequent study, Altabet~\emph{et al.}~\cite{AltabetPNAS2017} used FFS to inspect the effect of plate flexibility on the kinetics of hydrophobic evaporation. This work was motivated by their earlier theoretical analysis, which had revealed that increasing plate flexibility will enhance the thermodynamic stability of the evaporated state (Fig.~\ref{fig:hydrophobic-evaporation}C)~\cite{AltabetJCP2014}. In order to tune the flexibility of the hydrophobic plates, they adjusted $K$, the strength of springs that connected each substrate atom to its nearest neighbors. For each flexibility, they utilized the forward-backward FFS method (Section~\ref{section:method:forward-backward-ffs})  to compute free energy  profiles, in addition to the evaporation and wetting rates. In general, they found evaporation to become faster at smaller  $K$'s (Fig.~\ref{fig:hydrophobic-evaporation}D). Indeed, the dependence of rate and free energy landscape on $K$ was so strong that even small changes in $K$ would sometimes switch the thermodynamic stability of the wetted and evaporated states.  The authors argued that tuning the thermodynamics and kinetics of  evaporation via changing material flexibility can be a universal route for controlling self-assembly and conformational rearrangements in biological systems. By analyzing the transition path ensemble, they found that for all gap sizes and flexibility parameters, critical nuclei were gap-spanning vapor tubes. The formation of such tubes was far easier for the most flexible walls and was sometimes observed even during the exhaustive sampling of the wetted basin. For more rigid walls, however, their formation was delayed until  later milestones, and was preceded by the emergence of sub-critical gap-spanning tubes which would then grow radially to become critical.  These two steps were identified as the two bottlenecks to evaporation, and were harder to overcome for more rigid walls. They also found that the formation of the vapor phase within flexible walls had a lower free energy barrier and resulted in the formation of  a more stable vapor phase. 

In their latest work, Altabet and Debendetti~\cite{AltabetJChemPhys2017} utilized FFS to systematically investigate the dependence of the evaporation barrier and rate, as well as the free energy difference between the confined liquid and vapor phases to $d$. According to CNT, the dependence of nucleation rate on $d$ is either quadratic or linear depending on whether the liquid-vapor surface tension, or the three-phase line tension is dominant~\cite{SumitPNAS2012}. Their computed rates, however, revealed the breakdown of this classical picture at intermediate separations, a fact that they attributed to a structural transition within the metastable liquid from a trilayer to a bilayer. This breakdown resulted in a non-monotonic dependence of rate and free energy barrier on $d$. The precise separation at which this transition occurred depended on flexibility but was always around 1.2~nm. For the reverse condensation transition, however, no $d$-dependent structural transition occurred within the vapor phase, and hence no non-monotonicity was observed. The computed wetting rates and barriers therefore agreed reasonably well with those predicted by classical theories.

\subsubsection{Phase Separation}
\noindent
Phase separation is a process in which a new phase (with distinct density, symmetry and/or composition) emerges within the existing phase. In addition to being a cornerstone of modern chemical and biological separation, phase separation is thought to play a key role in biological self-assembly, such as the formation of membraneless organelles~\cite{HymanRev}. Consequently, its thermodynamics and kinetics have been extensively studied using a wide variety of theoretical, experimental and computational approaches~\cite{FalahatiSoftMatter2019}. However, the ability to accurately predict the timing and extent of phase separation in complex multi-component system still remains elusive. This problem is particularly daunting in the case of \emph{active matter}, or materials comprised of self-propelling particles or "agents`` that convert energy into mobility~\cite{RamaswamyAnnRevCondensMatPhys2010}. Examples of active matter are  diverse and include entities such as bacteria~\cite{WensinkPNAS2012}, flocks of birds~\cite{CavagnaPNAS2010} and a large variety of actively interacting entities within living cells \cite{RamaswamyJSM2017}. Due to their out-of-equilibrium nature, it is conceptually challenging to determine whether a particular assembly process is a thermodynamically driven phase separation~\cite{FalahatiCurrBiol2016, FalahatiPNAS2017}. An interesting subclass of active matter are active colloids~\cite{ZottiJPhysCondensMat2016}, which have potential application in areas such as self-assembly, \emph{in vivo} drug delivery, and lab-on-a-chip microfluidics~ \cite{EbbensCOCIS2016}. As a result, numerous computational studies of self-assembly in active matter  have been conducted, revealing important similarities between phase separation in active and equilibrium systems~\cite{ButtinoniPRL2013, RednerPRL2013, StenhammarPRL2013, StenhammarSoftMatter2014, SpellingsPNAS2015, MalloryJACS2019}. Probing phase separation under lower saturations, however, requires utilizing advanced sampling techniques  in order to access the timescales relevant to the nucleation of the emerging structures and patterns.

\begin{figure}
\centering
\includegraphics[width=0.43\textwidth]{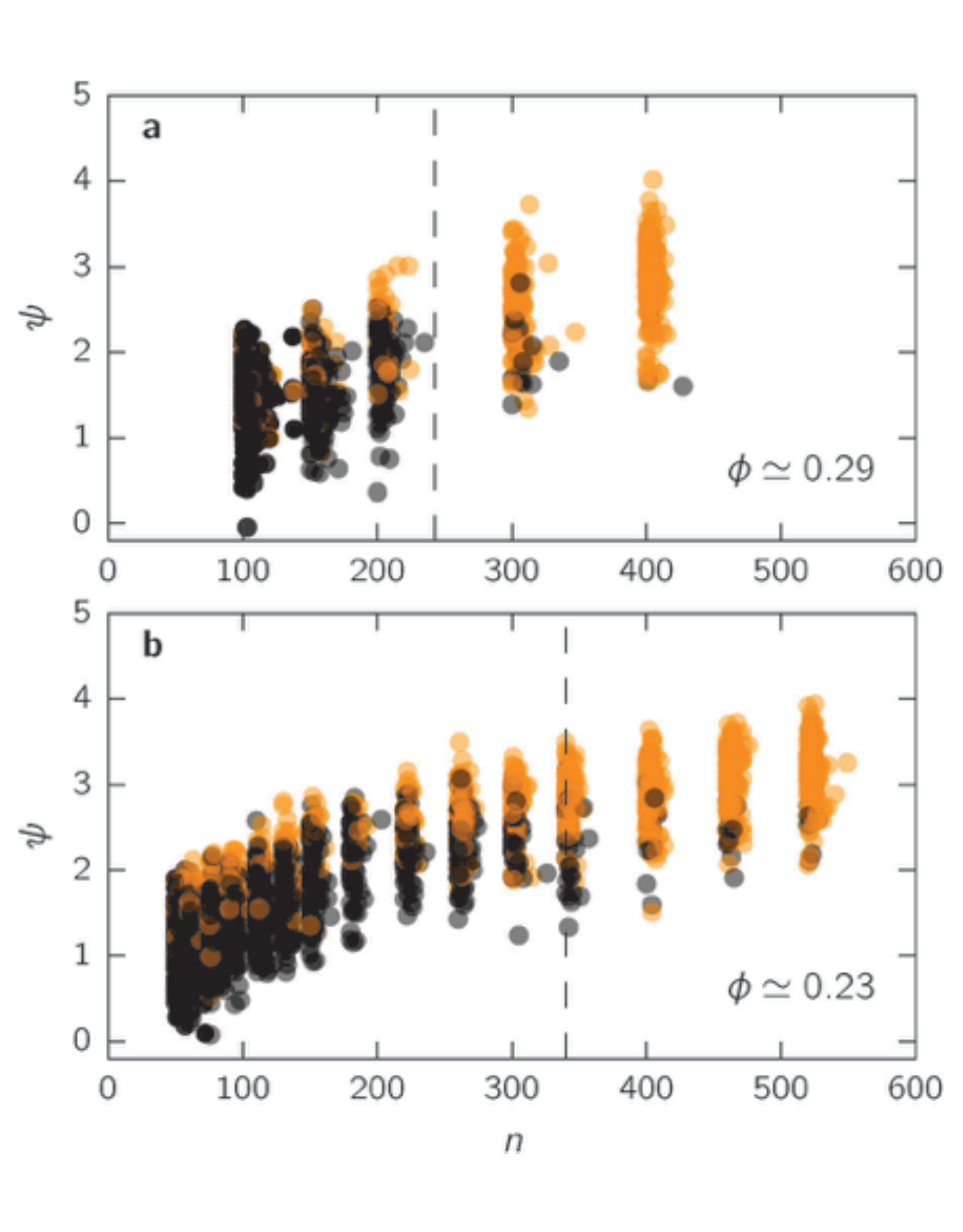} 
\caption{(Reproduced from Ref.~\citenum{RichardSoftMatter2016}) The color-coded distribution of per-configuration transition probabilities for configurations with different $n$ and $\psi$ values at (a) $\phi=0.29$ and (b) $\phi=0.23$. The dotted lines specify the critical $n$ at each packing fraction. Lighter colors correspond to larger transition probabilities.}
\label{fig:phase_1}
\end{figure}

Utilizing FFS to study phase separation is new. Recently, Richard~\emph{et al.}~\cite{RichardSoftMatter2016} utilized FFS alongside BD simulations  to probe the kinetics of phase separation as a function of packing fraction ($\phi$) in a system of active Brownian discs. Similar to crystal nucleation, they utilized $n$, the size of the largest dense liquid cluster, as the FFS order parameter. While phase separation was observed to readily occur during conventional BD at sufficiently large packing fractions, no nucleation event was observed for $\phi<0.29$ in brute-force BD.  Overall, nucleation rates  exhibited  an exponential dependence on $\phi$, suggesting the presence of an activated nucleation-like scenario similar to equilibrium phase separation processes. They also analyzed the transition states by computing committor probability distributions for three reaction coordinates: $n$ (which was also the FFS OP), $\rho$ (density) and $\psi$ (polarization), which quantifies whether the active discs within the largest cluster are moving towards or away from its center.  Analyzing the transition states revealed that none of these three reaction coordinates can accurately represent the transition state, since their committor probability distributions were not peaked around $50\%$. Performing maximum likelihood estimator (MLE) analysis~\cite{SwetlanaMolPhy2013} on linear combinations of $n$, $\rho$ and $\psi$, however, yielded reaction coordinates with narrowly peaked committor probability distributions. In particular, a linear combination of $n$ and $\psi$ was the best reaction coordinate. The authors interpreted this finding by noting that the stability of a dense liquid cluster is not only determined by its size, but also by the polarization of its surface particles. More precisely, if the surface particles are oriented away from the center of the cluster, they will destabilize it by moving outwards, while if they point inwards, their  push will be resisted by the center of the cluster, and the cluster will remain stable.  Indeed, the transition probability at any given $n$ was always higher for larger $\psi$'s as depicted in Fig.~\ref{fig:phase_1}.

Another interesting class of systems that can undergo phase separation are  \emph{metallic alloys}, which  are multi-component non-stoichiometric metallic crystals. When a metallic alloy falls out of its thermodynamic stability window, e.g.,~due to  changes in temperature or pressure, it typically phase separates into two coexisting crystals. The ensuing transition, however, can proceed via nucleation and growth and spinodal decomposition for small and large thermodynamic driving forces, respectively. In probing the kinetics of phase separation in alloys, it is both necessary to utilize advanced sampling techniques such as FFS, and to develop alternative algorithms such as kinetic MC to capture structural rearrangements in crystals. Recently, FFS was used for studying Cu precipitation within a BCC Fe-Cu alloy. Even though this process has been extensively studied in experiments~\cite{OthenPhilMagLett1991},  there are still important outstanding questions about the composition of Cu-rich precipitates~\cite{SchoberApplPhysA2010}. Recently, Qin~\emph{et al.}~\cite{QinActaMaterialia2018} combined FFS with rigid lattice Monte Carlo (LMC), a kinetic MC scheme that allows for vacancy jumps, which are critical to structural relaxation in crystals. The correspondence between MC steps and time was established by matching the vacancy jump timescales to experiments~\cite{WarczokIntJMaterRes2011}. They studied phase separation in alloys with 1 and 1.5\% Cu, and found critical nuclei to have between 10-40 atoms. They found the smaller nuclei to have a substantial Fe contact and to be anisotropic in shape, but the Fe content decreased drastically upon aging.

\subsubsection{Coalescence}\label{section:coalescence}
\noindent
The ability to control the  size distribution of  dispersed phase droplets within emulsions is critical to many applications, such as oil extraction~\cite{EowChemEngJ2002}, and the preservation of food~\cite{RoosFoodEngRev2016}, pharmaceutical~\cite{WangPharmRes2019} and cosmetic~\cite{CastelFoodHydrocolloids2017} products. Of particular relevance to all such applications is the timescale of droplet coalescence, a thermodynamically driven process that proceeds through the collision of droplets followed by the rupture of the liquid-liquid interface. In systems without surfactants, the probability of surface rupture after collision is usually very high, and the kinetics of coalescence is therefore determined by the rate of collisions between droplets.  The likelihood of surface rupture can, however, be considerably decreased upon introducing surfactants. Under such circumstances, droplet coalescence will become a rupture-limited rare event, and will only occur if droplet collision is followed by the rupture of the droplet surface and nucleation of a liquid bridge between the two droplets. Therefore, the kinetics and mechanism of droplet coalescence in the rupture-limited regime can only be probed using advanced sampling techniques such as FFS. 

\begin{figure}
\centering
\includegraphics[width=0.4\textwidth]{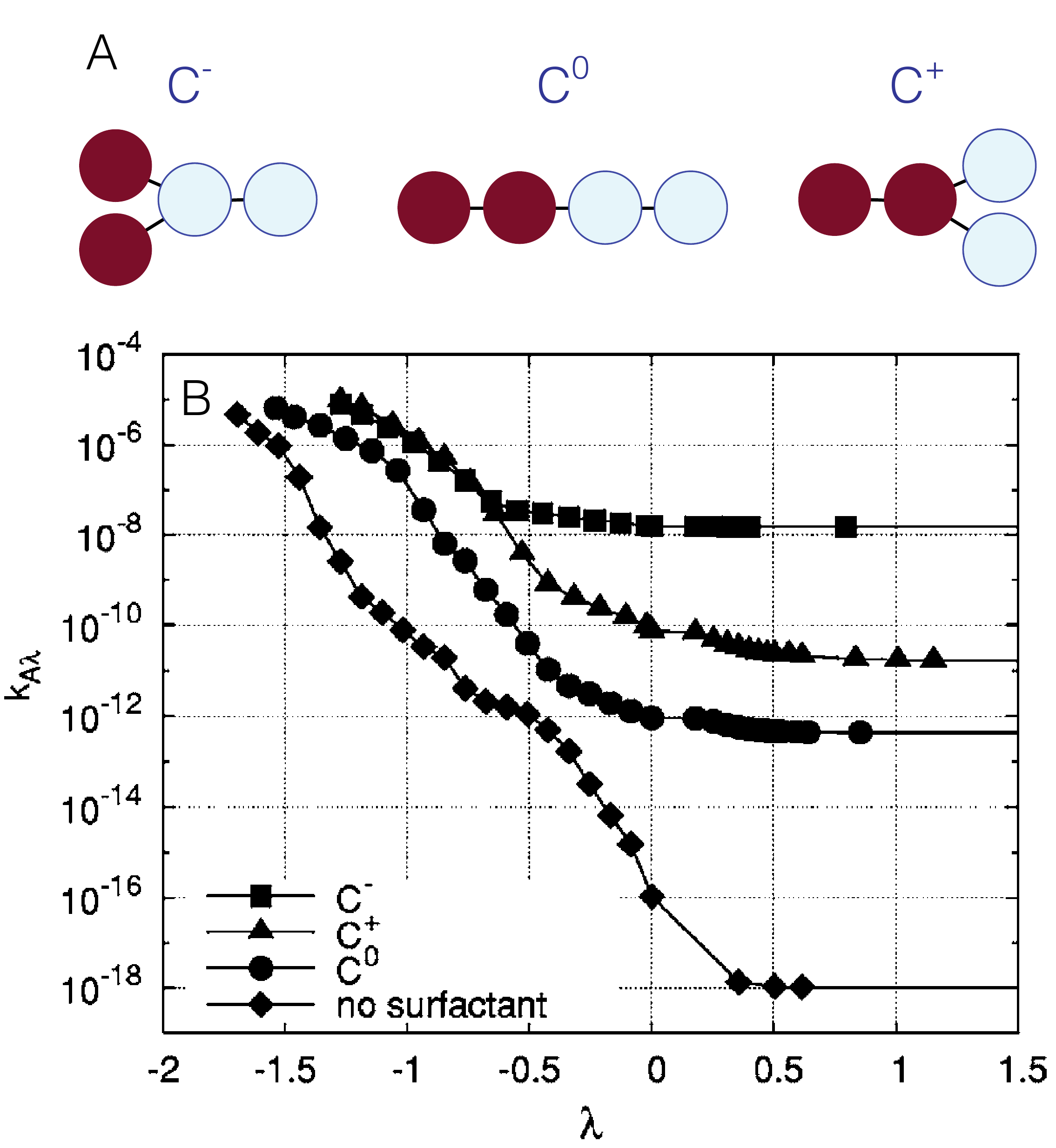} 
\caption{(A) Surfactants with negative, zero and positive curvatures pre definition of Israelachvili~\cite{IsraelachviliColloidSurfaceA1994}. Dark red and light blue beads correspond to hydrophobic and hydrophilic groups, respectively. (B) (Reproduced from Ref.~\citenum{RekvigJCP2007}) Partial coalescence rates as a function of the order parameter for a system with no surfactant and for systems with surfactants with different  curvatures.}
\label{fig:coalescence}
\end{figure}

Rekvig and Frenkel~\cite{RekvigJCP2007} were the first to use FFS  to study droplet coalescence in a coarse-grained model of an oil-water emulsion. They considered surfactant-free systems, as well as three types of surfactants with positive, zero, and negative natural curvatures (as defined by  Israelachvili~\cite{IsraelachviliColloidSurfaceA1994} and depicted in Fig.~\ref{fig:coalescence}A). They also focused on $\mu$m-scale and larger droplets, which enabled them to represent the contact region as a water film sandwiched between two flat oil-water interfaces. In order to simulate larger systems with longer time steps, they utilized the  coarse-grained dissipative particle dynamics (DPD) method~\cite{GrootJChemPhys1997}.  They chose the radius of the connecting bridge as the FFS order parameter. For configurations without such a bridge, the negative of the minimum distance between the two oil films was chosen as the OP.  

Their most remarkable-- and counter-intuitive-- finding was that the surfactant-free system is more stable, and undergoes coalescence at a rate five orders of magnitude smaller than the next most stable system (Fig.~\ref{fig:coalescence}B). They attributed this counter-intuitive behavior to the inability of DPD to capture capillary evaporation and nanobubble formation, which both play pivotal roles in oil-water-oil coalescence. They also probed the dependence of coalescence kinetics and mechanism on  film separation in the surfactant-free system. Apart from the expected increase in stability for films that are further apart, they found a qualitative difference between the  mechanism of coalescence for thinner and thicker films. While thinner films coalesced after the emergence of sufficiently large thickness fluctuations, thicker films only coalesced after the formation of a critical bridge connecting the two oily droplets. 

They also explored the effect of the natural curvature of the surfactant on the coalescence kinetics. According to channel nucleation theory of Kabalnov and Wennerstr\"{o}m~\cite{KabalnovLangmuir1996}, the free energy barrier to coalesce will be smaller if the bending of the droplet surface during rupture results in a favorable curvature for surfactant molecules. This will imply that a surfactant with positive natural curvature will better stabilize an oil-water-oil system. Indeed, they found the positive-curvature surfactants to be better stabilizing agents than their negative-curvature counterparts. They, however, observed that zero-curvature surfactants are  even better at stabilizing the mixture (Fig.~\ref{fig:coalescence}B). They explained this finding by suggesting that zero-curvature surfactants will lead to higher surface tensions and bending rigidities, and therefore reduced interfacial fluctuations.

\subsubsection{Wetting}
\begin{figure}
\centering
\includegraphics[width=.45\textwidth]{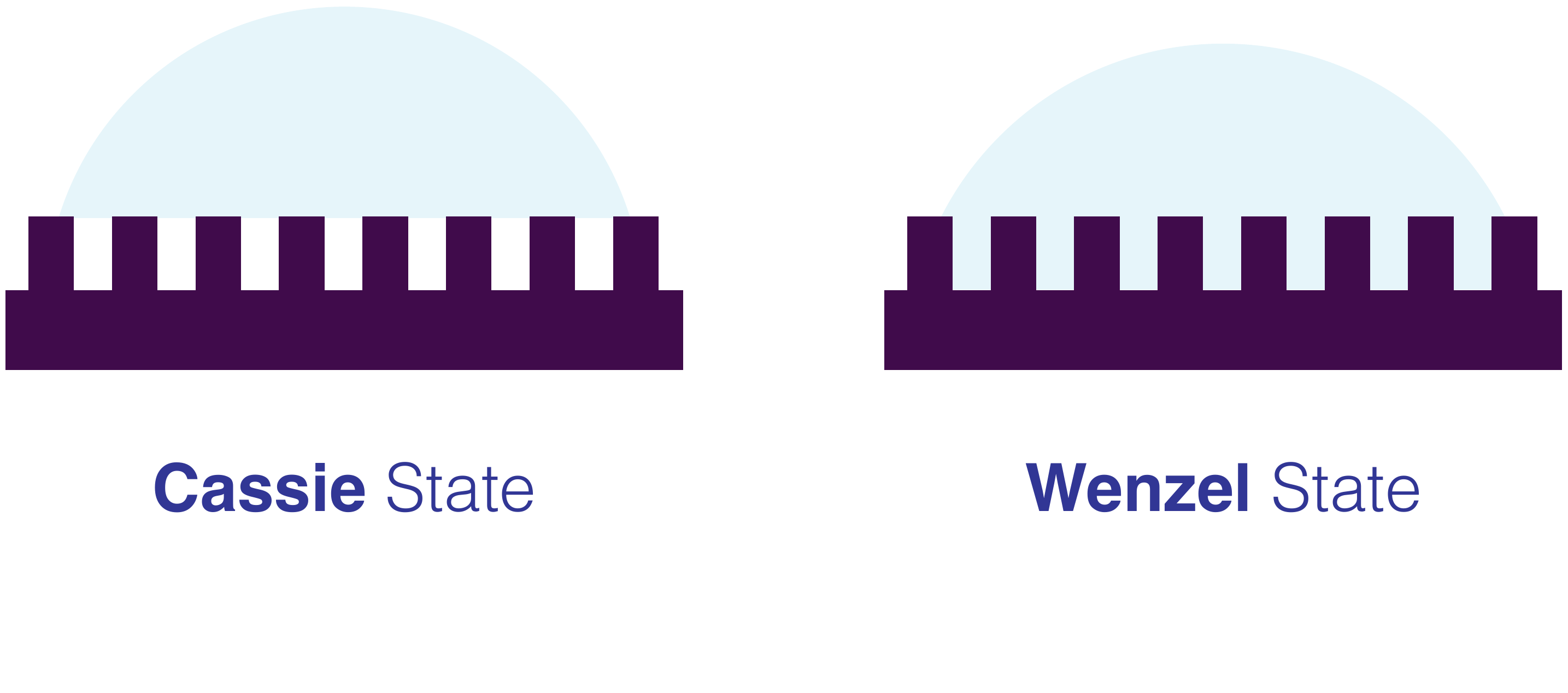}
\caption{\label{fig:wenzel-Cassies} Wenzel and Cassie states in wetting of topographically rough surfaces.}
\end{figure}

\noindent
Wetting is a process that can occur in a system of two coexisting phases $1$ and $2$, which are in contact with an external surface $s$.  According to classical thermodynamics, the equilibrium behavior of such a system will depend on a dimensionless parameter called  \emph{wettability}~\cite{HajiAkbariJCP2017}, which is given by the Young equation:
\begin{eqnarray}
\zeta &=& \frac{\gamma_{2s}-\gamma_{1s}}{\gamma_{12}}
\label{eq:Young-generalize}
\end{eqnarray}
Here, $\gamma_{\alpha\beta}$ is the surface tension between phases $\alpha$ and $\beta$. Full wetting of $s$ by phases $1$ and $2$ will occur for $\zeta\ge1$ and $\zeta\le-1$, respectively. For $|\zeta|<1$, wetting will be partial and both phases will be in contact with $s$ wth a contact angle $\theta_c=\cos^{-1}\zeta$. If the phase that is expected to fully or partially wet $s$ is not originally in contact with it, however, the process of reaching equilibrium might involve crossing a free energy barrier, similar to what was discussed for coalescence in Section~\ref{section:coalescence}. A scenario that is even more interesting is the wetting of topographically inhomogeneous surfaces for which multiple (meta)stable wetted states might exist with differing levels of microscopic exposure of the valleys to the wetting phase. For chemically uniform but topographically rough surface, for instance, two such states exist, namely the Wenzel~\cite{WenzelIndEngChem1936} and Cassie~\cite{CassieTransFaradaySoc1944} states with fully wetted and largely de-wetted valleys, respectively (Fig.~\ref{fig:wenzel-Cassies}). Since the free energy barriers that separate such metastable states can be very large, a surface might remain at a lower level of wetting (e.g.,~the Cassie state) even when full wetting is thermodynamically favored. This provides an avenue for designing phobic surfaces particularly for low-surface tension fluids for which surface chemistry alone might not be sufficient to induce solvophobicity~\cite{TutejaScience2007}. Understanding the kinetics and mechanism of wetting on rough surfaces can therefore pave the way to design surfaces with enhanced kinetic stability (i.e.~larger wetting barriers). More often than not, however,  achieving this goal requires utilizing advanced sampling techniques due to prohibitively long transition times.
In recent years, several computational studies of surface roughness for wetting transitions have been conducted using a wide variety of advanced sampling techniques~\cite{KoishiPNAS2009, SavoyLangmuir2012, SavoyLangmuir2012b, GiacomelloLangmuir2012, ShahrazLangmuir2014, RenLangmuir2014, PrakashPNAS2016} including FFS~\cite{SavoyLangmuir2012, ShahrazLangmuir2014}.

The first investigation of wetting using FFS was conducted by Savoy~\emph{et al.}~\cite{SavoyLangmuir2012}, who inspected the kinetics and mechanism of Cassie-to-Wenzel transition in a model system of droplets comprised of LJ tetramers placed in contact with a rough surface with equally spaced nails (also comprised of LJ particles). They utilized the constrained branched growth variant of FFS~\cite{VelezVJCP2010} with the number of LJ monomers below the nail tops as the order parameter. By systematically tuning $\epsilon_{12}$, the strength of LJ interactions between LJ monomers and surface particles, they tuned the solvophobicity of the substrate. Their rate calculations and analysis of the TPE revealed that wetting rates were higher for philic droplets (i.e.,~those with larger $\epsilon_{12}$ values), and for shorter posts. They also found the transition to occur faster for smaller droplets, as they need to undergo less deformation from their spherical shape during the wetting process. By analyzing TPE, they found the transition state to be comprised of  configuration in which a droplet  touches the bottom surface for the first time. The authors argue that the utilized order parameter is sub-optimal, as it undergoes unphysical fluctuations when larger droplets fill multiple cavities. By conducting committor analysis on  linear combinations of a few reaction coordinates, they identified $\lambda + \gamma^2$ as a better order parameter, with $\gamma$ being the density of particles within the most filled cavity. They argued that this combined OP correctly captures the extent of wetting of a cavity, and the necessary geometric distortions in the droplet. Upon conducting FFS using this optimized OP, they obtained similar rates but with an increased computational efficiency.

A more recent FFS study of wetting was conducted by Shahraz~\emph{et al.}~\cite{ShahrazLangmuir2014}, who utilized constrained branched growth FFS~\cite{VelezVJCP2010} to compute the rates of Wenzel-Cassie  (W$\rightarrow$C)  and Cassie-Wenzel (C$\rightarrow$W) transitions for an LJ  droplet in contact with a grooved surface. They systematically assessed the importance of groove width ($G$) and height ($H$) on the rate and free energy profile of each transition. Similar to Ref.~\citenum{SavoyLangmuir2012}, they found both rates  to decrease upon increasing $H$ and the size of the droplet. Increasing $G$, however, stabilized the Wenzel state. While the C$\rightarrow$W transition was equally sensitive to changes in $H$ and $G$, the W$\rightarrow$C transition was more sensitive to changes in $G$. They utilized their forward and backward FFS calculations (Section~\ref{section:method:forward-backward-ffs}) to also compute the free energy profiles, and found good agreement with their theoretical model.

\subsection{Collective Conformational Rearrangements in Biomolecular Systems}\label{section:conformation-bio}


\noindent
Biomolecules are naturally occurring organic molecules with critical biological functions, and are synthesized biochemically in living cells. For instance, proteins-- or polypeptides-- catalyze biochemical reactions, transport ions and molecules through membranes, transduce signals, and play structural roles within the cell. Nucleic acids, such as DNA and RNA, however, act as storers and transmitters of genetic information. The ability of these biomolecules to perform such complex functions is usually linked to their "native`` structure(s) (e.g.,~tertiary and quaternary structures of proteins), or transitions between multiple conformations (e.g.,~between single- and double-stranded DNA  during replication and transcription). Therefore, investigating the conformational rearrangements that result in the emergence or dissociation of such native structures is essential for understanding the origin of their stability and function.  This understanding can be leveraged to identify strategies for treating diseases that arise due to undesirable conformations of biomolecules, such as Alzheimer's, Parkinson's, Huntington's, and cataract.  Biomolecules can also be used for nano-engineering of novel structures via manipulating systems such as polymers and nanoparticles. Consequently, there has been considerable interest in using experimental and computational techniques to explore conformational rearrangements in biomolecules, which has been an active area of research for decades.

From early days of computational biology, molecular simulations have been at the forefront of exploring different aspects of biomolecular structure and function~\cite{KarplusNatStructMolBiol2002}. Among these, major conformational rearrangements are the most challenging to study as their occurrence usually requires crossing free energy barriers. Capturing the kinetics and mechanism of such transformations therefore requires utilizing advanced sampling techniques~\cite{ScheragaAnnRevPhysChem2007} such as FFS. Compared to the other rare events considered thus far, conformational rearrangements of biomolecules need to be described using more sophisticated structural order parameters, which can be further optimized using MLE-based approaches~\cite{PetersJCP2006, BorreroJCP2007, LechnerJCP2010,LechnerPRL2011}.

\subsubsection{Protein folding and Aggregation}

\noindent
As mentioned above, proteins perform a wide variety of complex biochemical functions.
The functionality of most proteins is due to their highly specific native structure,  which makes them uniquely suitable for their respective biological activity. The inability of a protein to form its native structures can therefore not only hamper its respective biological function, but can also result in  pathological aggregation of misfolded peptides, which has been linked to several diseases such as the Alzheimer's, Parkinson's, Huntington's, and cataract.  Understanding protein folding has thus been an important frontier of biochemical research for many decades. However, protein folding almost always involves crossing a free energy barriers, and can occur over timescales not accessible to conventional MD. Indeed, even though the first\cite{McCammonNature1977} MD simulation of a protein was conducted in 1977, it took an additional 21 years for the folding pathway of a protein to be explored in an atomistic explicit-solvent MD simulation~\cite{DuanScience1998} of villin headpiece subdomain, a 36-residue peptide~\cite{McKnightJMB1996}. Historically, several approaches have been pursued to circumvent this timescale problem. One approach has been to drastically curb the computational cost of the underlying simulations by focusing on simplified models of proteins, such as on-lattice models, coarse-grained models and implicit-solvent models. In other words, using such models enables faster integration of equations of motion, or faster exploration of the configuration space.  Another approach is to use advanced sampling techniques. Indeed, numerous studies of protein folding have been conducted using  replica exchange MD (REMD) ~\cite{ZhouPNAS2003}, TPS\cite{JuraszekPNAS2006}, TIS\cite{BolhuisPNAS2003, JuraszekBiophysicalJ2008} and FFS\cite{BorreroJCP2006, JuraszekBiophysicalJ2008, BorreroBioPhy2010, SmitJPhysChemB2017}. It must, however, be noted that even though these two approaches can be pursued concurrently, using advanced sampling techniques will provide increased flexibility to choose realistic force-fields that accurately represent a peptide's interactions with solvent molecules. As discussed above, FFS has been utilized with models with different levels of coarse-graining. Here, we discuss the applications of FFS to study protein folding in various systems\\

\begin{figure}
\centering
\includegraphics[width=.38\textwidth]{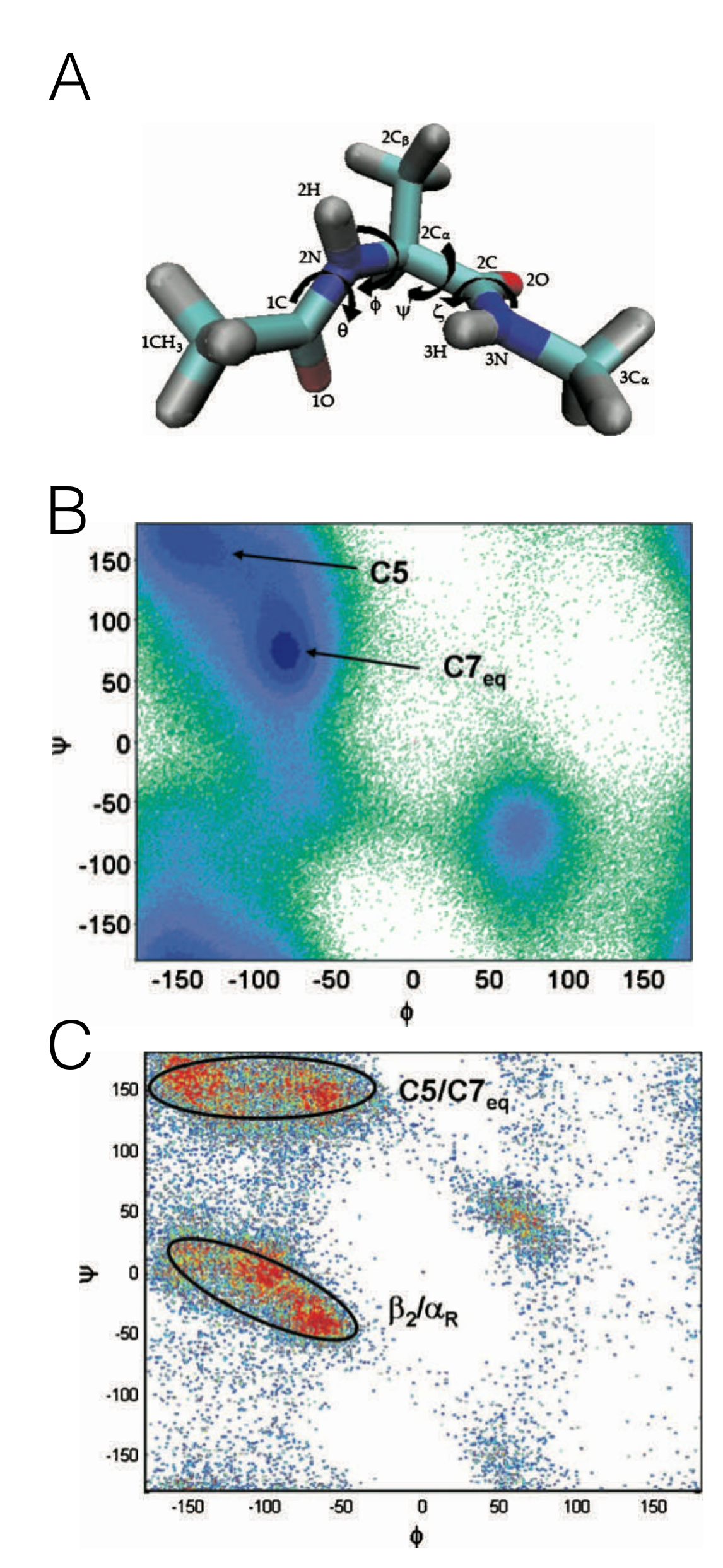}
\caption{\label{fig:alanine-dipeptide}(Reproduced from Ref.~\citenum{VelezJCP2009}) (A) Molecular structure of alanine dipeptide. (B-C) Free energy landscape of the alanine dipeptide system in (B) vacuum and (C) aqueous environments at 300~K.}
\end{figure}

\paragraph{Short Peptides:} Even short peptides can possess rough free energy landscapes with multiple local minima. Transitions among such minima are typically rare events as they require surmounting free energy barriers. Yet, they are usually frequent enough to be captured during computationally tractable MD and MC simulations. Short peptides are therefore excellent systems for validating and benchmarking advanced sampling techniques. Among those, the alanine dipeptide system (Fig.~\ref{fig:alanine-dipeptide}A) has been extensively studied computationally~\cite{BrooksChemRev1993}. In vacuum, the alanine dipeptide system has two major minima, namely the C7$_{\text{eq}}$ and C5 conformations (Fig.~\ref{fig:alanine-dipeptide}B). Solvated dipeptide, however, can occupy multiple minima in the free energy landscape (Fig.~\ref{fig:alanine-dipeptide}C). For instance, Velez-Vega~\emph{et al.}~\cite{VelezJCP2009} utilized the FFS-LSE approach to identify important reaction coordinates for the C7$_{\text{eq}} \Rightarrow$C5 transition in vacuum, and the $\beta_2/\alpha_R\Rightarrow\text{C5/C7}_{\text{eq}}$ transition in the solvated peptide, and found excellent agreements with experiments. By utilizing FFS-LSE, they identified an extra dihedral angle (not commonly used in standard verification approaches) to be important in the C7$_{\text{eq}} \Rightarrow$C5 transition in vacuum, and several overlooked dihedral angles in the $\beta_2/\alpha_R\Rightarrow\text{C5/C7}_{\text{eq}}$ transition. In a later paper, DeFever and Sarupria validated the contour FFS algorithm in  the alanine dipeptide system~\cite{DeFeverJChemPhys2019}.
\\

\paragraph{Qualitative Coarse-grained Models:} In these models,  each protein is represented as  a connected graph of generic interaction sites each representing a single amino acid. Such models can be  both on- and off-lattice. In on-lattice models, interaction sites occupy the vertices of a lattice, and the system evolves using on-lattice MC moves that maintain peptide connectivity. In off-lattice models, however, interaction sites move continuously in space and connectivity is maintained via traditional bonded interactions, such as bonds, angles, dihedrals, etc. In general, on-lattice models are  faster computationally than their off-lattice counterparts. Their major disadvantage, however, is the lack of a rigorous framework for mapping MC steps to real time in experiments or off-lattice simulations. Furthermore, such on-lattice models usually lack specificity, which makes them imperfect for capturing the folding kinetics of a particular protein.  Regardless, these qualitative models offer the highest level of coarse-graining and  constitute the least computationally demanding representations of proteins, and despite their shortcomings, can provide valuable insights into the underlying physics of protein folding and conformational rearrangements. As far as protein folding is concerned, it has been shown that peptides represented by such models can 'fold` into  'native` states upon energy minimization. This has created increased interest in using qualitative coarse-grained models to study protein folding~\cite{BaumketnerJMB2003, JewettPNAS2004, ContrerasMartinezBiotech2006}. 

FFS was first used by Borrero~\emph{et al.}~\cite{BorreroJCP2006} to study the folding kinetics of a 48-residue lattice protein represented using the Miyazaka-Jernigan  contact energy potential~\cite{MiyazawaMacromolecules1985} in bulk and confined geometries. Studying protein folding under confinement is biologically relevant as many chaperonins facilitate protein folding by engulfing the corresponding peptide in a nano-cage~\cite{HayerHartlTrendsBiochemSci2016}. Borrero~\emph{et al.} utilized  the fraction of native contacts present in a partially-folded configuration as the FFS order parameter, and computed folding rates using branched growth FFS. They concluded that confinement enhances the folding kinetics up to a point. Too strong of a confinement, however,  was found to be detrimental to folding since it traps the protein in local free-energy minima. They also found that the folding process proceeded via the formation of a nucleus of native contacts in the transition state. The importance of such nuclei of "native`` contacts was further established in another work by the same group in which Borrero~\emph{et al.}~\cite{BorreroBioPhy2010} used FFS to study the reassembly of fragments of a split lattice protein. They found that the kinetics of folding and reassembly was much faster for fragments containing equally split core nuclei, as compared to fragments in which one fragment had a higher content of the core residues.\\

\paragraph{Trp-cage:} Trp-cage \cite{NeidighNatureStructBio2002} is a 20-residue peptide that readily folds into its native structure within the timescales accessible to conventional MD. It has therefore been extensively studied  via direct MD~\cite{SnowJACS2002, SimmerlingJACS2002, OtaPNAS2004, DingBiophysicalJ2005} and advanced sampling techniques such as REMD~\cite{ZhouPNAS2003, KimPNAS2016,  UralcanJPhysChemB2018, UralcanJPhysChemLett2019}, TPS~\cite{JuraszekPNAS2006} and TIS~\cite{JuraszekBiophysicalJ2008}. The average folding time for Trp-cage is estimated to be $\approx 4.1~\mu$s, and can, in principle, be computed using conventional MD. Such brute-force calculations, however, will still be computationally expensive, and collecting quality statistics about the folding mechanism, in particular, will be extremely costly. 

\begin{figure}
\centering
\includegraphics[width=.34\textwidth]{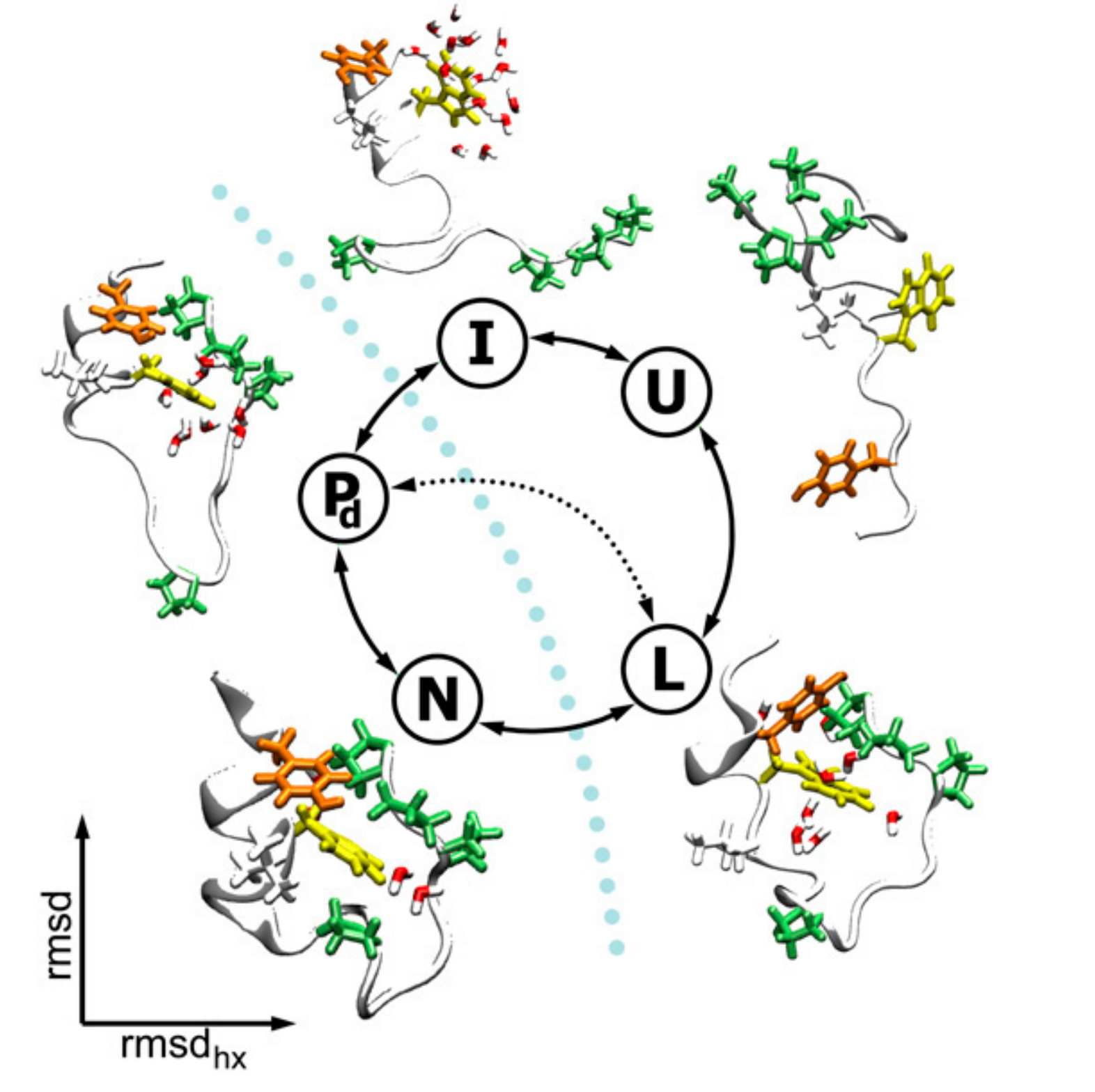}
\caption{\label{fig:trp-cage}(Reproduced from Ref.~\citenum{JuraszekBiophysicalJ2008}) Two distinct folding pathways for the Trp-cage protein.}
\end{figure}

The first application of FFS to study conformational rearrangements in Trp-cage was conducted by Juraszek and Bolhuis~\cite{JuraszekBiophysicalJ2008}. This was a follow-up to their earlier TPS calculation in which they had identified two major folding pathways for Trp-cage (Fig.~\ref{fig:trp-cage}), namely the U$\rightarrow$L$\rightarrow$N pathway in which the tertiary contact forms before the $\alpha$-helix, and the  U$\rightarrow$I$\rightarrow$P$_d\rightarrow$N pathway in which the helix forms prior to the establishment of the tertiary contact~\cite{JuraszekPNAS2006}. In Ref.~\citenum{JuraszekBiophysicalJ2008}, they explored the more likely pathway, namely U$\rightarrow$L$\rightarrow$N, and used TIS to compute both folding and unfolding rates while employing FFS only to inspect the unfolding process. They utilized $\text{rmds}_{\text{hx}}$, the root mean square deviation of $\alpha$-helical residues from the native structure, as the order parameter for both TIS and FFS. The unfolding rate computed from FFS was around 80 times smaller than that of TIS, corresponding to a  $\approx4k_BT$ discrepancy in the free energy barrier. They attributed this discrepancy to the fact that $\text{rmds}_{\text{hx}}$ is a sub-optimal OP. FFS was found to be more sensitive to a subpar OP than TIS, likely due to insufficient sampling in the basin and between milestones. \\

\paragraph{Protein Rupture:} When a tensile force is exerted on the two  ends of a folded peptide, it can result in a process called \emph{rupture} in which the peptide loses its folded structure and becomes fully extended~\cite{TskhovrebovaNature1997}. The precise sequence of molecular-level events that result in protein rupture, however, are not fully understood. In general, the average time needed for protein rupture decays exponentially with the applied force. For certain proteins, however, a more counterintuitive behavior is observed in which the rupture time exhibits a non-monotonic dependence on force~\cite{MarshallNature2003}. Jiang~\emph{et al.}~\cite{JiangPCCP2014} utilized FFS to explore the kinetics and mechanism of rupture in a coarse-grained G\={o}-type representation~\cite{KaranicolasProetinSci2002} of the L protein~\cite{KimActaCrystallogr2001}. When the applied tensile forces were small, the peptide ruptured via a one-step mechanism. For large tensile forces, however, the protein got trapped in a partially unfolded intermediate. While the transition into this intermediate was faster than the one-step full denaturation, the subsequent unfolding was considerably slower, resulting in a non-monotonic dependence of rupture time on tensile force. 
\\

\paragraph{Protein Aggregation:} Under suitable thermodynamic conditions, many peptides and proteins can aggregate into distinct phases.  Protein aggregation can fulfill important biological functions. A notable example is a family of proteins called \emph{intrinsically disordered proteins (IDPs)}~\cite{TOMPA2012509} that do not possess any native folded structure, and function  by dynamically interacting with their target molecules, such as ligands and other IDPs. As has been demonstrated in numerous publications in recent years, IDP aggregation plays an important role in many important biological processes, such as the formation of membraneless organelles~\cite{FalahatiCurrBiol2016, FalahatiPNAS2017}. In the meanwhile, protein-- and IDP-- aggregation can result in the formation of pathological bodies within biological cells, and cause a wide variety of neurodegenerative and cognitive diseases, such as Alzheimer's and cataract. For instance, Alzheimer's disease is thought to be primarily caused by the aggregation of $\beta$-amyloid precursor protein (APP) and the microtubule-associated protein tau (MAPT)-- also known as \emph{tau}~\cite{HardyTrendsPharmacolSci1991}. Among these, MAPT is an IDP whose aggregation can result in fibril formation.  Historically, computational studies of protein aggregation have  mostly focused on fibril and amyloid formation in neurodegenerative diseases, which is critical to systematic design of effective therapeutic interventions.

One of the earliest investigations of peptide aggregation using FFS was conducted by Luiken and Bolhuis~\cite{LuikenJPhysChemB2015}, who investigated the effect of the hydrophobicity of short amyloidogenic peptides on the kinetics and mechanism of $\beta$-fibril formation. In particular, they considered dilute solutions of three peptide fragments that they had previously studied in REMD simulations~\cite{LuikenPCCP2015} using the mid-resolution coarse-grained force-field of Ref.~\citenum{BereauJCP2009}. Since complex processes such as protein aggregation are very difficult to characterize at a molecular level, such REMD simulations are valuable-- and sometimes indispensable-- tools in identifying the final state of the system which, in turn, is necessary for designing suitable order parameters for methods such as FFS. Based on their observations in Ref.~\citenum{LuikenPCCP2015}, the authors utilized the number of in-registry residues in the two largest peptide clusters as the order parameter.  For more hydrophobic fragments, they found the nucleation process to be comprised of two steps; the formation of a dense protein-rich liquid followed by the nucleation of $\beta$-fibrils. More hydrophilic fragments, however, turned into a fibril structure via single-step nucleation.

\begin{figure}
\centering
\includegraphics[width=0.45\textwidth]{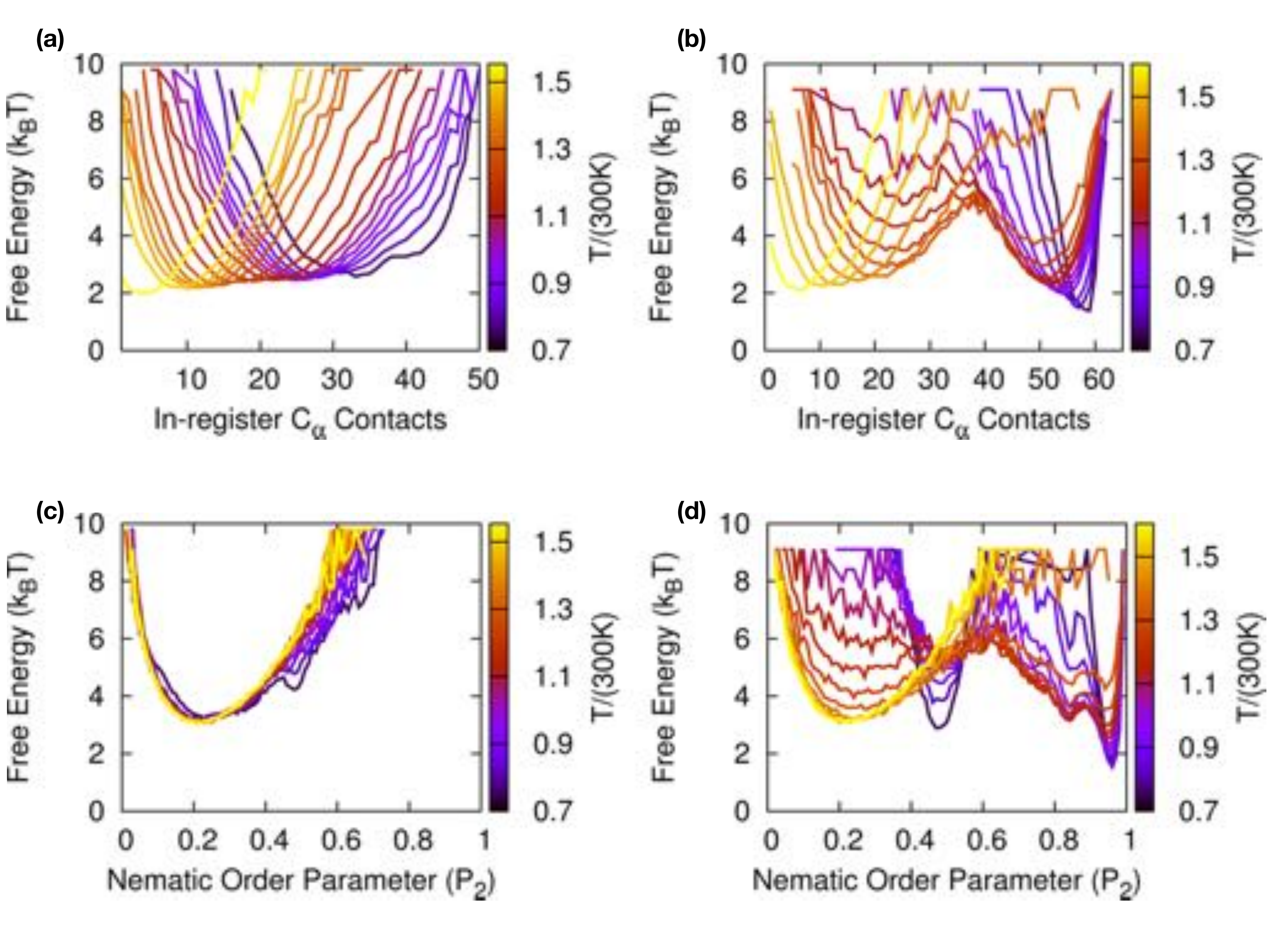} 
\caption{(Reproduced from Ref.~\citenum{SmitJPhysChemB2017}). Free energy profiles for the aggregation of PHF6* (left) and PHF6 (right) as a function of the number of $C_{\alpha}$ contacts (a, b) and the nematic order parameter ($P_2$) (c-d)}
\label{fig:amyloid_1}
\end{figure}

In a recent paper from the same group, Smit~\emph{et al.}~\cite{SmitJPhysChemB2017} utilized FFS to study amyloid formation by PHF6 and PHF6*,  two aggregation-prone 6-amino acid  fragments of  \emph{tau}, which have been shown to play pivotal roles in the formation of tau oligomers. They represented these peptides using a re-parameterization of the  force field given in Ref.~\citenum{BereauJCP2009}. They first performed REMD simulations of both PHF6 and PHF6* to understand the nature of the free-energy landscape and the types of secondary structures formed within the aggregates. Based on these REMD simulations that showed the formation of $\beta$-fibril sheets, the authors decided to use the number of in-register $C_{\alpha}$ contacts between two largest peptide clusters as the order parameter. The REMD simulations also revealed that only PHF6 was able to form  ordered $\beta$-fibril sheets and PHF6* formed disordered oligomers. The free energy landscape for aggregation of PHF6 and PHF6* as a function of $C_{\alpha}$ contacts and the nematic order parameter (which quanties the extent of alignment between fibers) showed a clear difference, with two stable states for PHF6 and only one for PHF6*~(Fig.~\ref{fig:amyloid_1}). This finding was confirmed by the rates computed from FFS, which were $\approx 15$ orders of magnitude smaller for PHF6* in comparison to  PHF6. Performing backward FFS showed that aggregated PHF6* fragments were indeed very unstable and quickly dissolved into the disordered state. These findings  support the  absence of any steric zipper crystal structures~\cite{WiltziusNature2009}, i.e.,~basic structural units of amyloid fibrils, for PHF6* fragments. They also support the experimental finding that the removal of PHF6 prevents all aggregation whereas some aggregation takes place even in the absence of  PHF6*.  \\

\subsubsection{Conformational Rearrangements in Nucleic Acids}

\noindent
Nucleic acids are linear polymers of a class of organic molecules called \emph{nucleotides}. There are two types of naturally occurring nucleotides in biological systems, ribonucleotides and deoxyribonucleotides, which are the building blocks of ribonucleic acids (RNAs) and deoxyribonucleic acids (DNAs), respectively. Unlike proteins whose folding and assembly is governed by complex intra- and inter-residue interactions, the assembly of nucleic acids  is primarily driven by the complementarity of their constituent nucleotides, i.e.,~the propensity of certain base pairs to form strong hydrogen bonds with one another (A/T and C/G in the case of DNA, and A/U and C/G in the case of RNA).  This feature makes nucleic acids ideal for unambiguous storage and transmission of information. 

DNA is the material that stores genetic information in biological cells and most viruses, and is usually comprised of two complementary strands hybridized together~\cite{WatsonNature1953}. RNAs are, however, mostly responsible for transmitting the information encoded in DNA to fulfill biological functions such as protein synthesis-- also known as translation. RNAs are usually single-stranded, but can hybridize with other RNAs. In addition, both DNA and RNA can self-hybridize to form  secondary structures such as hairpins and origamis.    The process through which DNA and RNA strands assemble into double-stranded complexes is called \emph{hybridization}, while the dissociation of such complexes into their constituent strands is referred to as \emph{denaturation}. Both processes have been extensively studied using a wide variety of experimental and theoretical approaches~\cite{AhsanBiophysJ1998, ChalikianPNAS1999, ClausenSchaumannBiophysJ2000, SchmittJChemPhys2011}. Experimental evidence points to a nucleation-zippering mechanism~\cite{PorschkeJMolBiol1971} for hybridization in which a critical nucleus of base pair contacts is formed, followed by the zippering of the remaining bonds. There is, however, considerable gap in our molecular-level understanding of nucleation-zippering and other probable mechanisms of hybridization. Understanding how different molecular processes contribute to the overall kinetics of hybridization is crucial for engineering self-assembly processes mediated by nucleic acids by modulating hybridization kinetics.  Molecular simulations augmented by advanced sampling techniques such as FFS are excellent for this purpose, as they can capture the kinetics and mechanism of hybridization over a wide range of conditions. Over years, advanced sampling techniques such as TPS~\cite{RadhakrishnanPNAS2004, WangBMCStructBio2007, HuPNAS2008, SambriskiJPCM2008, HoefertSoftMatter2011}, replica exchange MC~\cite{AraqueJChemPhys2011} and MD~\cite{LiJChemPhys2014}, metadynamics~\cite{ElderBiomacromolecules2015} and umbrella sampling~\cite{DicksonJChemTheoryComput2011, SchmittJChemPhys2013} have been utilized to study DNA hybridization in the absence~\cite{SambriskiJPCM2008, HoefertSoftMatter2011, SchmittJChemPhys2013, LiJChemPhys2014, ElderBiomacromolecules2015} and presence~\cite{RadhakrishnanPNAS2004, HuPNAS2008, WangBMCStructBio2007} of proteins.

Historically, FFS investigations of nucleic acid assembly have been conducted using coarse-grained models. Most of these studies have utilized the oxDNA~\cite{OuldridgeJCP2011}  and oxRNA~\cite{SulcJChemPhys2014} force-fields, and have explored different aspects of nucleic acid assembly, such as  DNA hybridization~\cite{OuldridgeNucleicAcidRes2013}, hairpin formation~\cite{MosayebiJPhysChemB2014, SchreckNucleicAcidRes2015}, and toehold-mediated DNA~\cite{SrinivasNucleicAcidRes2013} and RNA displacement~\cite{SulcBiophysJ2015}. 

One of the first oxDNA-based studies was conducted by Ouldridge~\emph{et al.}~\cite{OuldridgeNucleicAcidRes2013}, who utilized direct and Rosenbluth FFS to probe the kinetics and mechanism of  hybridization between two complementary 14-base strands with non-repetitive and repetitive sequences. Due to the complicated nature of DNA hybridization, especially in the case of repetitive sequences, they utilized a composite order parameter based on the minimum distance of hybridizing strands and the number of favorable complementary base pairs in the assembled structure. They found the rate of duplex formation to depend on the strength of initial contacts, and the ease of internal displacement of initially misaligned bonds to form the correct structure at later stages. As for repetitive sequences, they identified two principal mechanisms for correcting misaligned contacts, which they called 'inchworm` and 'pseudoknot` displacements. Both these corrective displacements require crossing a free energy barrier, and yet occur at sufficiently large rates to increase the number of possible pathways for hybridization. As a result, strands with repetitive sequences were found to hybridize at higher rates. Nonetheless, the emerging metastable states persist for long enough that the authors could divide the overall transition process into two stages and sample each through a separate FFS calculation. The first calculation was aimed at forming the metastable intermediate while the second calculation probed mismatch correction and the formation of the thermodynamically stable duplex. 

\begin{figure}
	\centering
	\includegraphics[width=.45\textwidth]{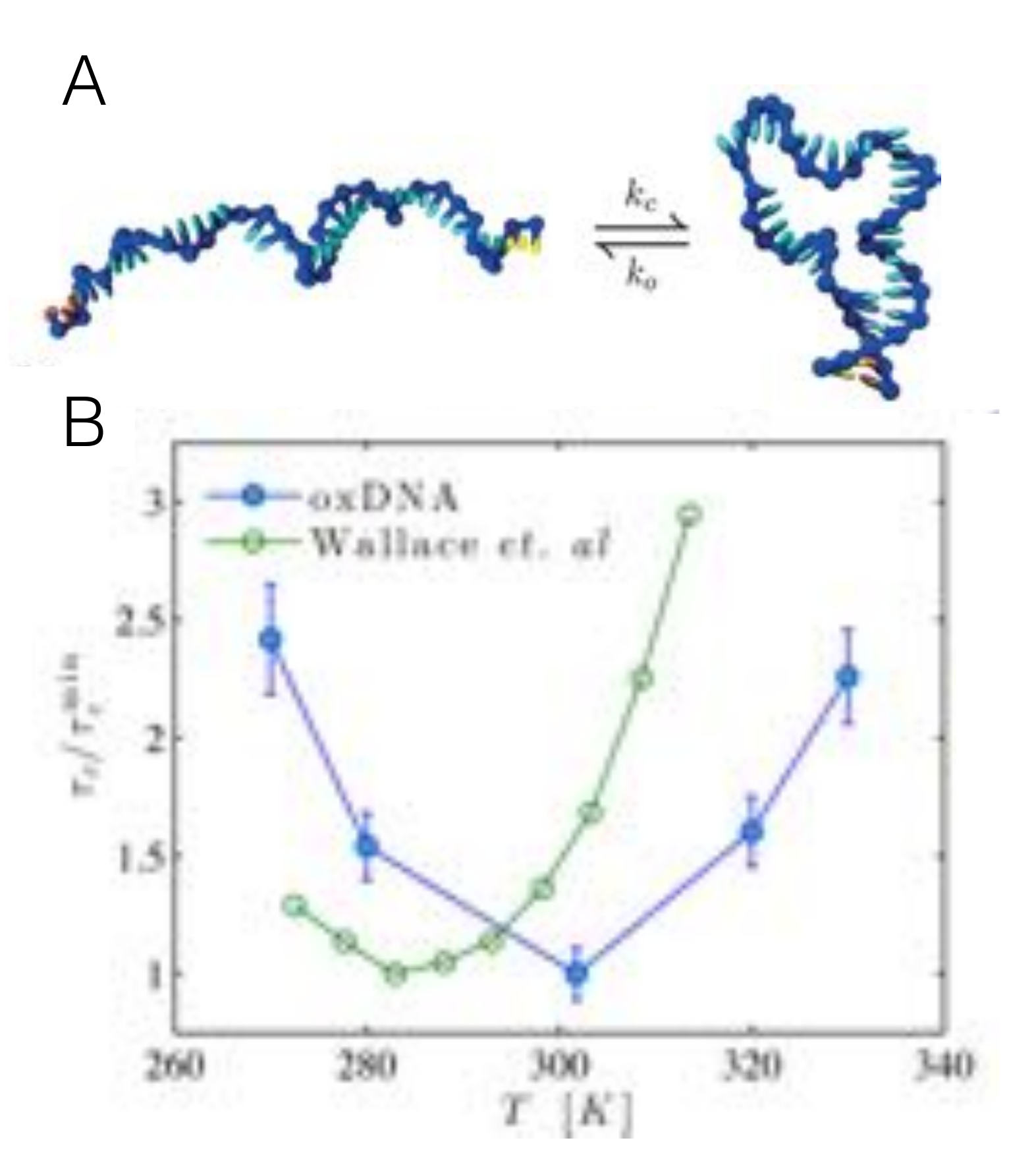}
	\caption{\label{fig:hairpin}(Reproduced from Ref.~\citenum{MosayebiJPhysChemB2014}) (A) A schematic representation of the self-hybridization transition studied in Ref.~\citenum{MosayebiJPhysChemB2014}. (B) Non-monotonic dependence of the hybridization time, $\tau_c$, on temperature. The experimental data is from Wallace~\emph{et al.}\cite{WallacePNAS2001}.
	}
\end{figure}

In two other papers from the same group~\cite{MosayebiJPhysChemB2014, SchreckNucleicAcidRes2015}, they utilized FFS to explore the kinetics and mechanism of hairpin formation. Hairpins are secondary structures that form when two parts of a single DNA (or RNA) strand with complementary sequences self-hybridize. For instance, Mosayebi~\emph{et al.}~\cite{MosayebiJPhysChemB2014} explored self hybridization in two hairpin forming 40-base-long DNA strands (Fig.~\ref{fig:hairpin}A), and observed a non-monotonic dependence of rate on temperature, with the rate reaching a maximum close to melting (Fig.~\ref{fig:hairpin}B). These findings are in qualitative agreement with earlier experimental observations~\cite{WallacePNAS2001}, and can be explained by noting that the free energetic penalty of bending the DNA strand becomes dominant at low temperatures. Indeed, this nonmonotonicity almost disappears upon decreasing this bending penalty. In another paper, Schreck~\emph{et al.}~\cite{SchreckNucleicAcidRes2015} investigated the relationship between hairpin formation propensity and full duplex formation. They observed that introducing self-hybridizing sequences within the two complimentary strands will decrease their hybridization rate, while considerably increasing the denaturation rate of double-stranded DNA. This is because the formation of enthalpically favored hairpin intermediates decreases the free energy barrier to denaturation.

Apart from their biological roles, nucleic acids have garnered increased interest in materials science, since Watson-Crick complementarity provides a rigorous paradigm for engineering the assembly of novel structures~\cite{YurkeNature2000, RothemundNature2006, MrikinNatMat2013}. One such possibility is toehold-mediated strand displacement (TMSD)~\cite{YurkeNature2000}, a process in which an 'invading` DNA or RNA strand $S$ that is complementary to a 'substrate` strand $S'$ displaces an 'incumbent` strand $T$ that is partially hybridized with $S'$. Despite being thermodynamically favored, such a  displacement  usually involves crossing a free energy barrier. Srinivas~\emph{et al.}~\cite{SrinivasNucleicAcidRes2013} used FFS alongside the oxDNA force-field to study how the kinetics of TMSD depends on toehold length-- or the number of excess bases within $S$ that are complementary to $S'$--, and observed the rate to increase exponentially with the toehold length for short toeholds before plateauing for sufficiently long (5 bases and up) toeholds. Their findings were in agreement with earlier experimental observation~\cite{ZhangJACS2009}. Their calculations also provided valuable mechanistic insight into toehold displacement, which they found to occur via a combination of 'branch migration via invasion` in which one base from the invading and incumbent strands each compete to bind their complementary base on the substrate, and 'branch migration via sequential disruption and formation of bonds` in which the contact bases of the invading and incumbent strands both detach from the substrate.  In a second paper from the same group, \~{S}ulc~\emph{et al.}~\cite{SulcBiophysJ2015} used the oxRNA model and FFS to investigate TMSD for RNA, and found it to occur faster if the toehold sequence is located at the $5'$ end of the invading sequence. Not surprisingly, they found the rate to decrease upon increasing temperature.

\begin{figure}
\centering
\includegraphics[width=0.5\textwidth]{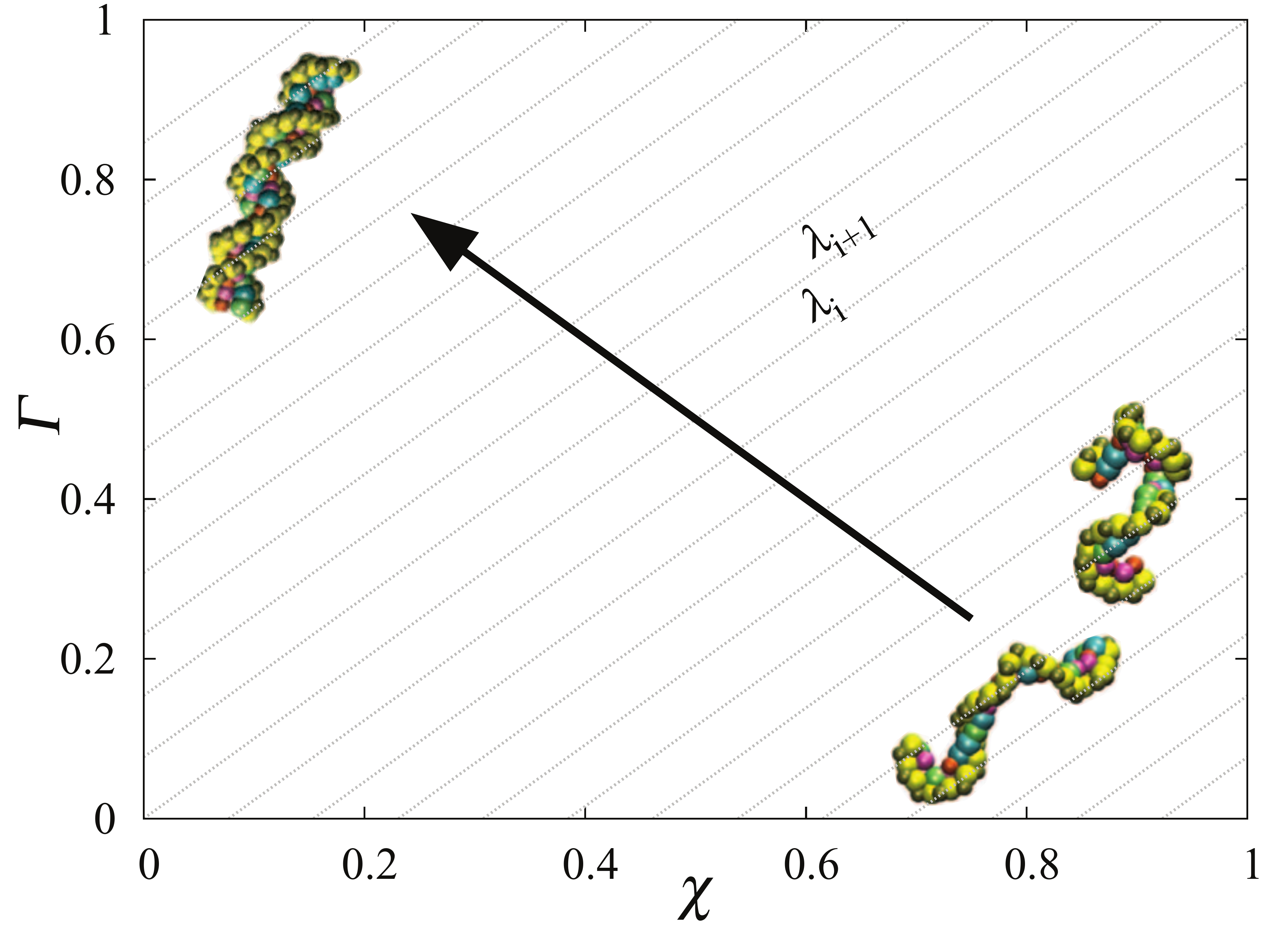} 
\caption{(Reproduced from Ref.~\citenum{HinckleyJChemPhys2013}) FFS interfaces are chosen  based on an order parameter which is a linear combination of $\chi$, i.e.,~the center of mass separation between the two strands, and $\Gamma$, the fraction of total possible Watson-Crick base pairs formed.}
\label{fig:OP-DNA} 
\end{figure}

Another coarse-grained force-field that has been used in FFS studies of hybridization is the 3SPN force-field developed by the de Pablo Group~\cite{KnottsJCP2007, SambriskiBiophysJ2009} in which each nucleotide is represented by three interaction sites. Unlike the oxDNA model, the 3SPN model contains explicit electrostatic interactions (handled in the context of Debye-Huckel theory), and allows for internal rearrangements of the individual nucleotides. Hinckley~\emph{et al.}~\cite{HinckleyJChemPhys2013} developed a re-parameterized version of 3SPN-- that they called 3SPN.2-- and utilized RBFFS to compute hybridization rates for four different sequences at two different ionic strengths. The order parameter utilized in their study was a linear combination of the separation between centers of mass of two strands, as well as the number of base pair contacts formed between them (Fig.~\ref{fig:OP-DNA}). Their computed rates were around 1-2 orders of magnitude larger than the experimentally measured rates under similar conditions~\cite{GaoNucleicAcidRes2006}, which they attributed to the coarse-grained nature of the 3SPN.2 model. 

In a second paper, Hinckley~\emph{et al.}~\cite{HinckleyJCP2014} employed their 3SNP.2 force-field to conduct a comprehensive investigation of the sensitivity of hybridization kinetics to features such as sequence, length, and ionic strength. Consistent with the findings of Ref.~\citenum{OuldridgeNucleicAcidRes2013}, they found homogeneous (poly A and poly AC) strands to hybridize at higher rates than strands with non-repetitive sequences. Analyzing the TPE revealed a zippering mechanism for heterogeneous strands, while for homogeneous strands, hybridization proceeds through a combination of inchworm and pseudoknot displacemnts. For the poly-A homopolymer, however, a 1D defect diffusion mechanism-- also called ``slithering``-- was observed. The authors also used FFS to compute hybridization rates as a function of strand length $N$, which enabled them to test their scaling theory. Under conditions at which  contact formation was the rate-limiting step, their theory agreed well with simulations, even though the uncertainty was high due to short ($N\le30$) sequences considered therein and large uncertainties in FFS rates. As for ionic strength, they observed the rates to be smaller at lower ionic strengths due to stronger repulsions between charged strands. Nonetheless, the hybridization mechanism remained unchanged. 

\subsection{Structural Relaxations and Conformational Rearrangements in Polymer Melts and Solutions}\label{section:polymer-relaxation}

\noindent
Polymers are an important class of materials with interesting thermodynamic, mechanical and transport properties.  Understanding the structure-property relationship in polymers, and optimizing the processes that result in the formation of polymeric materials with desirable structural properties usually requires a fundamental-level knowledge of a wide variety of rare events, such as first-order phase transitions, conformational rearrangements, and structural relaxation processes. While mean-field theories based on scaling arguments have been remarkably successful in predicting the thermodynamics, structure and dynamics of polymer melts and solutions, their limited ability to account for correlations and collective phenomena makes them unsuitable for capturing the kinetics and mechanism of rare events. Molecular simulations augmented with advanced sampling techniques, however, can provide useful mechanistic information not otherwise attainable from such mean-field representations. In recent decades, a wide variety of advanced sampling techniques  such as FFS~\cite{HuangJChemPhys2008, HernandezOrtizJChemPhys2009, CaoJCP2015, ZhuJCP2017, RezvantalabMacromolecules2018}, TPS~\cite{tenWoldePNAS2002}, umbrella sampling~\cite{GoelJPhysChemB2008, JamadagniLangmuir2009}, the string method~\cite{MillerPNAS2007} and metadynamics~\cite{BocahutMacromolecules2016} have been utilized for studying rare events in polymeric systems. FFS in particular has been utilized for studying a wide variety of problems such as translocation~\cite{HernandezOrtizJChemPhys2009}, structural relaxation~\cite{CaoJCP2015, ZhuJCP2017} and conformational rearrangements~\cite{HuangJChemPhys2008, RuzickaPCCP2012, RezvantalabMacromolecules2018}.

\subsubsection{Polymer Translocation}

\noindent
Polymer translocation is a process in which a polymeric molecule traverses a pore, and understanding its kinetics and mechanism is of considerable interest in various applications~\cite{WanunuPhysLifeRev2012}. In order to enter and traverse the pore, a polymer might need to overcome energetic and/or entropic barriers, which can turn translocation into a rare event. In recent years, advanced sampling techniques have been utilized for studying the kinetics of polymer translocation. For instance, HernÃ¡ndez-Ortiz~\emph{et al.}~\cite{HernandezOrtizJChemPhys2009} utilized FFS to study the translocation of a generic polymer through a $\mu$m-scale square pore. They represented the polymer using the coarse-grained bead-spring model, while solvent molecules were treated implicitly within the framework of Brownian dynamics. They also studied the role of hydrodynamic interactions in translocation by employing a method similar to Stokesian dynamics (SD)\cite{BradyAnnRevFluidMech1988}. For translocation, they utilized the $z$ coordinate of the polymer center of mass as the FFS order parameter. They validated their FFS calculation for the translocation of a single bead, and found excellent agreements between FFS and brute-force BD and SD simulations. They found that  hydrodynamic interactions affect the translocation of a single bead and a polymer chain differently. While for a single bead, the translocation rate was lower in the presence of hydrodynamic effects (presumably due to the effect of confinement on Stokes flow within the pore), the collective motions of the polymer chain arising from hydrodynamics-- and lacking in regular BD-- facilitate its translocation. Due to the importance of such collective motions, the rate was found to be more sensitive to chain length in the presence of hydrodynamic interactions, changing by 5 orders of magnitude upon increasing the chain length by a factor of 10.

\subsubsection{Structural Relaxation in Polymers}

\noindent 
Polymeric melts and solutions are unique in the sense that structural relaxation within them usually involves processes occurring over a wide range of length scales, a feature that confers on them interesting viscoelastic properties~\cite{FerryJApplPhys1955, PearsonMacro1984}. In the case of entangled branched polymer melts, for instance, structural relaxation involves nonlinear processes such as reptation (snake-like motion of the polymer chain), contour-length fluctuations (e.g.,~arm retractions) and constraint release (release and renewal of entanglement constraints). These processes govern the spatiotemporal evolution of microstructures within a melt, and understanding them is critical to elucidating the structure-property relationship in polymer melts, and to designing polymeric materials with desired mechanical and transport properties. The standard theoretical framework for understanding polymer relaxation is the tube model~\cite{deGennesJCP1971, DoiOxford1986, McLeishMacromolecules2002} in which the constraints on chain crossings are represented as virtual tubes around individual chains so that only the motions that are curvilinear to the tubes are allowed. The tube model is based on an earlier representation of polymers called the Rouse model~\cite{RouseJCP1953, ZimmJCP1956} in which each polymer is represented as Brownian beads connected to each other via harmonic strings. Within the tube model, the mobility of each Rouse bead is confined into a time-dependent one-dimensional tube. Over the years, several theoretical extensions of the tube model~\cite{DoiJPS1980, PearsonMacro1984, BallMacro1989, MilnerMacro1997, MilnerPRL1998} have been developed for describing the behavior of  branched entangled polymers such as star polymers.  Such mean-field descriptions are, however, not always quantitatively predictive~\cite{ChavezPRL2010}, and even though atomistic and bead-spring molecular simulations have been previously utilized for studying branched polymers~\cite{GrestMacromolecules1987, ChremosJChemPhys2015}, they are usually too computationally expensive to capture collective  phenomena relevant to structural relaxation in entangled polymers. 

\begin{figure}
\centering
\includegraphics[width=0.45\textwidth]{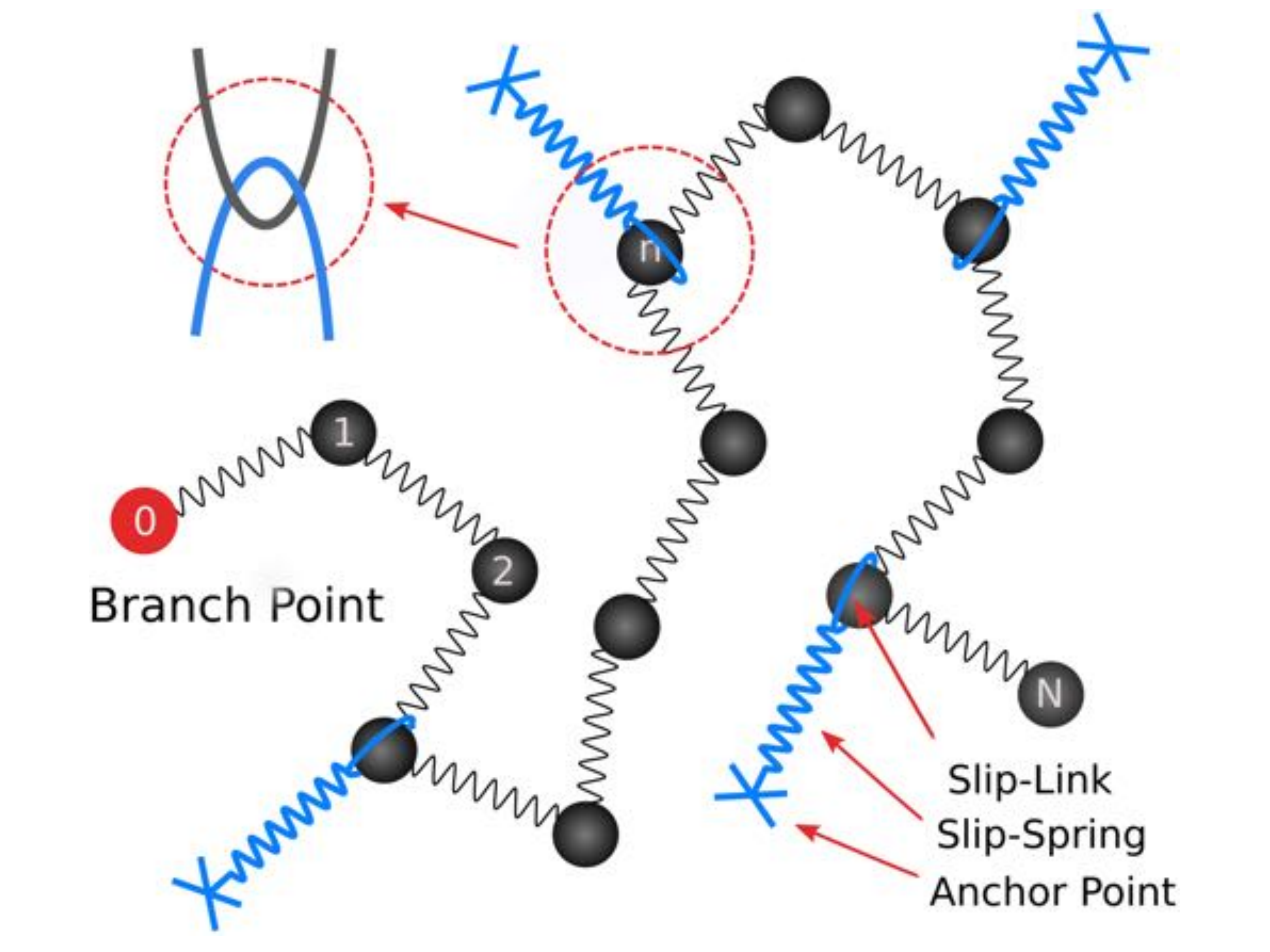} 
\caption{(Reproduced from Ref.~\citenum{ZhuJCP2017}) Schematic representation of the slip-spring model for one arm of a star polymer with the branch point represented as 0.}
\label{fig:slip-spring} 
\end{figure}

An alternative approach is to use mean-field theories such as the tube model to formulate more accurate dynamical representations, which, despite being analytically unsolvable, can  be numerically integrated to capture the relaxation kinetics.  One such coarse-grained representation is the slip-link (SL) or slip-spring (SS) model~\cite{LikhtmanMacromolecules2005} in which each branch is represented by several Rouse beads whose mobility is constrained by a handful of slip links that move along the chain and define the topology of the tube (Fig.~\ref{fig:slip-spring}). Further technical details about the SS model can be found in several excellent reviews~\cite{MasubuchiAnnRevChem2014, SchieberAnnRevChem2014}. It is, however, necessary to note that the SS model incorporates the basic idea of the tube model along with further explicit details about the polymer chain and its entanglement. Different relaxation processes can then be represented by altering the boundary conditions applied to the Rouse beads and the slip links. Despite the  simplicity and computational efficiency of the SS model, its simple time integration might still be inefficient in capturing certain long-range structural rearrangements such as chain extension and arm retraction, and advanced path sampling techniques such as FFS need to be used for capturing such rare events.

One of the fundamental questions recently addressed using FFS is to estimate $\tau(z)$, the average time that it takes for a Rouse chain to reach an end-to-end distance of $z$ for the first time.  
For instance, Cao~\emph{et al.}~\cite{CaoJCP2015} utilized direct simulations and FFS to determine $\tau(z)$ for the one-dimensional Rouse model.  There is no analytical expression for $\tau(z)$ even for the 1D Rouse model, and even though $\tau(z)$ can be estimated numerically, direct simulations are inefficient when $N$ and $z$ are large. Historically, $\tau(z)$ has been approximated using the Milner-McLeish theory~\cite{MilnerMacro1997, MilnerPRL1998} in which each chain within the tube is replaced with a single bead harmonically tethered to the origin. According to the Milner-McLeish theory, $\tau(z)$ is given by:
\begin{eqnarray}
\tau_{\text{MM}}(z) &\approx& \frac{\pi^2\tau_R}4\sqrt{\frac\pi6}\frac{\sqrt Nb}{z}\exp\left[\frac{3z^2}{2Nb^2}\right]\label{eq:tau-MM}
\end{eqnarray}
Here, $N$ is the number of Rouse beads within the chain, $b$ is the statistical segment length, and $\tau_R=4\xi_0N^2b^2/4\pi^2k_BT$ is the  relaxation time for a Rouse chain with one fixed and one free end with $\xi_0$ being the friction coefficient of each bead. The main objective of Cao~\emph{et al.}~\cite{CaoJCP2015} was to test the validity of Eq.~(\ref{eq:tau-MM}). They considered a chain with one end fixed at the origin, and used the free end position as the order parameter. They indeed found the Milner-McLeish theory to overestimate the scale-free mean-passage time $s\tau(s)\tau_R^{-1}\exp(-3s^2/2)$ by a factor of 10 and larger, with deviations becoming larger for longer chains. Here $s:=z/b\sqrt{N}$, is the end-to-end distance normalized by the number of beads, $N$.  The authors proposed alternative theories to predict the limiting behaviors of $\tau(s)$ for small and large $s$'s, which, despite being able to predict the qualitative dependence of $\tau(s)$ on $s$, failed to correctly predict the pre-factors. The authors attribute this failure to the breakdown of the hidden Markov assumption that they invoked in their derivation.

\begin{figure}
\centering
\includegraphics[width=0.45\textwidth]{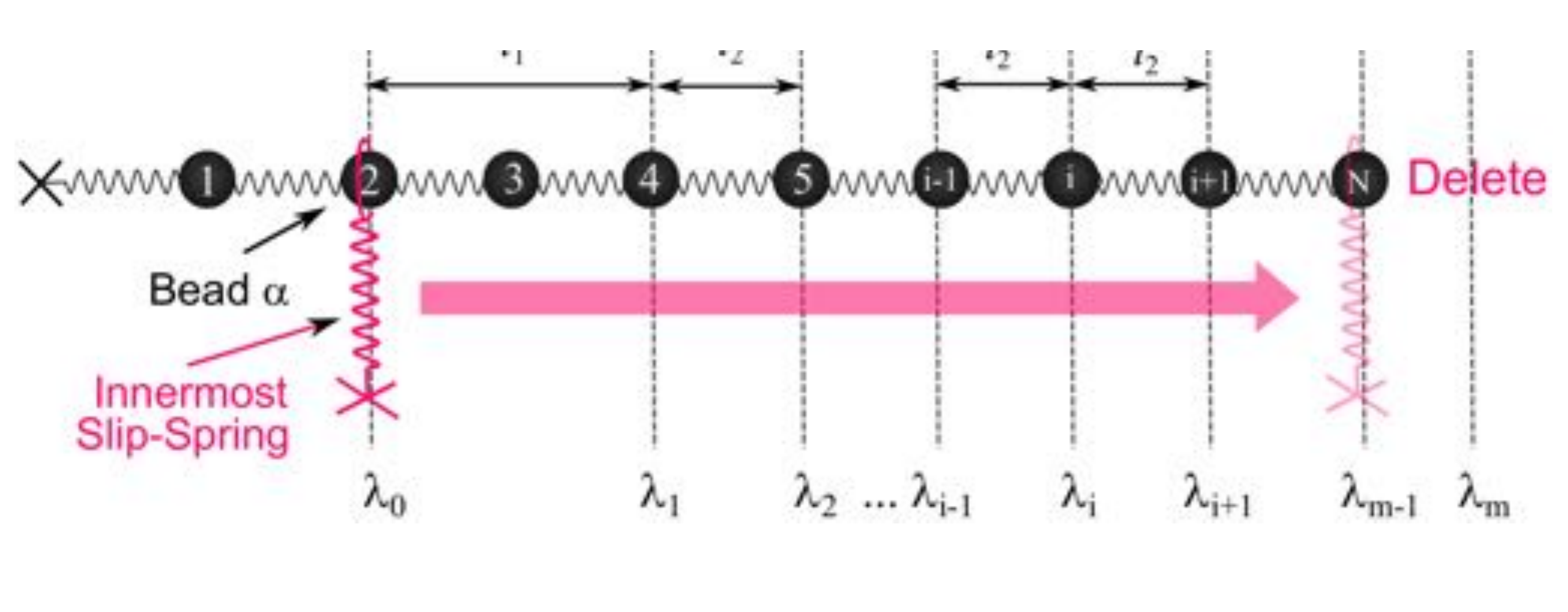} 
\caption{(Reproduced from Ref.~\citenum{ZhuJCP2017}) The FFS interfaces utilized in Ref.~\citenum{ZhuJCP2017} chosen based on the index of the innermost slip-link as the order parameter.}
\label{fig:OP-CR-Polymer} 
\end{figure}

In a second paper~\cite{ZhuJCP2017}, three of the same authors employed FFS, this time with the SS model, to estimate the relaxation time for arm retraction as a function of entanglement, in the absence and presence of constraint release. This was conceptually opposite to the problem of arm extension in Ref.~\citenum{CaoJCP2015}, and involved capturing $\tau_d$, the mean first time that the chain's free end returns to its fixed end. In the absence of constraint release, they utilized the index of the monomer containing the innermost slip-link, shown in~Fig.~\ref{fig:OP-CR-Polymer} as the FFS order parameter, and observed good agreement with  relaxation times calculated from direct simulations of mildly entangled polymers (8 entanglements). For well-entangled chains (with 16 entanglements), $\tau_d$'s can only be estimated using FFS, and agree reasonably well with the theoretical predictions for the exponential dependence of $\tau_d$ on arm length. The authors further extended the applicability of FFS to systems with constraint release (CR), by modifying the order parameter to track the index of the innermost \emph{surviving} original slip link, as some slip-links are released during the process. The computed $\tau_d$ were again in good agreement with those computed from direct simulations.

\subsubsection{Conformational Rearrangements in Polymers}

\noindent
Isolated polymer chains can undergo large-scale conformational transformations. In addition to the widely known coil-to-globule transformation~\cite{ChanAnnRevBiophysBiophysChem1991}, polymer chains can undergo a myriad of disorder-order and polymorphic transitions~\cite{ZhouPhysRevLett1996}. For instance, Zhou~\emph{et al}.~\cite{ZhouPhysRevLett1996} and Taylor~\emph{et al}.~\cite{TaylorJChemPhys2009} investigated a solvated polymer chain comprised of attractive hard sphere beads tethered together via an infinite square well potential. Depending on temperature, the range of attractive non-bonded interactions, and the width of the square well potential, the polymer can fold into a 'crystal`,~i.e.,~a structure in which individual beads are arranged into a crystalline lattice. Recently, R\r{u}\v{z}i\v{c}ka~\emph{et al.}~\cite{RuzickaPCCP2012} used FFS to probe the kinetics of liquid-to-crystal and crystal-to-liquid transition. They utilize potential energy as their order parameter, and evolved the system using the discontinuous molecular dynamics (DMD) method~\cite{ChapelaMolPhys1994}. They constructed a kinetic phase diagram for the liquid-to-solid transition by defining the temperature at which the forward and reverse rates match as the kinetic melting temperature. The constructed phase diagram qualitatively agrees with one determined from MC simulations. The authors attributed small systematic differences between FFS and MC phase diagrams to the lower efficiency of DMD in sampling the configuration space, as well as potential systematic errors in FFS. They also analyzed the transition ensemble and found the largest eigenvalue of the Laplacian contact matrix as a good reaction coordinate for describing crystallization.

In addition to conformational rearrangements in the bulk, polymer chains can undergo conformational changes in confinement. For instance, a linear polymer chain can reverse within a pore, a process that can be a rare event for sufficiently long chains and sufficiently small pores. Huang~\emph{et al.}~\cite{HuangJChemPhys2008} utilized FFS, alongside with the transition state theory and the Kramers' theory to estimate reversal rates for a flexible polymer within a cylindrical pore. For FFS, they utilized $z_N-z_1$, the difference between the $z$ coordinates of the chain ends as the order parameter. In general, they found the rates computed from FFS to be $1-2$ orders of magnitude smaller than those predicted from TST and the Kramers' theory. The dependence of those rates on pore radius and chain length, however, followed the trends predicted by those theories.

In a more recent study, Rezvantalab and Larson~\cite{RezvantalabMacromolecules2018} used FFS to investigate loop-to-bridge transition in a telechelic polymer confined between two surfaces that strongly attract the chain ends. Under such a scenario, four arrangements are possible, namely a loop (with the two ends attached to the same surface), a bridge (with each end attached to one surface), dangling (with one attached and one free end), and free (when both ends are free). Among these, loop and bridge arrangements tend to have lower energies. However, any transition between them will involve crossing a free energy barrier (i.e.,~the breakage of one connection). Rezvantalab and Larson~\cite{RezvantalabMacromolecules2018} considered models with different levels of details (including the Rouse model, and the freely joined chain (FJC) model~\cite{JainMacromolecules2008}) by utilizing the $z$ coordinate of one end as the FFS order parameter. They found that the bridge formation process can be broken  into two components, namely the escape of the chain end from the potential energy minimum at the original surface, and the stretching of the original chain. These two processes were found to be mostly decoupled in the Rouse model to the extent that proper normalization of the transition time with the escape time resulted in a perfect agreement with Ref.~\citenum{CaoJCP2015}. In the case of the FJC model, however, these two steps were found to be more correlated.

\subsection{Ion Transport through Semipermeable Membranes}\label{section:filtration}

\noindent 
A semipermeable membrane only allows for unimpeded transport of certain molecules and ions while impeding or fully blocking the passage of other entities. This can potentially create a separation of timescales between  the transport of the favored and impeded components. Characterizing such a separation is key to designing efficient and economically affordable semipermeable membranes for application such as water desalination~\cite{ShannonNature2008}, as well as chemical separation of gases~\cite{KorosJMembSci1993, HutchingsNanoscaleHoriz2019}, ions~\cite{Cheng2018, Razi2019} and organic solvents~\cite{MarchettiChemRev2014}. In recent years, a wide variety of simulation techniques have been utilized to study water permeability and solute selectivity in nanoporous membranes~\cite{CohenNanoLett2012, RichardsSmall2012, RichardsPCCP2012, KonathamLangmuir2013,  HeiranianNatComm2015, CohenNanoLett2016}.

\begin{figure}
\centering
\includegraphics[width=0.45\textwidth]{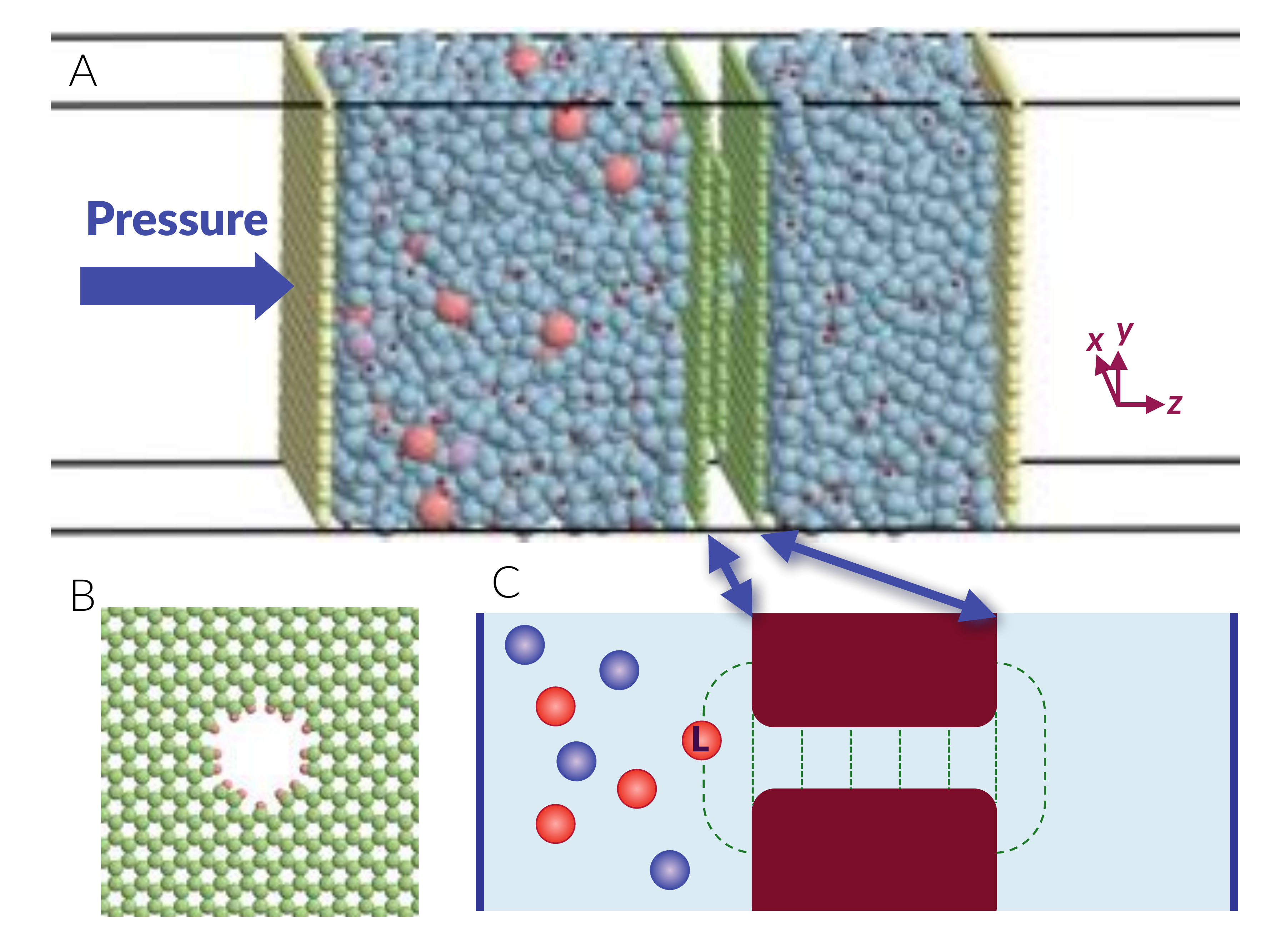} 
\caption{(Reproduced from Ref.~\citenum{Malmir2019Arxiv}). (A) Schematic representation of a graphite-based semi-permeable membrane with membrane (green), piston (yellow), water (blue), sodium ions (light purple) and chloride ions (peach). (B) Nanopore cross section  with carbon atoms (green) and passivating hydrogens (red) (C) Schematic representation of the directed curved distance from the pore mouth which is used as the OP in FFS. The level sets of the OP are shown with dotted green lines.}
\label{fig:filtration} 
\end{figure}

The first FFS investigation of ion transport through nanoporous membranes was done by Malmir~\emph{et al.}~\cite{Malmir2019Arxiv}, who used jFFS~\cite{AmirHajiAkbariJCP2018} and non-equilibrium MD~\cite{CohenNanoLett2012, NguyenJCTC2013, CohenNanoLett2016} to study the transport of sodium and chloride ions across a nanoporous graphite membrane in the presence of an external hydrostatic pressure gradient~(Fig.~\ref{fig:filtration}A-B). The authors utilized the directed curved distance of the leading ion from pore mouth as their order parameter~(Figure: \ref{fig:filtration}C). By computing  mean first passage times for the solvent and the solute, they accurately computed the salt rejection rate to be $\approx99.99\%$, which corresponds to an ion passage ratio of one ion per 10,000 water molecules. By analyzing the TPE, they observed that the first ion to traverse the pore is always a chloride, which they attributed to the partial positive charge of hydrogen atoms passivating the pore interior.  Interestingly, the critical state atop the free-energy barrier found from committor analysis was present just outside the pore exit and not somewhere within the expected bottleneck of pore interior. They explained this intriguing result by calculating the average net force on the leading ion and found it to be non-zero even after the ion had exited the pore. They showed that such asymmetry was arising due to the accumulation of $\text{Na}^{+}$ ions at the pore mouth, exerting a restraining force on the $\text{Cl}^{-}$ ion. The authors also explored the transport of sodium ions and found passage times close to a millisecond.

\subsection{Rare-Switching Events}\label{section:rare-switching}

\noindent
All the rare events discussed so far take place at atomistic and mescoscopic scales. The intrinsic dynamics of the underlying systems can thus be described using different flavors of particle-based molecular simulation methods. There are, however, a wide variety of important rare events that occur in systems with differing intrinsic dynamics.  Since advanced sampling techniques such as FFS are indifferent towards the intrinsic dynamics of the underlying system, they can in principle be used for probing the kinetics and mechanism of such rare events. In this section, we will focus on two such processes studied using FFS, namely rare switching events in  biochemical  networks~\cite{AllenPRL2005, FrenkelFFS_JCP2006, MorelliBiophysJ2008, MorelliJChemPhys2008, MorelliPNAS2009, MorelliBiophysJ2011, ZhangCJChemPhys2012} and magnetic domains~\cite{VoglerPRB2013, VoglerJApplPhys2015, DesplatArxiv2019}. Both these systems can exist at multiple stable states separated by free energy barriers whose crossing requires the emergence of improbable fluctuations. Similar to other rare events discussed so far, such switching events can be characterized using suitable order parameters that can be constructed from mechanical observables of the underlying systems. Here, a mechanical observable refers to any property that can be unambiguously defined for any given configuration of the underlying system along a dynamic trajectory.

\subsubsection{Biochemical Networks and Genetic Switches}

\noindent 
Biochemical networks are typically comprised of multiple proteins or RNAs that can interact with some controlling DNA sequences, and can exist at multiple steady states with distinct biochemical phenotypes.  Characterizing the kinetics and mechanism of switching  between such states is critical to a molecular-level understanding of cell function. It is, in principle, possible to simulate such networks using particle-based methods. Due to the presence of a large number of interacting agents, however, such calculations will be insurmountably expensive, even if highly coarse-grained mesoscopic representations are utilized. As a result, alternative phenomenological representations are constructed to mimic the intrinsic dynamics of such networks. As mentioned in Section~\ref{section:validation}, such phenomenological descriptions can be combined with advanced sampling techniques such as FFS to probe the kinetics of switching in such networks. Indeed, FFS was historically developed for and validated by  accurately predicting the behavior of a simple biochemical network, namely a bistable genetic switch in which the expression of two genes that encode two proteins is controlled by the binding of the dimers of the same proteins to a controlling sequence, such as a promoter (Fig.~\ref{fig:validation-systems}A). In subsequent years, different aspects of this simple network were investigated using FFS. 

In one of the first standalone studies of  bistable genetic switches using FFS,  Morelli~\emph{et al.}~\cite{MorelliBiophysJ2008} investigated the effect of exclusivity on the kinetics and mechanism of switching. In particular, they considered two switch types, a \emph{general} switch in which the binding of each dimer type to the promoter does not block the attachment of the other type, and an \emph{exclusive} switch in which such secondary binding is not permitted. For both switch types,  switching  always becomes faster upon increasing the dimerization rate and decreasing the rate of dimer-DNA binding.  They, however, demonstrated that exclusive switches flip more frequently than general switches, and their stationary order parameter distributions (i.e.,~their free energy profiles) only depend on the dimer-DNA binding rate and does not change with altering the protein dimerization rate. In another paper~\cite{MorelliJChemPhys2008}, the same authors conducted a systematic investigation of \emph{coarse-graining}, or reformulating the network dynamics with the aim of including faster reactions implicitly, and demonstrated that while fast protein-protein interactions can be safely coarse-grained,  explicitly  including protein-DNA interactions are critical for maintaining the correct kinetics and mechanism of switching.  In upcoming years, the same authors published two more papers on genetic switches, with the aim of assessing the role of DNA looping in the lyzogen-to-lysis transition in  bacteriophage $\lambda$~\cite{MorelliPNAS2009}, and identifying strategies to represent the effect of macromolecular crowding~\cite{MorelliBiophysJ2011}. Bistable genetic switches have also been studied by other researchers as well in recent years~\cite{ZhangCJChemPhys2012}.

\subsubsection{Magnetic switches}

\begin{figure}
\centering
\includegraphics[width=0.45\textwidth]{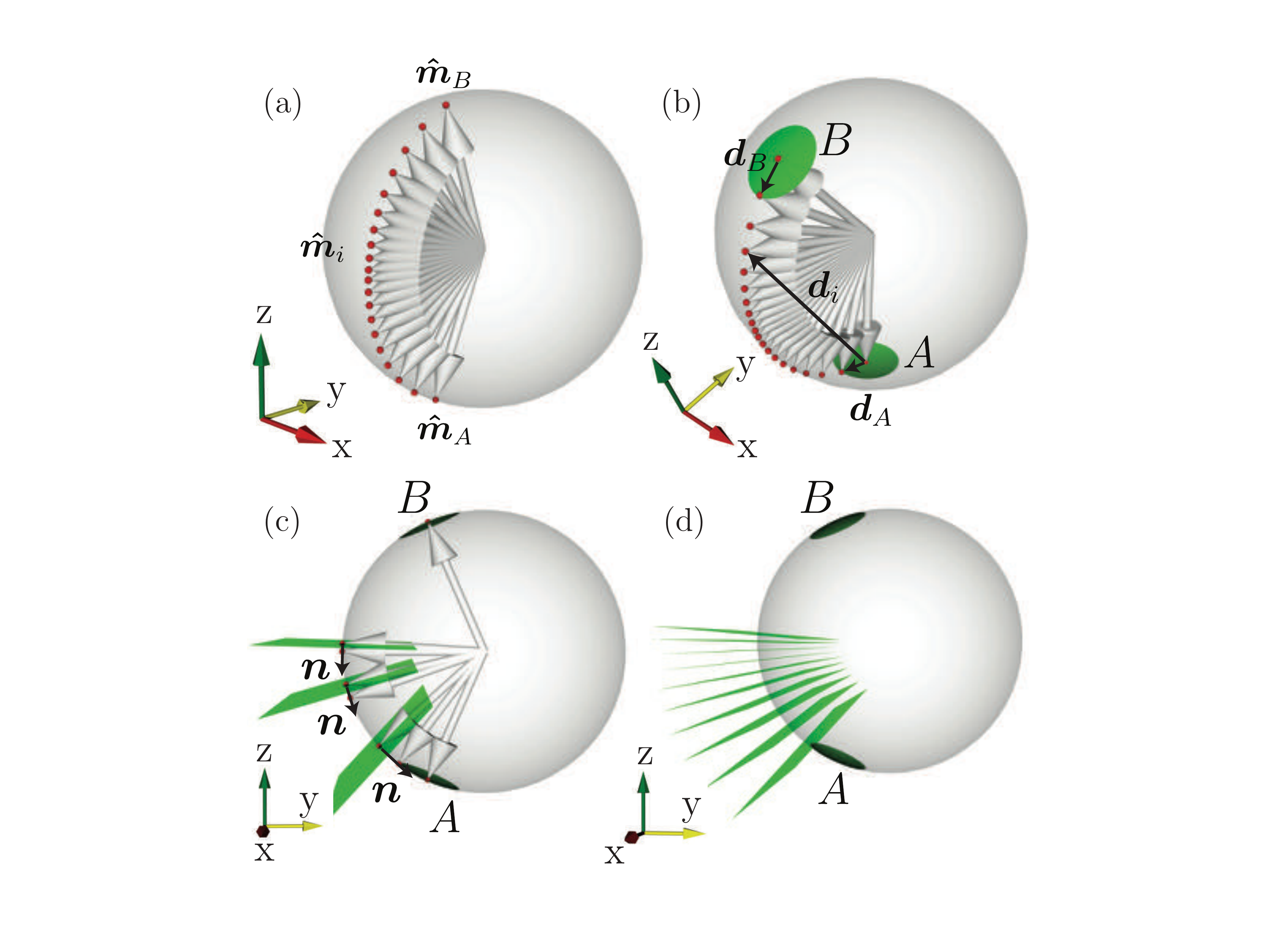} 
\caption{(Reproduced from Ref.~\citenum{VoglerPRB2013}) Schematic of (a) the minimum free energy path for switching from state $\mathbf{\hat{m}_A}$ to $\mathbf{\hat{m}_B}$ in a single macrospin particle with intermediate states $\mathbf{\hat{m}_i}$ obtained from NEB. Definition of (b) the stable basins A and B and (c) interfaces containing the magentization vector $\mathbf{\hat{m}_i}$, defined by normal $\mathbf{n_i}$. (d) Representative FFS interfaces.}
\label{fig:magnetic_switch} 
\end{figure}

\noindent The ability to predict and control the stability of magnetic systems is of considerable importance to many applications in which sustained magnetization is necessary, from magnetic recording devices to permanent magnets~\cite{CoffeyJAP2012}. In principle, magnetic materials are comprised of nano-domains with aligned magnetic spins. Such domains can rearrange as a result of external fields, or due to thermal fluctuations that can generate an effective instantaneous thermal magnetic field. The temporal evolution of such rearrangements is governed by the Landau-Lifshitz-Gilbert (LLG) equation with thermal fluctuations incorporated using Langevin dynamics~\cite{ScholzJMagnMagnMater2001}. In stable magnets, rearrangement times can be too long to be accessible to  conventional LLG simulations. As a result, several groups have utilized FFS to estimate switching timescales between different magnetization states of nm-scale magnetic domains simulated using the LLG framework. The first such study was by Vogler~\emph{et al.}~\cite{VoglerPRB2013}, who considered switching in a super-paramagnetic macrospin  nanoparticle, as well as a graded nanoparticle. In both systems, they utilized the nudged elastic band (NEB) method~\cite{HenkelmanJChemPhys2000} to identify the minimum energy path, which they then used for constructing a suitable reaction coordinate, as shown in Fig.~\ref{fig:magnetic_switch}.
 In the case of a single macrospin, they considered the switching of magnetization from the $-z$ to the $+z$ direction, in the presence of an external magnetic field along the $-y$ direction.At any given time $t$, they infer the crossing of a milestone $\mathbf{\hat{m}_{i}}$ from the sign of the scalar product $\mathbf{n_i}.[\mathbf{\hat{m}(t)} - \mathbf{\hat{m}_i}]$, where the normal $\mathbf{n_i}$ is defined as $\mathbf{n_i} = \mathbf{\hat{m}_i} - \mathbf{\hat{m}_{i-1}}$,~(Fig:~\ref{fig:magnetic_switch} A, C). The transition times computed from FFS were in excellent agreement with those estimated from the Kalmykov theory~\cite{KalmykovJAP2004} and direct Langevin simulation.  They further employed FFS for a more complex anisotropic domain and demonstrated the ability of FFS to efficiently sample transitions with barriers as high as $50k_BT$. In a subsequent paper, Volger~\emph{et al.}~\cite{VoglerJApplPhys2015} used the same methodology to investigate the effect of grain anisotropy on magnetic stability and concluded that anisotropy can enhance  kinetic stability by 12 orders of magnitude. A more recent study~\cite{DesplatArxiv2019} has utilized FFS to study magnetic collapse rates in skyrmions, and found excellent agreement with the transition state theory.

\section{Summary and Future Outlook}\label{section:conclusion}

\noindent In this perspective, we focus on the forward-flux sampling method, and its application to study a wide variety of rare events.  We  discuss the implementation details of the original FFS algorithm in Section \ref{section:ffs-overview}, and provide a detailed overview of more recent methodological developments in Section~\ref{section:variants:new}. Such newer FFS variants/extensions can be classified into four categories, those that expand the applicability of FFS to new systems and order parameters (Section~\ref{section:variants:new:expand}), those that provide recip\'{e}s for constructing optimal order parameters (Section~\ref{section:variants:new:op}), those that allow for extracting free energy profiles from FFS simulations (Section~\ref{section:variants:new:free-energy}), and those that enhance the efficiency and statistical precision of FFS (Section~\ref{section:variants:new:efficiency}). Finally, we discuss various software engineering efforts to streamline and automate the implementation of FFS variants in Section~\ref{section:implementation}.

After discussing the methodological aspects of FFS, we dedicate Section~\ref{Impact of FFS} to a detailed discussion of the plethora of rare events that have been studied using FFS. We broadly classify these numerous applications into several broad categories, namely nucleation (Section~\ref{section:nucleation}), collective conformational rearrangements in biomolecular systems (Section~\ref{section:conformation-bio}), structural relaxation and conformational rearrangements in polymers (Section~\ref{section:polymer-relaxation}), solute and ion transport in membranes (Section~\ref{section:filtration}) and rare switching events (Section~\ref{section:rare-switching}). Each broad theme is then broken into smaller categories, and constitutes a rich array of disparate phenomena that can be studied using path sampling techniques such as FFS. 
 For each application, we discuss the rationale for using a path sampling techniques such as FFS and the kinetic information and mechanistic insights obtained from such calculations. We also provide a brief discussion of important implementation details, such as the choice of FFS variants and order parameters. These examples reveal the depth and breadth of information that can be learnt from FFS, and provide a roadmap for expanding the application of FFS-like methods to new systems and rare events.

What makes FFS so broadly applicable is that it is agnostic towards the  the intrinsic dynamics of the underlying system as long as it is Markovian. This sets FFS apart even from  closely related path sampling techniques such as TIS that can only be used with reversible MD integrators. The accuracy and predictive power of any FFS calculation, however, will depend heavily on the physical fidelity and  efficiency of the employed intrinsic dynamics. In the context of molecular simulations, for instance, utilizing a physically realistic force-field that can be efficiently simulated for a sufficiently large number of relaxation times is critical for the accuracy of rates and mechanisms obtained from FFS. It is therefore necessary to note that the recent manyfold increase in applying FFS  to such a diverse range of rare events is due to several extrinsic factors, such as an increase in computational power,  development of accurate and efficient MD and MC engines, and  introduction of more accurate and efficient coarse-grained force-fields. Thanks to these developments, FFS has successfully contributed to a better understanding of several important rare events in materials science, soft condensed matter physics, and biology. This trend is only expected to accelerate in upcoming years, and as mentioned in Section~\ref{section:rare-switching}, there is no reason to believe that FFS cannot be applied to studies of rare events with unconventional dynamics, such as stock market crashes, earthquakes and extreme weather events. 

An emerging area of materials science and biology that can benefit a lot from FFS is active matter. Due to the irreversibility of employed dynamical algorithms, and the lack of a proper notion of a free energy, most standard advanced sampling techniques cannot be readily applied to active systems. FFS, however, can be easily applied to non-equilibrium systems, and can provide valuable kinetic and mechanistic information about rare events that occur under out-of-equilibrium conditions. Such processes are ubiquitous in materials science and biology, and understanding their kinetics is a new frontier that is yet to be fully explored. 

Despite these strengths, FFS still suffers from several important methodological shortcomings, and more work is needed to address all such issues. For instance, assessing the accuracy of FFS is still nontrivial, especially when it comes to multi-milestone correlations and precritical bottlenecks. The ensuing errors are particularly difficult to quantify since  FFS is typically applied under conditions at which the corresponding rare event cannot be readily  observed, e.g.,~in a computationally tractable MD simulation, and applying other advanced sampling methods might not always be feasible.  Another alternative is to conduct multiple independent replicas of the same calculation and conduct simple averaging. For many applications, however, even a single properly conducted FFS calculation is already too expensive~\cite{HajiAkbariPNAS2015, SossoJPhysChemLett2016, HajiAkbariPNAS2017, SossoChemSci2018}. There is therefore an urgent need to develop alternative rigorous approaches to characterize such correlations in FFS calculations.

Another algorithmic issue that has not been systematically studied is the factors that govern the  numerical convergence of an FFS calculation. As discussed in Section~\ref{section:FFS:temporally-cg}, it has been previously shown that altering parameters such as sampling window can hamper the convergence of FFS by promoting unphysical false positive crossing events. There are, however, no comprehensive set of heuristics for determining the factors that assure-- or doom-- the convergence of an FFS calculation. 

Another potential area of further exploration is the task of identifying and constructing quality order parameters and reaction coordinates for complex rare events. As discussed extensively throughout this perspective, choosing a good order parameter becomes challenging in systems with complex free energy landscapes, particularly when multiple independent  reaction pathways are available. Failing to construct good reaction coordinates can generally result in inaccurate rate estimates, and identification of incorrect or unrealistic mechanisms.  While several schemes\cite{BorreroJCP2007, PetersJCP2006, LechnerJCP2010, LechnerPRL2011} have been proposed for optimizing order parameters (such as the FFS-LSE method of Section~\ref{section:variants:new:op}), they all rely on starting from a collection of candidate collective variables, and combining them to construct a quality reaction coordinate. Selecting such a starting list is, however, usually based on physical intuition, which might be inadequate when the corresponding transition is poorly understood. This can in turn result in missing important CVs and the construction of suboptimal reaction coordinates. An alternative approach that has gained increased traction in recent years is to use big data and machine learning approaches, such as deep neural networks, to ``learn`` important-- but previously unidentified-- CVs. A recent work by Boattini~\emph{et al.}~\cite{EmanueleMolPhy2018} has explored the use of neural networks for estimating order parameters to classify structural motifs in binary hard sphere crystals. With the increase in computing power and the improvement of machine learning algorithms, this area of research is expected to garner further interest in upcoming years, and the plethora of trajectories and configurations collected during FFS can prove valuable in 'mining` such hidden-- but important-- collective variables. 

A key assumption of the original FFS algorithm is that the system can only access two basins within the free energy landscape. As discussed in numerous places throughout this perspective, this assumption is violated in many rare events, from nucleation to protein folding and DNA hybridization. This problem can, in principle, be remedied by carefully designing isolated FFS calculations between successive (meta)stable states if a single transition pathway exists between every two adjacent basin.  This approach will, however, be ineffective if the basins of interest are connected via a complex network of transition pathways.  This  very important possibility, however, had not been explored until recently~\cite{VijaykumarJCP2018} and further work is needed to generalize FFS-like approaches to such multi-state systems.

Finally, these prospective methodological developments will not result in broader application of FFS-- and other advanced sampling techniques-- to new systems and/or rare events unless sustainable and scalable workflows are developed for their straightforward implementation. In the area of conventional MD, this transformation has occurred over the last two decades, as researchers have moved away from developing their own MD code to using widely available community-based open-source MD packages, a transition that has not only increased efficiency, but has considerably enhanced the reproducibility of molecular simulations. One can argue that the same needs to happen for advanced sampling techniques, not only to expand their usage, but also to avoid  pitfalls and controversies~\cite{PalmerNature2014, PalmerJChemPhys2018} emerging from poor coding and data processing practices. As discussed in Section~\ref{section:implementation}, numerous efforts are already underway in this regard. But this will be the challenge of molecular simulations community in the upcoming decade, to incorporate all these emerging tools into a scalable intuitive workflow.

\begin{acknowledgments}
A.H.-A. gratefully ac- knowledges the support of the National Science Foundation CAREER Award (Grant No. CBET-1751971). 
\end{acknowledgments}

\bibliographystyle{apsrev}
\bibliography{References}

\end{document}